\documentclass[aps,prc,twocolumn,showpacs,floatfix,nofootinbib,preprintnumbers,superscriptaddress,amsmath,amssymb,longbibliography]{revtex4-1}
\usepackage{graphicx}
\usepackage{float}
\usepackage{dcolumn}
\usepackage{multirow} 
\usepackage{hyperref}
\usepackage{nicefrac}
\graphicspath{{.}}

\newcommand{\be}{\begin{equation}}
\newcommand{\ee}{\end{equation}}
\newcommand{\ba}{\begin{eqnarray}}
\newcommand{\ea}{\end{eqnarray}}

\begin{document}

\title{Pion-less effective field theory for atomic nuclei and lattice nuclei}

\author{A.~Bansal} \affiliation{Department of Physics and
   Astronomy, University of Tennessee, Knoxville, TN 37996, USA}

\author{S.~Binder} \affiliation{Department of Physics and
   Astronomy, University of Tennessee, Knoxville, TN 37996, USA}
 \affiliation{Physics Division, Oak Ridge National Laboratory, Oak
   Ridge, TN 37831, USA}

\author{A.~Ekstr\"om} \affiliation{Department of Physics, 
Chalmers University of Technology, SE-412 96 G\"oteborg, Sweden}
 \affiliation{Physics Division, Oak Ridge National Laboratory, Oak
   Ridge, TN 37831, USA}

 \author{G.~Hagen} \affiliation{Physics Division, Oak Ridge National
   Laboratory, Oak Ridge, TN 37831, USA} \affiliation{Department of
  Physics and Astronomy, University of Tennessee, Knoxville, TN
   37996, USA}
   
 \author{G.~R.~Jansen}
\affiliation{National Center for Computational Sciences, Oak Ridge National
Laboratory, Oak Ridge, TN 37831, USA}
\affiliation{Physics Division, Oak Ridge National
Laboratory, Oak Ridge, TN 37831, USA}

 \author{T.~Papenbrock} \affiliation{Department of Physics and
   Astronomy, University of Tennessee, Knoxville, TN 37996, USA}
 \affiliation{Physics Division, Oak Ridge National Laboratory, Oak
   Ridge, TN 37831, USA}

\begin{abstract}
We compute the medium-mass nuclei $^{16}$O and $^{40}$Ca using
pionless effective field theory (EFT) at next-to-leading order
(NLO). The low-energy coefficients of the EFT Hamiltonian are adjusted
to experimantal data for nuclei with mass numbers $A=2$ and $3$, or
alternatively to results from lattice quantum chromodynamics (QCD) at an
unphysical pion mass of 806~MeV. The EFT is implemented through a
discrete variable representation in the harmonic oscillator
basis. This approach ensures rapid convergence with respect to the
size of the model space and facilitates the computation of medium-mass
nuclei. At NLO the nuclei $^{16}$O and $^{40}$Ca are bound with
respect to decay into alpha particles.  Binding energies per nucleon
are $9-10$~MeV and $30-40$~MeV at pion masses of 140~MeV and 806~MeV,
respectively.
\end{abstract}


\maketitle 

\section{Introduction}

Pionless EFT is widely employed to describe the structure and
reactions of the lightest
nuclei~\cite{bedaque2002,hammer2010,griesshammer2012,konig2016}.
Variants of this EFT have also been applied to describe halo
nuclei~\cite{bertulani2002,higa2008,hammer2011,ryberg2014}, and dilute
Fermi gases~\cite{hammer2000}. Lattice nuclei, i.e. nuclei computed
from lattice QCD~\cite{beane2013}, can also be described in pionless
EFT~\cite{barnea2015,contessi2017}. In that approach, the relevant
low-energy coefficients (LECs) of the EFT are adjusted to data of
light nuclei computed with lattice QCD, and predictions are made for
heavier nuclei. Although present-day lattice QCD calculations of
nuclei use unphysically large pion masses, one might expect that
advances in that field will eventually allow us to tie nuclear
structure to QCD.

$^{16}$O is the heaviest nucleus computed in pionless EFT so far and
it was found to be unstable against break up into four $^4$He nuclei
at leading order (LO)~\cite{contessi2017}.  We are aware of only a few
applications of pionless EFT to nuclear structure calculations beyond
mass number $A\ge 4$: \citeauthor{platter2005} found that no
four-nucleon force is needed to describe $^4$He at LO. This result was
confirmed at NLO by \citeauthor{kirscher2010}; studies of heavier
helium isotopes are presented in
Refs.~\cite{kirscher2009,kirscher2015a}. Very recently,
\citeauthor{lensky2016} studied $^{3,4}$He at next-to-next-to-leading
order (N2LO).  \citeauthor{stetcu2007} computed $^6$Li at LO and found
it to be less bound than $^4$He.

In contrast to pionless EFT, chiral
EFT~\cite{vankolck1994,epelbaum2009,machleidt2011} has been used to
compute heavy nuclei up to the mass number $A=100$
region~\cite{binder2013,lahde2014,hagen2014,hagen2015,hergert2016,hagen2017,morris2017}. We
can only speculate about this discrepancy between chiral and pionless
EFTs. On one hand, it might be a concern that pionless EFT -- with a
breakdown scale around the pion mass $m_\pi\approx 140$~MeV -- cannot
be used to describe heavy nuclei with Fermi momentum $k_F\approx
270$~MeV.  On the other hand, the pion is still very massive compared
to the Fermi energy of about 40~MeV.  We also note that there could be
a mismatch in infrastructure. Many of the powerful nuclear quantum
many-body
solvers~\cite{dickhoff2004,navratil2009,barrett2013,hagen2014,hergert2016}
start from interactions in the harmonic oscillator basis, and matrix
elements for interactions from chiral
EFT~\cite{entem2003,nogga2004,hebeler2011,ekstrom2013,ekstrom2015} are
readily available in this basis. No similar and well established
infrastructure seems to exist for pionless EFT.

This paper has two goals. First, we want to study heavier nuclei such
as $^{16}$O and $^{40}$Ca with pionless EFT. We will adjust the LECs
of the EFT to both experimental data of light nuclei and to data from
lattice QCD. Second, we want to formulate pionless EFT directly in the
harmonic oscillator basis. This project was started
by~\citeauthor{stetcu2007} (with several applications to harmonically
trapped systems~\cite{stetcu2010,rotureau2010,tolle2011}), and a
formulation involving energy-dependent potentials is pursued by Haxton
and coworkers~\cite{haxton2000,haxton2007,haxton2008}.  Recently,
\citeauthor{binder2016} and~\citeauthor{yang2016} used the $J$-matrix
approach~\cite{heller1974,shirokov2004} to directly construct EFT
potentials in the oscillator basis.  Here, we follow and extend the
work of Ref.~\cite{binder2016} and formulate pionless EFT as a
discrete variable representation
(DVR)~\cite{harris1965,dickinson1968,light1985,baye1986}. A hallmark
of the present work is that the finite oscillator space itself becomes
the regulator, and no external regulator functions are
employed. Similar to nuclear lattice EFT~\cite{lee2009}, this
implementation tailors the EFT to the employed basis and thereby
facilitates the computations of Hamiltonian matrix elements and
nuclei.

Unfortunately, the computation of light nuclei in lattice QCD is not
without controversy, and there is no consensus whether nuclear
binding increases or decreases with increasing pion mass. The
calculations in Refs.~\cite{beane2012,beane2013,orginos2015,chang2015} infer
bound-state energies from plateaus in the time propagation and find
that nuclear binding increases with increasing pion mass. In contrast,
the calculations in Refs.~\cite{ishii2007,aoki2010,aoki2012} construct a
potential from a Bethe-Salpeter wave function and find that lattice
nuclei (computed at unphysically large pion masses) are less bound
than real nuclei~\cite{inoue2015}. Both approaches have been used as
input for the computation of increasingly heavier
nuclei~\cite{contessi2017,mcilroy2017}. In this work, we follow
Refs.~\cite{barnea2015,contessi2017} and use the lattice QCD results
of Ref.~\cite{beane2013} as input to constrain the LECs of our EFT.

This paper is organized as follows. In Sect.~\ref{Vdetails} we tailor
pionless EFT interactions to the harmonic oscillator basis using a
DVR.  In Sect.~\ref{sec:LECs} we discuss the fitting procedure used to
constrain LECs to data and lattice data, and present results for
$A=3,4$ nuclei for a range of ultraviolet (UV) cutoffs.  We use the
NLO interactions to compute atomic and lattice $^{16}$O and $^{40}$Ca
nuclei in Sect.~\ref{sec:heavierA}.  A summary of this paper is given
in Sect.~\ref{sum}. The formulation of the EFT in the harmonic
oscillator basis involves several technical elements and many
checks. For the purpose of readability this information is presented
in a number of Appendices.

\section{Pionless effective field theory in the oscillator basis}
\label{Vdetails}

\subsection{Pion-less EFT} 
We briefly introduce pionless EFT and refer the reader to the
reviews~\cite{vankolck1999,bedaque2002,hammer2013} for details on
this extensive subject. In pionless EFT, neutrons and protons are the
relevant degrees of freedom, and the breakdown scale is given by 
the pion mass. Using naive dimensional
analysis, nucleon-nucleon ($NN$) interactions in momentum space are
\begin{align*}
V_{NN}^{\rm{LO}}(\vec{p}',\vec{p}) &= {C}_{S} + {C}_{T} \vec{\sigma_{1}}\cdot\vec{\sigma_{2}}\\
V_{NN}^{\rm{NLO}}(\vec{p}',\vec{p}) &= {\rm{C}}_{1}{q}^{2} + {\rm{C}}_{2}{k}^{2} \\
&+ (C_3 {q}^{2} +C_4 {k}^{2})\vec{\sigma_{1}}\cdot\vec{\sigma_{2}} \\
&-i C_5\frac{\vec{\sigma}_1 +\vec{\sigma}_2}{2}\cdot\left(\vec{q}\times\vec{k}\right)\\
&+C_6 (\vec{\sigma_{1}}\cdot\vec{q})(\vec{\sigma_{2}}\cdot\vec{q})\\
&+C_7 (\vec{\sigma_{1}}\cdot\vec{k})(\vec{\sigma_{2}}\cdot\vec{k}) .
\end{align*}
Here $\vec{p}'$ and $\vec{p}$ are the outgoing and incoming relative
momenta, respectively, and we use the shorthands $\vec{q} = \vec{p} -
\vec{p'}$, $\vec{k} = (\vec{p'} + \vec{p})/2$ for the momentum
transfer and the average momentum, respectively. The LECs are denoted
as $C_i$.

Large scattering lengths in the singlet and triplet $S$ waves, due to
a weakly bound deuteron and almost bound di-neutron, reflect the
existence of another small momentum scale denoted by $\aleph \approx
40$~MeV, and lead to the Kaplan-Savage-Wise (KSW) power
counting~\cite{kaplan1998}. In the singlet and triplet $S$ partial
waves the LO LECs are proportional to the respective scattering
lengths, i.e. they scale as $1/\aleph$ instead of $1/m_\pi$, which was
expected otherwise as the pion mass sets the breakdown scale.  The
unnatural size of both $S$ wave LECs (with respect to the expected
scaling $1/m_\pi$) results in a different treatment of their NLO
correction. Therefore, remaining interactions that enter at NLO in
naive dimensional analysis get demoted to N2LO in KSW counting. At NLO
pionless EFT involves only $S$ waves with the LO potentials
\begin{align}
V_{NN}^{\rm LO}({^1{S}_{0}}) &= \tilde{C}_{^1S_{0}} =  C_S - 3{C}_{T} , \nonumber \\
V_{NN}^{\rm LO}({^3{S}_1}) &= \tilde{C}_{^3{S}_1} =  C_S  + C_T \nonumber ,
\end{align}
and the NLO potentials 
\begin{align}
  \label{potNLO}
V_{NN}^{\rm NLO} ({^1{S}_0}) &= C_{^1{S}_0}(p^2 + {p'}^2) , \nonumber\\
V_{NN}^{\rm NLO} ({^3{S}_1}) &= C_{^3{S}_1}(p^2 + {p'}^2) . 
\end{align}
Pionless EFT can be used to reproduce the deuteron binding energy
and the effective range expansion for $NN$ scattering
\begin{equation}
  \label{ere}
k\cot\delta_{0}(k) = -\frac{1}{a_{0}} + \frac{1}{2}r_{0}k^{2}  + \ldots .
\end{equation}
This defines the $S$-wave scattering length $a_{0}$ and effective range
$r_{0}$. Pionless EFT yields the scattering length
at LO, and the effective range at NLO.

To renormalize the three-nucleon system, the three-nucleon force ($NNN$)
is promoted to LO~\cite{bedaque1999}. There are many equivalent ways
to write this contact~\cite{vankolck1994,epelbaum2002}, and we use
\begin{equation*}
V_{NNN} = \frac{c_E}{F_{\pi}^4 \Lambda_{\chi}} \sum\limits_{j\ne i}\vec{\tau_{i}}\cdot\vec{\tau_{j}} .
\end{equation*}
Here $\Lambda_{\chi} = 700$ MeV and $F_{\pi} = 92.4$ MeV are constants
(employed in chiral EFT) that make $c_E$ dimensionless; we include these
for convenience only.  Summarizing, the complete LO interaction is given by
\be
\label{potLO}
V_{\rm{LO}} = V_{NN}^{\rm LO}({^1{S}_0}) +  V_{NN}^{\rm LO}({^3{S}_1}) + V_{NNN} .
\ee

The full NLO potential consists of the terms~(\ref{potNLO}) added to
the LO potential~Eq. (\ref{potLO}). We will solve the NLO potential
with a non-perturbative method, as done previously, for instance, in
Ref.~\cite{kirscher2010,lensky2016}. The reason is as follows. At LO,
nuclei such as $^{6}$Li~\cite{stetcu2007} and
$^{16}$O~\cite{contessi2017} are unbound with respect to
$\alpha$-particle emission. Thus, no finite-order perturbation theory
will yield bound-state wave functions. The applications of
non-perturbative method might be valid only for UV cutoffs that are
not too large, and we will limit the range of cutoffs to up to about
700~MeV. Larger cutoffs are discussed in App.~\ref{app-wigner} in
connection with the Wigner bound.  For a perturbative treatment of the
three-nucleon systems, we refer the reader to Ref.~\cite{konig2016}.

In this work, we compute nuclei such as $^{16}$O and $^{40}$Ca. This
requires us to be judicious about the basis we want to employ. Very
recently, \citeauthor{binder2016} showed that EFTs can be formulated
in the harmonic oscillator basis, and they performed converged
calculations for heavy nuclei based on $NN$ interactions alone. In
what follows, we briefly review the essential ingredients of this
approach.

\subsection{Discrete Variable Representations}

A finite harmonic oscillator basis imposes infrared (IR) and UV
cutoffs~\cite{stetcu2007,hagen2010b,jurgenson2011,coon2012,furnstahl2012}.
These correspond to hard-wall boundary conditions in position and
momentum space, respectively. They depend on the maximum number of
oscillator quanta $N$ included in the basis and on the oscillator
length
\be
\label{bosc}
b \equiv \sqrt{\hbar/(\mu\omega)} . 
\ee
Here, $\mu$ is the reduced mass for two-nucleon system, 
and $\hbar\omega$ is the oscillator spacing.
In position space, the effective
hard wall is located at the radius~\cite{more2013}
\be
\label{L}
L = \sqrt{2(N+3/2+2)}b ,
\ee
while in momentum space the radius $\Lambda$ defining the UV cutoff is
given by~\cite{konig2014}
\be
\label{lambdaUV}
\Lambda =  \sqrt{2(N+3/2+2)}\hbar/b .
\ee
For many-body systems, similar expressions were derived in
Refs.~\cite{furnstahl2014b,wendt2015}. The effective hard wall in
position space modifies the asymptotic tail of bound-state wave
functions and introduces -- akin to \citeauthor{luscher1985}'s formula
-- a correction to bound-state energies and other
observables~\cite{furnstahl2012,odell2016,acharya2017}.

We will formulate pionless EFT in a spherical harmonic oscillator
basis. The radial basis functions at orbital angular momentum $l$ are
\begin{eqnarray}\label{harmonic oscillatorwavefunction_r}
  \lefteqn{\psi_{n,l}(r) = }\nonumber\\
  &&(-1)^{n}\sqrt{\frac{2n!}{\Gamma{(n+l+3/2)}b^{3}}}{\bigg{(}\frac{r}{b}\bigg{)}}^{l}e^{-\frac{1}{2}\frac{r^{2}}{b^{2}}}L_{n}^{l+\frac{1}{2}}\bigg{(}\frac{r^{2}}{b^{2}}\bigg{)}
\end{eqnarray}
in position space, and 
\begin{equation}\label{harmonic oscillatorwavefunction}
\tilde{\psi}_{n,l}(k) = \sqrt{\frac{2n!b^{3}}{\Gamma{(n+l+3/2)}}}{(kb)}^{l}e^{-\frac{1}{2}k^{2}b^{2}}L_{n}^{l+\frac{1}{2}}(k^{2}b^{2})
\end{equation}
in momentum space. Here, $L_n^{l+ 1/2}$ denotes the generalized
Laguerre polynomial. The finite basis consists of all states with $2n+l \le N$.
At fixed $l$, we employ the shorthand
\be
N_l \equiv (N -l)/2 
\ee
for the maximum radial quantum number.

\begin{figure}[H]
\includegraphics[width=0.48\textwidth]{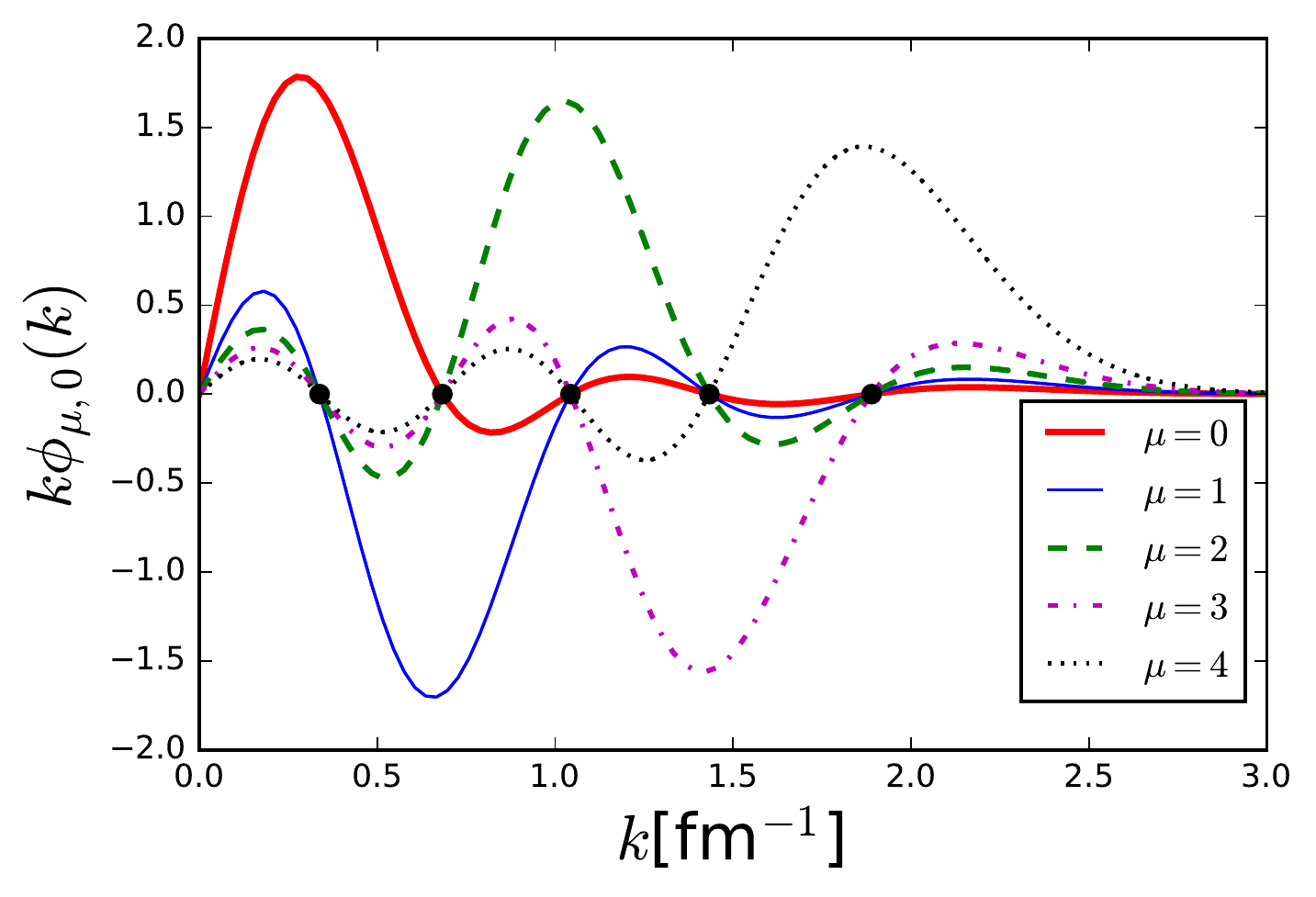}
\caption{(Color online) The $S$ wave eigenfunctions $\phi_{\mu,0}(k)$
  of the squared momentum operator [plotted as $k\phi_{\mu,0}(k)$] for
  $\mu = 0,1,...N_0$) corresponding to discrete momentum eigenvalues,
  shown as a function of momentum for a finite harmonic oscillator
  basis with $N = 8$, $\hbar\omega = 22$~MeV. The solid black dots on
  the $x$ axis indicate the discrete momentum eigenvalues.}
\label{fig1}
\end{figure}

For EFT applications in a finite harmonic oscillator basis it is
useful to replace the oscillator basis functions by the eigenfunctions
$\phi_{\mu,l}(k)$ of the squared momentum operator, because the latter
constitute a DVR.  For an introduction to DVRs we refer the reader to
some of the original
works~\cite{harris1965,dickinson1968,light1985,baye1986} and to the
reviews~\cite{littlejohn2002,light2007}. In the present paper, we
follow the notation of Ref.~\cite{binder2016}. Figure~\ref{fig1} plots
the $S$-wave DVR basis functions $\phi_{\mu,0}(k)$ with $\mu =
0,1...,N_{0} $ for the oscillator model space $N = 8$, $\hbar\omega =
22$~MeV.

The salient feature of a DVR is that the orthogonal basis functions
are localized around their corresponding eigenvalues and zero at other
eigenvalues.  The discrete momentum eigenvalues $k_{\mu,l}$, shown as
dots in Fig.~\ref{fig1}, fulfill $L_{N_l+1}^{l+
  1/2}(k^{2}_{\mu,l}b^{2})=0$.  In App.~\ref{app-dvr} we consider
other DVRs in the oscillator basis that are based on a different set
of discrete momentum points.

There are many ways to express the DVR wave functions. The expression
\be
\label{phimu}
\tilde{\phi}_{\mu,l}(k) = \langle k,l|\phi_{\mu,l} \rangle = \frac{k_{\mu,l}/b}{{k_{\mu,l}^2} - {k^{2}}} \tilde{\psi}_{N_l+1,l}(k)
\ee
immediately reflects the key DVR property
\begin{equation}
\label{key}
  \tilde{\phi}_{\nu,l}(k_{\mu,l}) = \delta_{\nu}^{\mu}{c_{\nu,l}^{-1}} .
\end{equation}
Here, 
\be
\label{cmu}
c_{\mu,l} \equiv \frac{k_{\mu,l }b}{\sqrt{(N_l+1)(N_l+l+3/2)}\tilde{\psi}_{N_l,l}(k_{\mu,l})}
\ee
is a normalization constant.  Alternatively, the expression
\begin{equation}
  \label{eqn:1}
\tilde\phi_{\mu,l}(k) = c_{\mu,l}\sum_{n = 0}^{N_l}\tilde{\psi}_{n,l}(k_{\mu,l})\tilde{\psi}_{n,l}(k) 
\end{equation}
exhibits the expansion in terms of the harmonic oscillator basis functions.

In the DVR, scalar products of wave functions $f(k)$ and $g(k)$ with
angular momentum $l$ are defined as
\be
\label{scapro-dvr}
\langle f|g\rangle_{\rm DVR} \equiv \sum_{\mu=0}^{N_l} c_{\mu,l}^2 f^*(k_{\mu, l}) g(k_{\mu,l}). 
\ee
This overlap results from employing $(N_l+1)$-point Gauss
Laguerre-quadrature in the computation of the exact scalar product
\be
\label{scapro-ex}
\langle f|g\rangle \equiv \int\limits_0^\infty dk k^2 f^*(k) g(k) .
\ee
Thus, $\langle f|g\rangle_{\rm DVR} = \langle f|g\rangle$ for
functions $f$ and $g$ that are spanned by the finite harmonic
oscillator space. In other cases, the scalar product in
Eq.~(\ref{scapro-dvr}) is an approximation of Eq.~(\ref{scapro-ex})
\cite{binder2016}. We note that this approximation is consistent with
EFT ideas as it neglects high-momentum contributions. In this paper,
we will frequently evaluate matrix elements of operators in the
DVR. In such cases, the subscript DVR will appear on the operator. As
we will see, the DVR yields simple expressions for matrix elements of
interactions and currents from EFT because the latter are usually
expressed in momentum space.

The DVR basis states $|\phi_{\mu, l}\rangle$ are related to the wave
functions~(\ref{harmonic oscillatorwavefunction_r}) and (\ref{harmonic
  oscillatorwavefunction}) via the definitions
\ba
\phi_{\mu,l}(r) &\equiv& \langle r,l |\phi_{\mu,l}\rangle , \nonumber\\
\tilde{\phi}_{\mu,l}(k) &\equiv& \langle k,l |\phi_{\mu,l}\rangle . 
\ea
Given the momentum-space matrix element $V(k',l';k, l)\equiv\langle k',
l'|\hat{V}|k, l\rangle$ in the partial-wave basis, we have in the DVR, 
\begin{equation}
  \label{eqn:2}
  \langle \phi_{\nu,l'}|\hat{V}_{\rm DVR}|\phi_{\mu,l} \rangle=
  c_{\nu,l'}c_{\mu,l} V(k_{\nu,l'},l';k_{\mu,l},l).
\end{equation}

Thus, the computation of matrix elements is very convenient in the DVR
basis (as it is merely a function call) once the EFT interaction is
available in the partial wave basis.  We also note that the momentum
space matrix elements of the DVR interaction ${\langle k', l'|V_{\rm
    DVR}|k,l\rangle}$ agree with the original interaction
$V(k',l';k,l)$ at the DVR momentum points $(k',k)=(k_{\mu, l'},k_{\nu,
  l})$ with $\nu=0,\ldots N_{l'}$ and $\mu=0,\ldots N_l$. One can
therefore ask to what extent does the resulting interaction, i.e., the
left-hand side of Eq.~(\ref{eqn:2}), preserve the low-momentum or IR
properties of $V(k',l';k,l)$? To explore this question we express the
momentum space matrix elements of the DVR interaction as
\ba
\label{dvr-kk}
\lefteqn{\langle k', l'|V_{\rm DVR}|k,l\rangle} \nonumber\\
&&=\sum_{\mu=0}^{N_l}\sum_{\nu=0}^{N_{l'}}
\langle \phi_{\nu,l'}|\hat{V}_{\rm DVR}|\phi_{\mu,l} \rangle
\tilde{\phi}_{\nu, l'}(k') \tilde{\phi}_{\mu, l}(k) \\
&&= \sum_{\mu=0}^{N_l}\sum_{\nu=0}^{N_{l'}} c_{\mu, l}c_{\nu,l'} \tilde{\phi}_{\nu, l'}(k') \tilde{\phi}_{\mu, l}(k) V(k'_\nu,l';k_\mu,l) .\nonumber
\ea
This shows that the low-momentum expression of the left-hand side is a
superposition of matrix elements. Though the IR cutoff of the basis is
$k_{0,l}$ at angular momentum $l$, the interaction does not vanish for
$k, k'<k_{0,l}$. In what follows, we will therefore improve its IR
behavior. Although the contribution from the interactions at low
momentum are reduced by the integration measure $dk k^2$ when it acts
on wave functions, the incorrect IR behavior raises questions
regarding the effective-range expansion of the DVR potential.

\subsection{IR improvement of the $NN$ interaction}

\begin{figure}[H]
\includegraphics[width=0.48\textwidth]{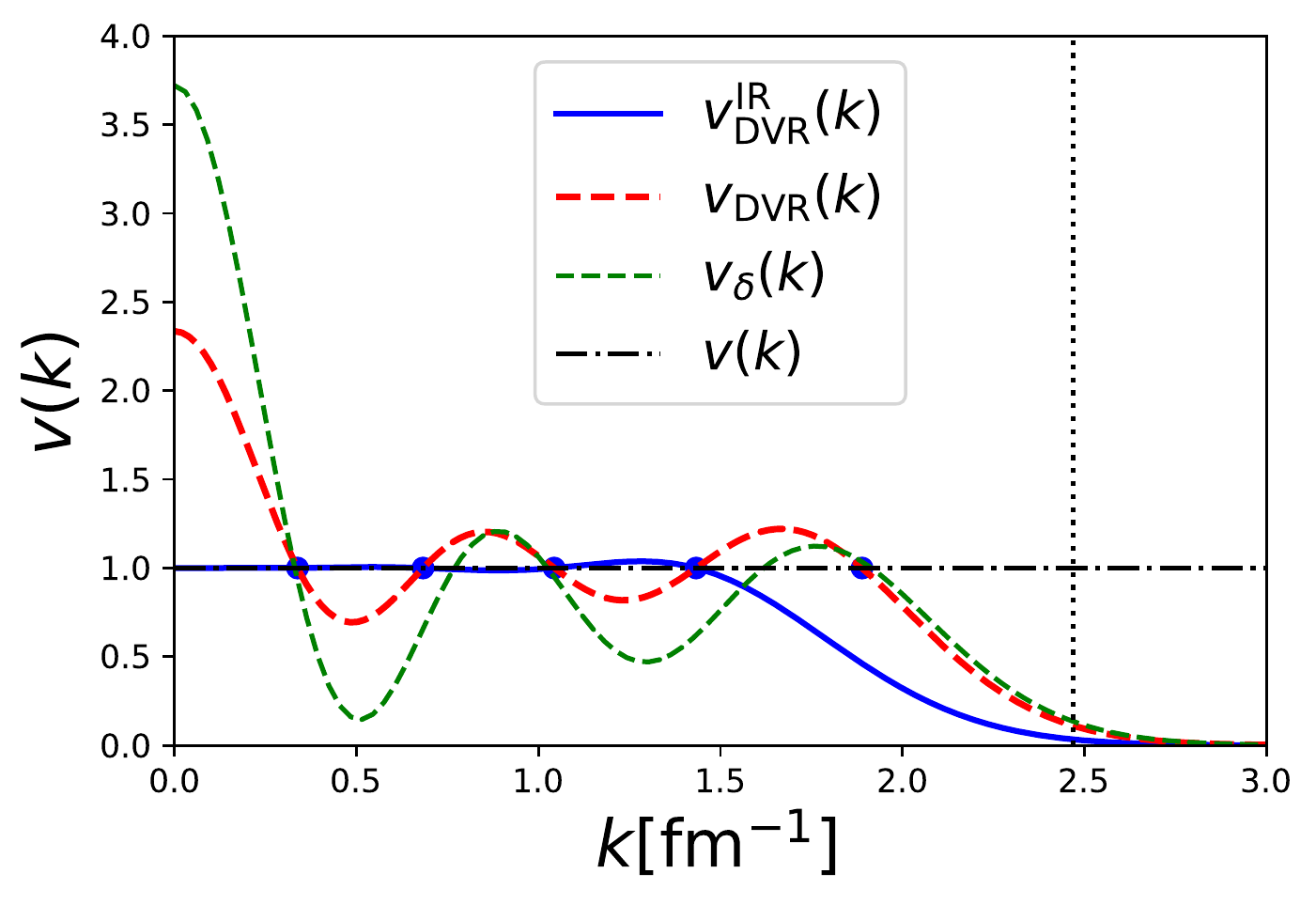}
\caption{(Color online) The dashed red (solid blue) curve shows the
  contact interaction in the DVR basis (with IR improvement) to be
  compared with the original momentum-space interaction $v(k)=1$ shown
  as a dash-dotted black line. The thin green dashed curve shows the
  contact (i.e. a $\delta$-function) in a finite harmonic oscillator
  basis with $N = 8, \hbar\omega = 22$~MeV, $l = 0$ for reference. The
  solid blue dots represent the DVR momenta.  The dotted black line
  marks the location of the UV cutoff introduced by the finite
  oscillator basis.}
\label{fig:4}
\end{figure}

Let us consider the case of a $NN$ contact
\begin{equation}
  \label{VLO}
V(k',l'=0;k,l=0) = C_{\rm{LO}} , 
\end{equation}
where $C_{\rm{LO}}$ is the coupling strength. The corresponding DVR interaction is
\be
\label{vlodvr}
\langle k', 0|V_{\rm DVR}^{\rm IR}|k,0\rangle =C_{\rm LO} v_{\rm DVR}(k') v_{\rm DVR}(k)
\ee
with
\be
\label{v-dvr}
v_{\rm DVR}(k) \equiv \sum_{\mu=0}^{N_0} c_{\mu, 0}\tilde{\phi}_{\mu, 0}(k) . 
\ee
Clearly, the DVR interaction differs from the original
potential~(\ref{VLO}), which we now rewrite as $C_{\rm LO}v(k)v(k')$
with $v(k)=1$. Figure~\ref{fig:4} shows $v(k)=1$ as the horizontal
dash-dotted line, and the DVR result $v_{\rm DVR}$ as the dashed red
line. The discrete DVR momenta are shown as solid dots. We see that
$v_{\rm DVR}$ coincides with the original $v(k)$ only at these
momenta, as expected for a DVR.  The $\delta$ function, evaluated
exactly in the oscillator basis, is shown as $v_\delta$.  It exhibits
the strongest oscillations (and particularly large deviations at small
momenta) from $v(k)=1$.  The vertical dotted line indicates the UV
cutoff in Eq.~(\ref{lambdaUV}); as expected $v_{\rm DVR}$ rapidly
vanishes here.  Regarding the IR properties of the DVR interaction, we
find that $v(k)$ and $v_{\rm DVR}(k)$ are indeed very different at
lowest momenta. This is not unexpected: The finite oscillator basis
introduces an IR cutoff (set by the smallest discrete momentum), and
thus one has no control for small momenta. We will correct this in
what follows.

To improve the IR behavior, we return to Eq.~(\ref{v-dvr}). This
function is a superposition of functions $\tilde{\phi}_{\mu, 0}(k)$
localized around $k\approx k_{\mu,0}$, and with weights $c_{\mu,
  0}$. The key idea is to force this function to have the value 1 at
$k = 0$ by altering the weight $c_{N_0,0}$ of the highest-momentum DVR
function $\tilde{\phi}_{N_0, 0}(k)$. This is in the EFT spirit,
because we improve the accuracy at low momentum at the cost of
possible loss of accuracy at high momentum. Thus, we define new
coefficients
\ba
\label{IRcmu}
\bar{c}_{\mu,0} &\equiv& c_{\mu,0} , \quad\mbox{for $\mu = 0,\ldots,N_0-1$} \nonumber\\        
\bar{c}_{N_0,0} &\equiv&\left(1 - \sum\limits_{\nu = 0}^{N_0-1} \tilde{\phi}_{\nu,l}(0)c_{\nu,0} \right)/{\tilde{\phi}_{N_0,0}(0)}
\ea
and consider the IR improved DVR potential
\ba
\label{v-dvr-lo}
v_{\rm DVR}^{\rm IR}(k) =
C_{\rm LO} \sum_{\mu=0}^{N_0} \bar{c}_{\mu,0}\tilde{\phi}_{\mu,0}(k).
\ea
By construction, it fulfills $v_{\rm DVR}^{\rm IR}(k)=1$ for discrete
momenta $k \in\{0,k_{0,0},\ldots,k_{N_0-1,0}\}$. The IR improved
contact is shown as the solid blue line in Fig.~\ref{fig:4}.  The IR
improvement at $k=0$ is obvious, and the oscillations are reduced
substantially.  In App.~\ref{app-er} we show that the curvature of
$v_{\rm DVR}^{\rm IR}(k)$ at $k=0$ decreases as $N^{-1}$ as the basis
size is increased. Thus, effective range corrections are suppressed,
as expected from a proper EFT. Summarizing, the IR improved DVR
contact interaction in momentum space is
\be
\langle k', 0|V_{\rm DVR}^{\rm IR}|k,0\rangle =C_{\rm LO}v_{\rm DVR}^{\rm IR}(k')
v_{\rm DVR}^{\rm IR}(k).
\ee

\begin{figure}[H]
\includegraphics[width=0.48\textwidth]{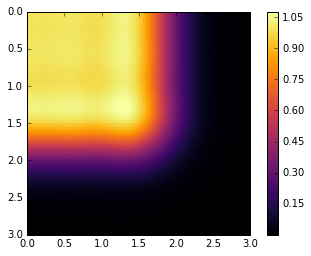}
\caption{(Color online) The IR improved DVR contact interaction $V_{\rm DVR}^{\rm IR}$
    plotted in momentum space. The axes represent momentum in units of 
  fm$^{-1}$.}
\label{fig:LOmplot}
\end{figure}

Figure~\ref{fig:LOmplot} shows this interaction as a matrix in
momentum space.  The interaction is very smooth and almost constant,
and rapidly approaches zero at the UV cutoff of the finite harmonic
oscillator basis with $N = 8$, $\hbar\omega = 22$~MeV. Thus, IR
improvement allows us to generate interactions with an accurate IR
behavior even for momenta that are much smaller than the IR cutoff
of the finite harmonic oscillator basis.

\begin{figure}[H]
\includegraphics[width=0.48\textwidth]{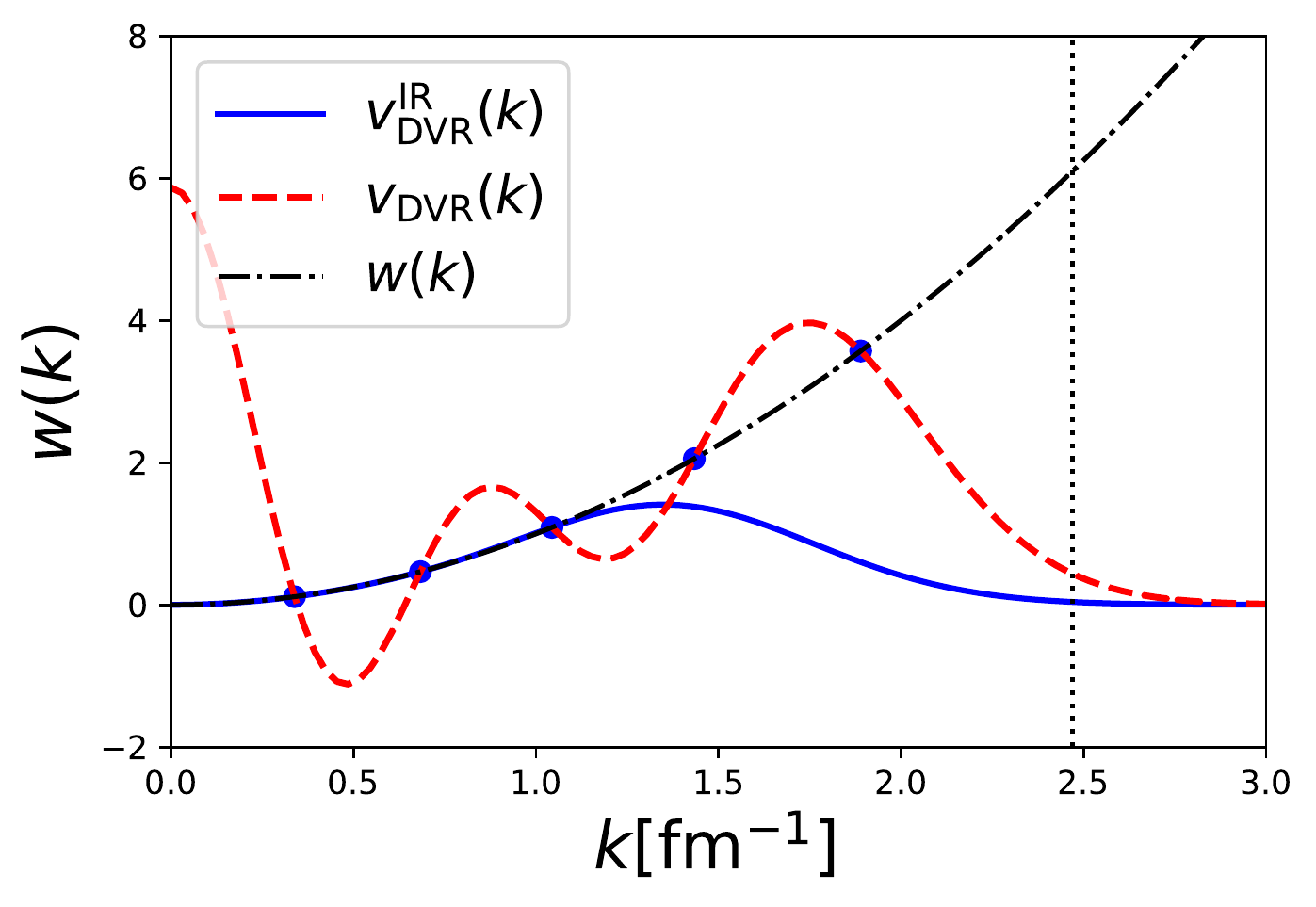}
\caption{(Color online) The solid blue (dashed red) curve shows the
  NLO interaction term tailored to finite harmonic oscillator basis
  through DVR with (without) IR improvement. The solid blue dots
  represent discrete momentum eigenvalues in the model space $N = 8,
  \hbar\omega = 22$~MeV, and $l = 0.$ The dotted black line depicts
  sharp cutoff $\Lambda$ introduced by finite harmonic oscillator
  basis and the dash-dotted black line plots the interaction in
  continuous momentum basis.}
\label{fig:5}
\end{figure}

We now turn to the IR improvement of the NLO interaction   
\begin{equation}
  \label{vnlo}
V(k',l'=0;k,l=0) = C_{\rm{NLO}}\left[w(k) +w(k')\right]
\end{equation}
with
\be
w(k)\equiv k^2. 
\ee
Here, $C_{\rm{NLO}}$ is the coupling strength. The interaction is no
longer separable, and the DVR interaction has momentum-space matrix
elements
\be
\langle k',0|\hat{V}_{\rm DVR}|k, 0\rangle  = C_{\rm{NLO}}\left[w_{\rm DVR}(k) +w_{\rm DVR}(k')\right] 
\ee
with
\be
\label{w-dvr}
w_{\rm DVR}(k) = \sum_{\mu=0}^{N_0} c_{\mu,0} k_{\mu,0}^2 \tilde{\phi}_{\mu,0}(k). 
\ee
Figure~\ref{fig:5} shows the functions $w(k)$ and $w_{\rm DVR}(k)$ as
the dash-dotted black and dashed red line, respectively. They coincide at the DVR
points (shown as dots). 

\begin{figure}[H]
\includegraphics[width=0.48\textwidth]{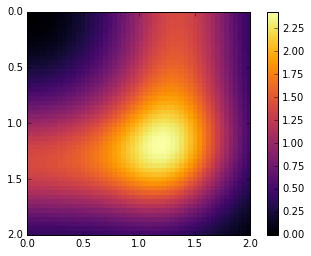}
\caption{(Color online) Momentum space matrix elements $V(k,k') = (k^2
  + {k'}^2)$ for $S$ waves in the model space $N = 8$ and $\hbar\omega
  = 22$~MeV as a function of the momenta $k$ and $k'$ after IR
  improvement. }
\label{fig:NLOmplot}
\end{figure}

It is clear that $w_{\rm DVR}$ has the wrong value and the wrong
curvature at $k=0$. This can be corrected by effectively changing the
values of the coefficients $c_{N_0-1,0}$ and $c_{N_0,0}$, i.e., the
DVR-improved function becomes
\be
\label{w-dvr-ir}
w_{\rm DVR}^{\rm IR}(k) = \sum_{\mu=0}^{N_0} \bar{c}_{\mu,0} k_{\mu,0}^2 \tilde{\phi}_{\mu,l}(k) 
\ee
with 
\ba
\label{cbarnlo}
\bar{c}_{\mu,0} &=&  
c_{\mu,0},  \quad\mbox{for $\mu = 0,\ldots, N_0-2$} \nonumber\\
\bar{c}_{N_0-1,0} &=&
\frac{\sum \limits_{\nu = 0}^{N_0-2} \tilde{\phi}_{\nu,0}(0) c_{\nu,0} \left(k_{N_0,0}^{2} - k_{\nu,0 }^2\right)  - k_{N_0,0}^2}{\tilde{\phi}_{N_0-1,0}(0) \left(k_{N_0-1,0}^2 - k_{N_0,0}^2\right)} \\
\bar{c}_{N_0,0} &=&\frac{\sum\limits_{\nu = 0}^{N_0-2} \tilde{\phi}_{\nu,0}(0) c_{\nu,0} \left(k_{N_0-1,0}^2 - k_{\nu,0 }^2\right) - k_{N_0-1,0}^2}{\tilde{\phi}_{N_0,0}(0)\left(k_{N_0,0}^2 - k_{N_0-1,0}^2\right)} \nonumber
\ea
The function $w_{\rm DVR}^{\rm IR}$ from Eq.~(\ref{w-dvr-ir}) is shown
as a solid blue line in Fig.~\ref{fig:5}. It agrees at $N-2$ DVR points with
$w(k)$ and has the correct IR behavior. The IR improved interaction
has matrix elements
\begin{equation}
  \langle \phi_{\mu, 0}|V_{\rm DVR}^{\rm IR}|\phi_{\nu,0}\rangle= C_{\rm NLO} \bar{c}_{\mu,0} \bar{c}_{\nu,0}
  \left(k_{\mu,0}^2 + k_{\nu,0}^2\right) , \nonumber
\end{equation}
and these are shown in Fig.~\ref{fig:NLOmplot}. It is clear that the
IR improvement can be extended to more general interactions.

\subsection{IR improvement of the  $NNN$ contact}
\label{subsec:3nf}
We consider the three-body contact
\begin{equation}
  \label{3nf}
V(k',p';k,p) = C_{NNN} 
\end{equation}
with its LEC $C_{NNN}$. The momenta $k,k'$ denote the incoming and
outgoing relative momentum between particles 1 and 2, respectively,
while $p,p'$ are the incoming and outgoing momentum of particle 3
relative to the center of mass of particles 1 and 2, respectively. We
note that for a contact interaction the corresponding orbital angular
momenta are zero; thus we ignore the orbital angular momentum label in
what follows. We also note that the matrix element~(\ref{3nf}) is not
fully antisymmetrized, but this is not relevant here.  In what
follows, we discuss two different non-local regulators in oscillator
basis.

\subsubsection{Cutoff in Jacobi momenta}
\label{sec:sq3NF}

\begin{figure}[H]
\includegraphics[width=0.45\textwidth]{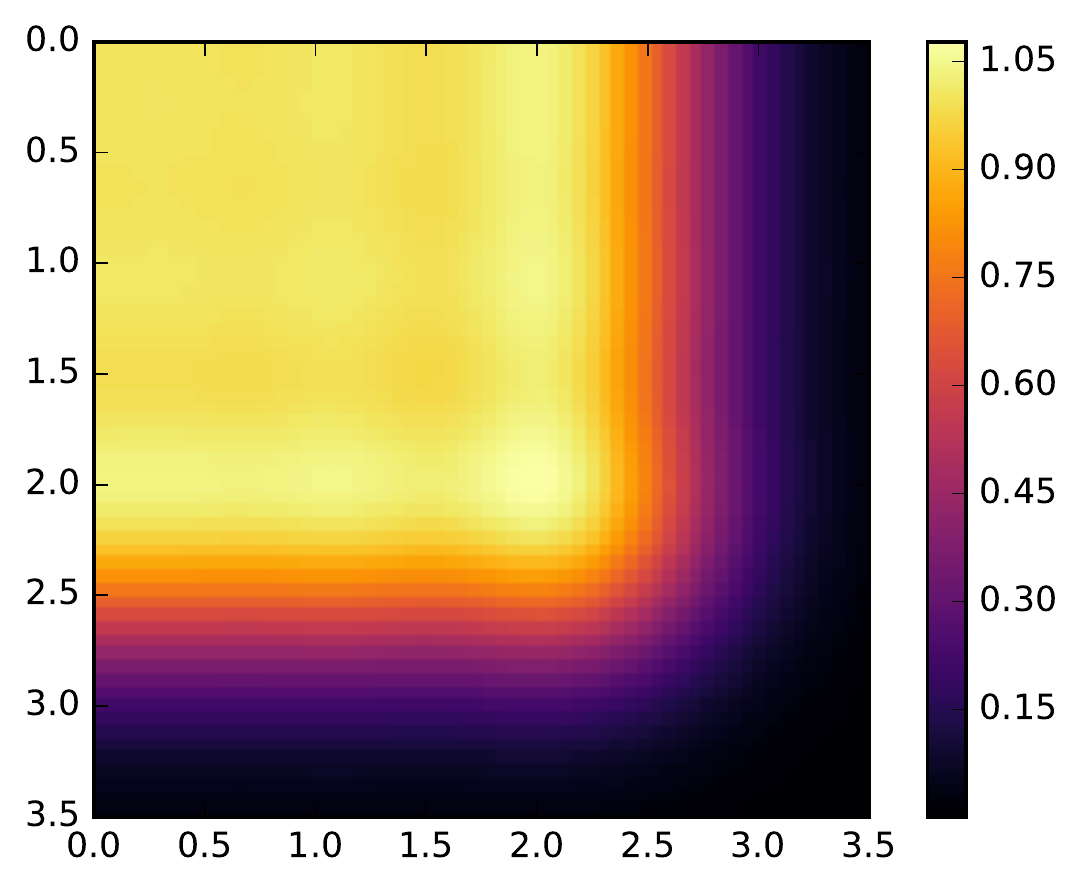}
\caption{(Color online) Momentum space matrix elements $v_{\rm
    DVR}(k)v_{\rm DVR}(p)$ in harmonic oscillator model space with $N
  = 8$ and $\hbar\omega= 22$~MeV as a function of the two incoming
  Jacobi momenta $k$ and $p$. }
\label{fig:sq3nf}
\end{figure}

One possibility is to regulate the incoming Jacobi momenta $k$ and $p$
individually (and similar for the outgoing Jacobi momenta). This
approach is somewhat unusual as it corresponds to regulator functions
$f(p)f(k)$ that are multiplied with the interaction. 
In this case, the DVR interaction  becomes
\ba
\label{3nfdvr}
\lefteqn{\langle k',p'|\hat{V}_{\rm DVR}^{\rm{sq}}|k,p \rangle =}\nonumber\\
&&C_{NNN}^{\rm{sq}} 
v_{\rm DVR}(k')v_{\rm DVR}(p')v_{\rm DVR}(k)v_{\rm DVR}(p) , 
\ea
and $v_{\rm DVR}$ is as in Eq.~(\ref{v-dvr}).  Thus, the IR
improvement of the $NNN$ contact is identical to the $NN$ contact
discussed above, and we have to replace $v_{\rm DVR}(k)$ in
Eq.~(\ref{3nfdvr}) by Eq.~(\ref{v-dvr-lo}). Figure~\ref{fig:sq3nf}
plots the function $v_{\rm DVR}(k)v_{\rm DVR}(p)$ for $S$ waves in
both Jacobi momenta in harmonic oscillator model space with $N = 8$ and $\hbar\omega=
22$~MeV. Note that we have renamed the LEC as $C_{NNN}^{\rm{sq}}$
in Eq.~(\ref{3nfdvr}) because of the square shape of the interaction
in the Jacobi basis.

\subsubsection{Hyperspherical cutoff}
Usually, the cutoff of the $NNN$ force is in the hyper momentum, see
Refs.~\cite{epelbaum2002,navratil2007} for examples.  We introduce the
hyperradial momentum $\rho$ and the hyperangle $\alpha$ as

\ba
\label{HH_r1r2}
k &=& \rho \cos\alpha , \nonumber\\
p &=& \rho \sin\alpha . 
\ea
The $NNN$ contact is isotropic in hyperspherical coordinates and only
depends on the hypermomentum $\rho$. We recall that the orbital
angular momenta corresponding to the Jacobi momenta vanish for the
$NNN$ contact, and so does the hyperspherical angular momentum. In
this special case, the hyperradial wave function of interest is the
eigenstate of a six-dimensional harmonic oscillator with vanishing
hyperangular momentum, i.e.,
\be
\label{HH_wf}
\tilde{\Psi}_{n}(\rho) = \bar{b}^3 \sqrt{\frac{2n!}{\Gamma{(n+3)}}} e^{-\frac{\rho^2 \bar{b}^2}{2} } L_n^{2} (\rho^2 \bar{b}^2)
\ee
and corresponds to the energy $(2n + 3)\hbar\omega$. Here, $\bar{b} =
\sqrt{\hbar/m\omega}$ is the oscillator length in terms of the nucleon
mass $m$ and differs from Eq.~(\ref{bosc}).

It is straightforward to derive the DVR for the hypermomentum. It is
based on the discrete momenta $\rho_{\mu}$ (with $\mu = 0,\ldots,N$),
which are the zeros of the Laguerre polynomial $L_{N+1}^{2}(\rho^2
\bar{b}^2)$. The momentum eigenfunction corresponding to eigenvalue
$\rho_{\mu}$ is
\be
\label{HH_Phi}
\Phi_{\mu}(\rho) = C_{\mu}\sum\limits_{n = 0}^{N} \tilde{\Psi}_{n}(\rho_{\mu}) \tilde{\Psi}_{n}(\rho) .
\ee
Here, $C_{\mu}$ is a normalization constant. Analogous to
Eq.~(\ref{cmu}), we find
\be 
\label{Cmu}
C_{\mu}= \frac{\rho_{\mu}\bar{b}}{\sqrt{(N + 3)(N + 1)}\tilde{\Psi}_{N}(\rho_{\mu}\bar{b})} .
\ee
The $NNN$ contact thus becomes
\be
\label{HH_dvr3nf}
\langle \rho'|U_{\rm DVR} | \rho \rangle = u_{\rm{DVR}}(\rho')u_{\rm{DVR}}(\rho)
\ee
with
\be
\label{u_DVR}
u_{\rm{DVR}}(\rho) = \sum\limits_{\mu=0}^{N}C_{\mu}\tilde{\Phi}_{\mu}(\rho) .
\ee

As before, this DVR interaction needs IR improvement. We generalize
the solution~(\ref{IRcmu}) to improve the low-momentum behavior of the DVR
interaction at hyperspherical radial momentum $\rho = 0$
\ba
\label{IRCmu}
\bar{C}_{\mu} &\equiv& C_{\mu} , \quad\mbox{for $\mu = 0,\ldots,N-1$} \nonumber\\   
\bar{C}_{N} &\equiv&\left(1 - \sum\limits_{\nu = 0}^{N-1} \tilde{\Phi}_{\nu}(0)C_{\nu} \right)/{\tilde{\Phi}_{N}(0)} 
\ea
and arrive at the IR improved function
\ba
\label{HH_IR}
u_{\rm DVR}^{\rm IR}(\rho) = \sum_{\mu=0}^{N} \bar{C}_{\mu}\tilde{\Phi}_{\mu}(\rho) .
\ea
Thus, the IR improved potential is
\be
\label{U_3nf}
\langle \rho'| U_{\rm DVR}^{\rm IR}|\rho \rangle = C_{NNN}^{\rm tr}u_{\rm DVR}^{\rm IR}(\rho')u_{\rm DVR}^{\rm IR}(\rho) .
\ee
Here $C_{NNN}^{\rm tr}$ is the corresponding coupling strength.

\begin{figure}[H]
\includegraphics[width=0.45\textwidth]{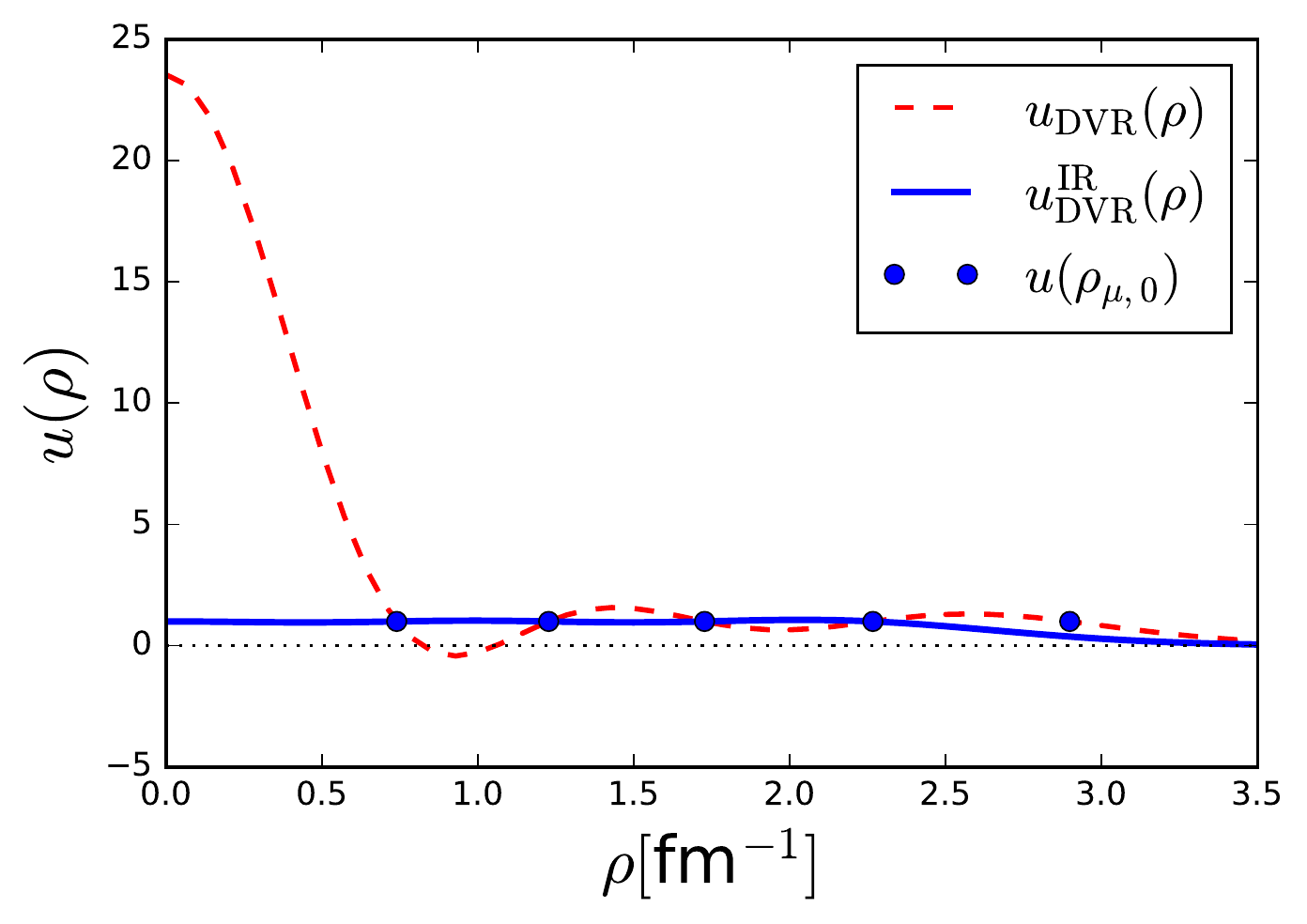}
\caption{(Color online) The solid blue and dashed red curves shows the
  three-nucleon contact in the DVR basis in hyperspherical coordinates
  with and without IR improvement. The former is close to $u(\rho)=1$
  at low momentum. The solid blue dots represent the DVR momenta
  $\rho_{\mu}$ for $\mu = 0, \ldots ,N$ when $N = 8$ and $\hbar\omega
  = 22$~MeV.}
\label{fig:rho_3NF}
\end{figure}

Figure~\ref{fig:rho_3NF} compares Eq.~(\ref{HH_IR}) of the IR improved
contact (the solid blue line) with the contact in Eq.~(\ref{u_DVR})
lacking IR improvement (the dashed red line). Note that the latter
exhibits particularly large deviations from a constant value typical
for a contact at small momenta below the IR cutoff. This is because
the integration measure $d\rho \rho^5$ suppresses low-momentum
deficiencies in the usual scalar product.

\begin{figure}[H]
\includegraphics[width=0.45\textwidth]{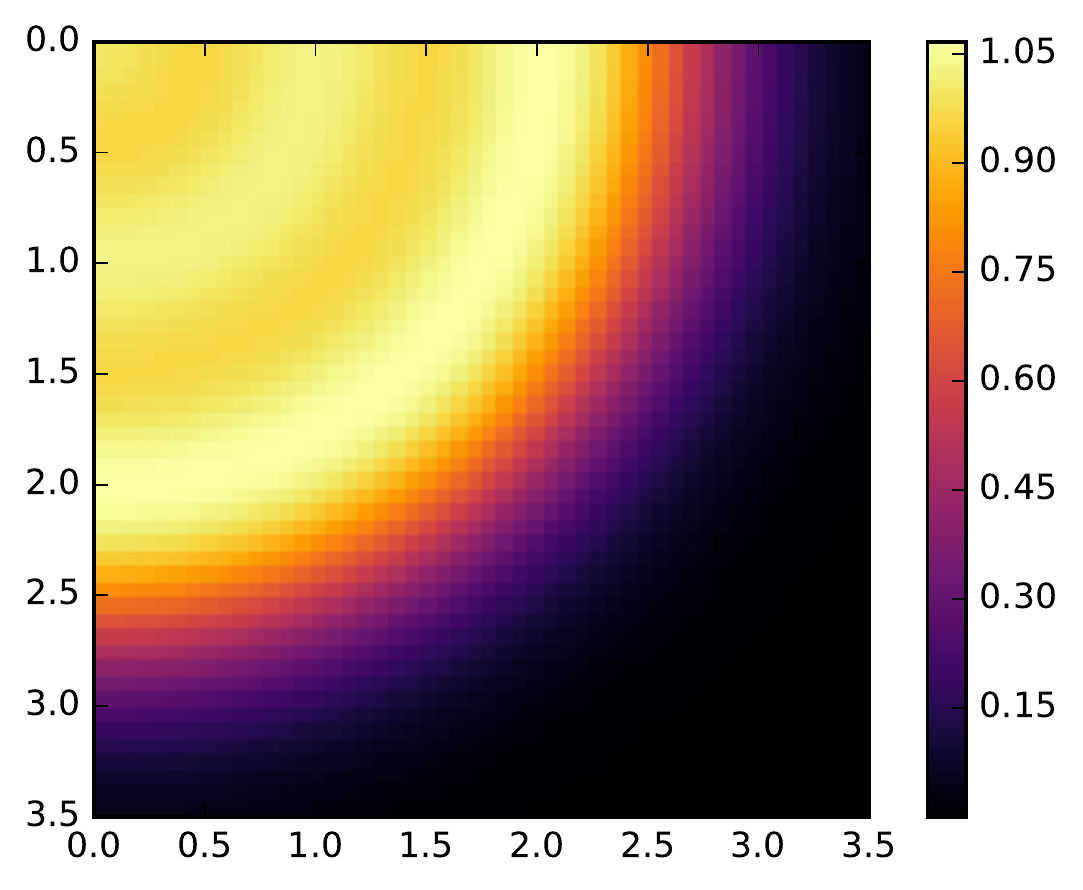}
\caption{(Color online) Size of momentum space matrix elements
  $\bar{u}(k,p) = c_{\mu,0}c_{\nu,0}u_{\rm DVR}^{\rm
    IR}\big{(}\sqrt{k^2 + p^2}\big{)}$ in harmonic oscillator model
  space $N = 8, \hbar\omega= 22$~MeV,$ l_1,l_2 = 0$. $x$- and $y$-
  axis represent Jacobi momenta in fm$^{-1}$.}
\label{fig:HH_3NF}
\end{figure}

Our computer codes use the $NNN$ potential in Jacobi coordinates as
input. For this reason, we need to transform the matrix elements in
Eq.~(\ref{U_3nf}) to the Jacobi basis. The DVR provides us with a very
simple and elegant solution to this problem. Recall that the DVR in
the Jacobi momenta provides us with a Gauss-Laguerre integration that
becomes exact for polynomials of degree $N$ in $k$ and in $p$. Thus,
the basis functions in Eq.~(\ref{HH_wf}) can be exactly integrated,
and
\be
\label{Vtr_3nf}
\langle \phi_{\nu',0}  \phi_{\mu',0}|\hat{U}_{\rm DVR}^{\rm IR}|\phi_{\mu,0}  \phi_{\nu,0} \rangle 
= C_{NNN}^{\rm tr} \bar{u}_{\mu' \nu'} \bar{u}_{\mu \nu}
\ee
with $\bar{u}_{\mu \nu} = c_{\mu,0}c_{\nu,0}u_{\rm DVR}^{\rm
  IR}\big{(}\sqrt{k_{\mu,0}^2 + p_{\nu,0}^2}\big{)} $.  We note that
the reduced mass is set to $m$ in calculating $c_{\mu,0}$ and
$k_{\mu,0}$ here. Figure~\ref{fig:HH_3NF} plots matrix elements of the
DVR interaction, given in Eq.~(\ref{Vtr_3nf}), in Jacobi momentum
space.

Closer inspection reveals that the overlap between the hyperradial
wave function $\Psi_N(\rho)$ and the radial wave functions
$\tilde{\psi}_{i,0}(k)\tilde{\psi}_{j,0}(p)$ vanishes for $i+j > N$.
Thus, the hyperspherical cutoff corresponds to a ``triangular'' cutoff
in the oscillator basis of the Jacobi coordinates. For this reason,
the LEC of the $NNN$ contact in Eq.~(\ref{U_3nf}) carries the subscript ``tr''. 
In what follows we will employ the hyperspherical formulation
of $NNN$ potential unless specified otherwise.

\subsection{Discussion}
Let us briefly summarize and discuss the main results of this
Section. We introduced a momentum-space DVR in the harmonic oscillator
basis as an efficient tool to implement an EFT. The DVR potential
agrees with the momentum-space potential only at a set of discrete
momenta. The low-momentum behavior of the DVR potential can be
corrected such that it agrees with the momentum-space potential at
zero momentum. We have shown how to implement these IR improvements
for $NN$ and $NNN$ potentials.

One may wonder whether the IR improvement is really
necessary. Clearly, if one aims at an EFT that is valid at lowest
momenta, the IR improvement cannot be avoided. As we will see below,
this is particularly so when LECs of the EFT potential are adjusted to
the effective range expansion. The works~\cite{binder2016,yang2016}
showed that a lack of IR improvement leads to oscillations in phase
shifts, which made it difficult to adjust the interaction to
data. However, it is not clear how much structure calculations of
nuclei~\cite{hagen2014,hergert2016} have been impacted by the use of a
finite harmonic oscillator basis without IR improvements. It could be
that observables such as ground-state energies and radii of well-bound
nuclei are not sensitive to the details of the underlying interaction
at lowest momenta. The argument is that the relevant momentum scale,
i.e. the momentum corresponding to the smallest separation energy,
often exceeds the IR cutoff of the oscillator basis, see
Refs.~\cite{konig2017,forssen2017} and App.~\ref{app-extra}.

Many details regarding the implementation of an EFT as a DVR in the
oscillator basis are presented in the Appendix of this work. There we
show that a DVR can be implemented in many ways (see
App.~\ref{app-dvr}), that the IR improvement is a systematic and
controlled approximation (see App.~\ref{app-er}), that there are
simple scaling laws for the resulting DVR interactions (see
App.~\ref{sec:scale}), that the Wigner bound is obeyed (see
App.~\ref{app-wigner}), and that regulator differences, i.e. different
combinations of $\hbar\omega$ and $N$ with the same UV cutoff
$\Lambda$, are higher-order effects (see App.~\ref{harmonic
  oscillatorEFT}). We also explore the effects of truncations of $NNN$
forces in App.~\ref{trunc-E3max}, and finally show in
App.~\ref{app-extra} that IR extrapolations work well in the DVR
approach.

\section{Calibration and results for $^3$H and $^{3,4}$He}
\label{sec:LECs}

\subsection{Atomic nuclei}
In this Section, we adjust the LECs in pionless EFT to data. For
atomic nuclei, we will use the deuteron binding energy, the
effective-range expansion of the $S$-wave phase shifts, and the phase
shifts of the CD-Bonn potential~\cite{machleidt2001} to constrain the
LECs of the $NN$ interaction. The $NNN$ contact will be adjusted to
reproduce the triton binding energy.

To compute phase shifts in the harmonic oscillator basis, we follow
Ref.~\cite{shirokov2004}, which is based on the $J$-matrix
approach~\cite{heller1974}. For the computation of binding energies we
proceed as follows. For the interaction we will employ a model space
with $N=8$. The Hamiltonian, i.e., the sum of kinetic energy and the
interaction, will be evaluated in model spaces of size $N=8, 10, 12,
\ldots$. For the interaction, the matrix elements between states with
$N>8$ are zero. Thus, UV convergence is achieved by construction.  The
increase of the model space for the kinetic energy yields IR
convergence, see Ref.~\cite{binder2016} for details.  In what follows,
we report virtually converged results for nuclei with mass numbers
$A=2,3,4$.  We vary the oscillator spacing to probe the cutoff
dependence of our results.

At LO we have two LECs associated with $NN$ contact interactions and
one for the $NNN$ contact. In the ${^3S}_{1}$ partial wave, the LEC is
adjusted to reproduce the deuteron binding energy. The coupling
strength in the singlet $S$ channel for the $NN$ contact is adjusted
to the neutron- proton ($np$) phase shifts of the CD-Bonn potential
for energies $E_{\rm{rel}} \in [0.01,0.1]$~MeV. The predicted value
for the triplet $S$ scattering length agrees with data within $30\%$,
which is what we expect from simple error estimates discussed below.
Table~\ref{tab:dataLO} shows the values of the LECs at LO for
potentials defined in model spaces with $N=8$ for different cutoffs.

\begin{table}[htb]
\caption{The leading order LECs $\Tilde{C}_{^3s_{1}}$ and
  $\Tilde{C}_{^1s_{0}}$ (both in $10^{-5}$MeV$^{-2}$), and ${c_{E}}$
  (dimensionless) for atomic nuclei (with nucleon mass $m=939$~MeV)
  for different momentum cutoffs $\Lambda$ (in MeV) obtained from
  varying the oscillator frequency $\hbar\omega$ (in MeV), for
  interactions in a model space with $N = 8$. }
\begin{tabular}{|D{.}{.}{0}|D{.}{.}{2}|D{.}{.}{6}|D{.}{.}{6}|D{.}{.}{6}|}
 \hline 
\hbar\omega& \Lambda &  \multicolumn{1} {c|}{$\Tilde{C}_{^3s_1}$} & \multicolumn{1} {c|} {$\Tilde{C}_{^1s_0}$}  & \multicolumn{1} {c|} {$c_E$} \\ \hline 
 5& 232.35  & -1.006988 & -0.597220 & -0.163306\\ 
10& 328.59  & -0.624098 & -0.431559 & -0.671882\\ 
22& 487.38  & -0.379465 & -0.296100 & -0.238514 \\ 
40& 657.19  & -0.266381 & -0.221703 & -0.091625  \\ \hline
\end{tabular}
\label{tab:dataLO}
\end{table}%

We note that the LECs $\Tilde{C}_{^3s_{1}}$ and $\Tilde{C}_{^1s_{0}}$
approximately obey the relation $C_{\rm LO}\propto
(\hbar\omega)^{-1/2}$. This is a consequence of the deuteron's weak
binding, see Appendix~\ref{sec:scale} for details.  We also note that
the LECs of the $NN$ interaction are consistent with analytical
results. To see this, we consider the LO potential
\begin{equation}
V(k',k) = C_0 v(k',\Lambda) v(k,\Lambda).
\end{equation}
Here, $v(k',\Lambda)$ is the regulator function and $\Lambda$ is the
cutoff. For the step-function regulator $v(k,\Lambda) = \Theta(\Lambda -
k)$ we have
\begin{equation}
C_0 \approx -\frac{2\pi^2}{m\Lambda} \frac{4\pi}{(2\pi)^3}, 
\end{equation}
valid for $\Lambda \gg \kappa,a^{-1}$ where $\kappa$ is the binding
momentum and $a$ the scattering length. Similarly, for a 
Gaussian regulator $v(k,\Lambda) =
e^{-\frac{1}{2}\frac{k^2}{\Lambda^2}}$ one has
\begin{equation}
C_0 \approx -\frac{4\pi\sqrt{\pi}}{m\Lambda} \frac{4\pi}{(2\pi)^3}, 
\end{equation}
under the same conditions~\footnote{These results are obtained in
  momentum space with the momentum integration $\int_0^\infty dk k^2$
  as used in the harmonic oscillator EFT. They differ by a factor
  $4\pi/(2\pi)^3$ from results obtained with the usual integration
  measure $\int d^3k /(2\pi)^3$.}. For $\Lambda= 487$~MeV and $m =
939$~MeV we find $C_{0} \approx -0.22\times 10^{-5}$~MeV$^{-2}$ for
the sharp cutoff, and $C_{0} \approx -0.25\times 10^{-5}$~MeV$^{-2}$
for the Gaussian regulator. These results are similar in size to what
is reported in Table~\ref{tab:dataLO} for the same cutoff. Thus, the
results from our EFT constructed in the harmonic oscillator basis are
fully compatible with expectations from a momentum-space EFT.

We now turn to the NLO potential. According to KSW counting we have three
LECs from LO contacts, and two additional LECs from the NLO $NN$
contact interaction in $S$ waves. We determine the LECs using
non-perturbative solvers for the $J$ matrix and the Hamiltonian
eigenvalues.  In the triplet $S$ channel the LECs are inferred from the
 deuteron binding energy and matter radius  (1.976 fm). 
 In the singlet $S$ channel the LECs are adjusted 
 to $np$ phase shifts of the CD-Bonn potential
for energies $E_{\rm{rel}} \in [0.01,0.1]$~MeV. The $NN$ interaction
at NLO determines the scattering lengths and the effective range $r_0$.
Once the $NN$ potential is fixed at NLO, the LEC for the $NNN$ contact
is adjusted to reproduce the triton binding energy. The results for
the LECs are presented in Table~\ref{tab:dataNLO}.

\begingroup
\squeezetable
\begin{table}[htb]

\begin{center}
\caption{The next-to-leading order LECs $\Tilde{C}_{^3s_{1}}$ and
  $\Tilde{C}_{^1s_{0}}$ (both in $10^{-5}$MeV$^{-2}$), and
  ${C}_{^3s_{1}}$ and ${C}_{^1s_{0}}$ (both in $10^{-10}$MeV$^{-4}$),
  and ${c_{E}}$ (dimensionless) for atomic nuclei (with nucleon mass
  $m=939$~MeV) for different momentum cutoffs $\Lambda$ (in MeV)
  obtained from varying the oscillator frequency $\hbar\omega$ (in
  MeV), for interactions in a model space with $N = 8$. }
\begin{tabular}{|r|c|D{.}{.}{6}|D{.}{.}{6}|D{.}{.}{6}|D{.}{.}{6}|D{.}{.}{6}|} 
 \hline 
$\hbar\omega$& $\Lambda$ & \multicolumn{1} {c|}{$\Tilde{C}_{^3s_1}$}&\multicolumn{1} {c|}{$C_{^3s_1}$} & \multicolumn{1} {c|} {$\Tilde{C}_{^1s_0}$} & \multicolumn{1} {c|}{$C_{^1s_0}$}& \multicolumn{1} {c|}{$c_E$} \\ \hline 
 5& 232.35 & -1.001248 &-0.039732 &-0.718772 & 1.124941 &  0.533367 \\ 
10& 328.59 & -0.919696 & 1.078144 &-0.588224 & 0.725705 & -0.274206\\ 
22& 487.38 & -0.809378 & 0.772254 &-0.612966 & 0.727724 & -0.008170 \\ 
40& 657.19 & -0.866529 & 0.689544 &-0.605710 & 0.590509 & -0.061330 \\ \hline
\end{tabular}
\label{tab:dataNLO}
\end{center}
\end{table}%
\endgroup

\begin{figure}[htb]
   \includegraphics[width=0.48\textwidth]{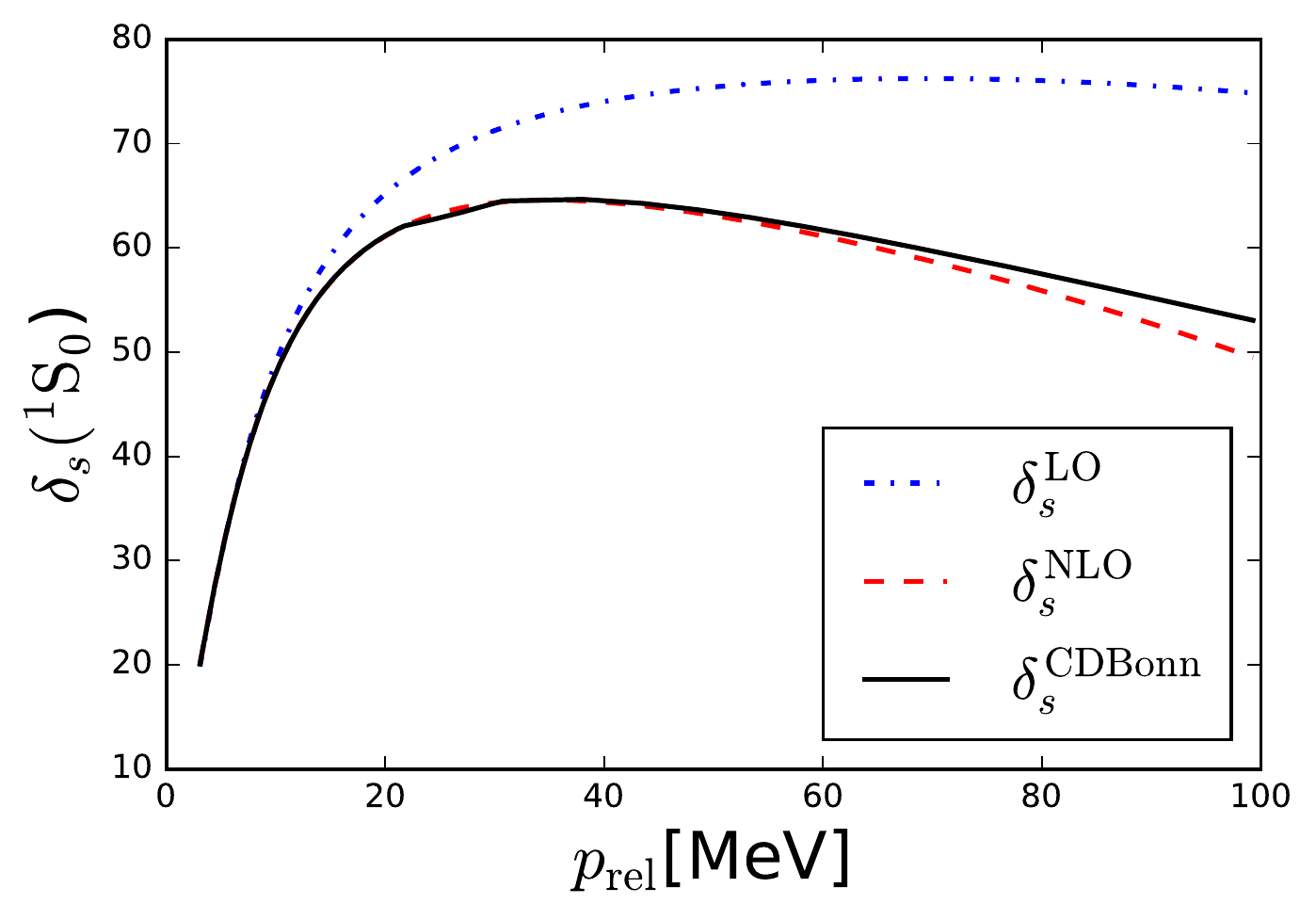}
   \includegraphics[width=0.48\textwidth]{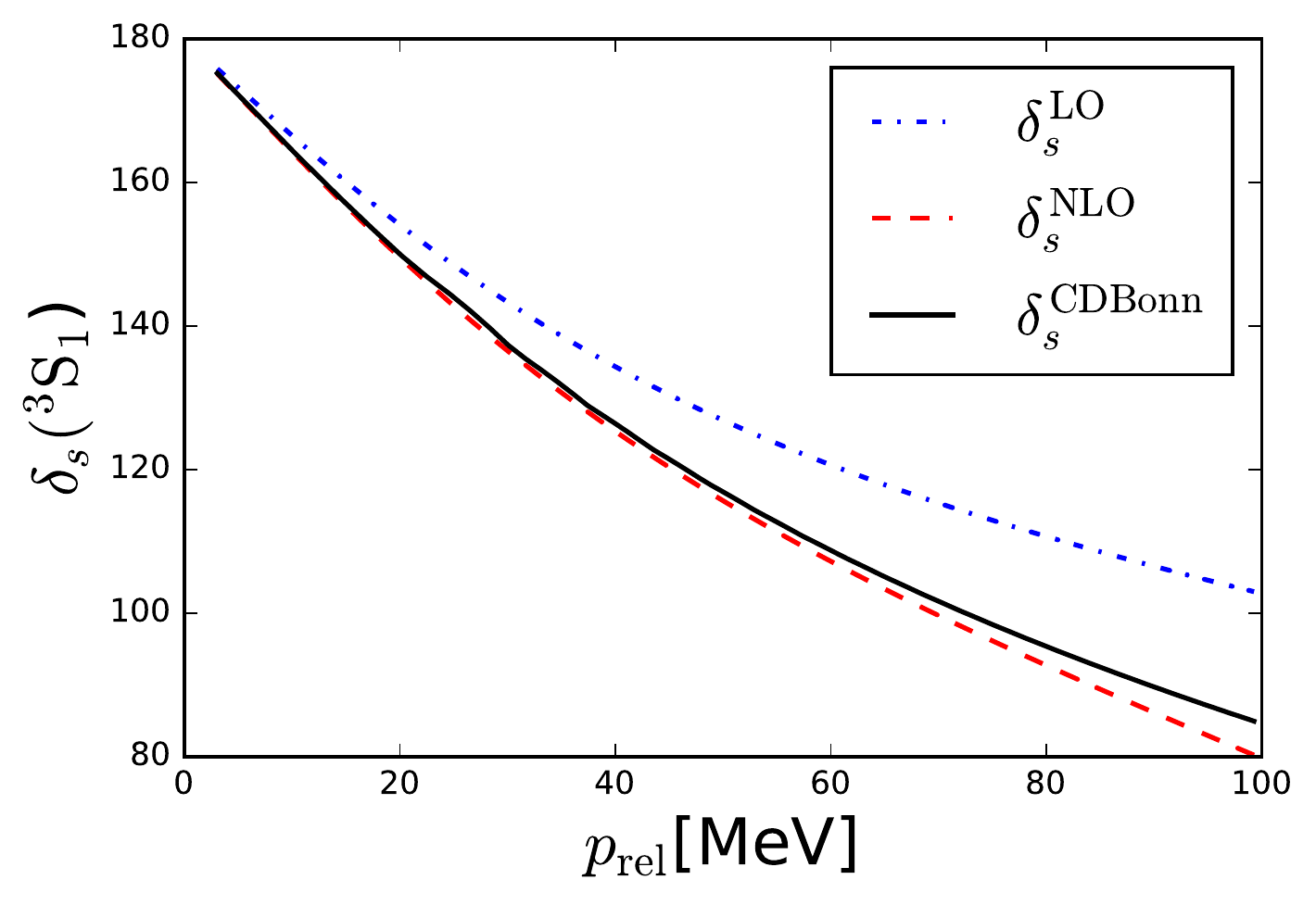}
\caption{(Color online) Phase shifts in the partial waves $^1S_0$
  (upper panel) and $^3S_1$ (lower panel) from IR improved potentials
  at NLO (red dashed) and LO (blue dot-dashed), respectively, in a
  model space $N = 8, \hbar\omega = 22$~MeV, $l = 0$. The black curves
  shows the neutron-proton phase shifts of the CD-Bonn potential.}
  \label{fig:LO_NLOphaseshifts}
  \end{figure}

Figure~\ref{fig:LO_NLOphaseshifts} shows the phase shifts from
pionless EFT at LO (blue dot-dashed line) and NLO (red dashed line),
and compares them to those of the CD-Bonn potential (black line).  The
LO potentials reproduce phase shifts for momenta~$p_{\rm rel} \lesssim
{a_{s,t}}^{-1}$, while the NLO interactions extends the range to
$p_{\rm rel}\lesssim {r_{s,t}}^{-1}$.  Results are consistent with our
expectation from EFT. The phase shift plots illustrate the quality of
the IR improved potentials. The oscillations that were observed in
Refs.~\cite{binder2016,yang2016} are much reduced.

Our LO results (binding energies and point-proton radii) for the
light nuclei $^{3}$H and $^{3,4}$He, computed with a translationally
invariant no-core-shell model~\cite{kamuntavicius2000}, are collected
in Table~\ref{tab:A34}.  The results for $NN$ interaction alone
exhibit a strong cutoff dependence. This dependence becomes much
weaker once the $NNN$ contact is included. At LO with $NNN$ forces
included, the nucleus $^4$He is underbound. This result is consistent
with the results reported by \citeauthor{kirscher2009}, also obtained
at lower cutoffs (though the authors expressed some doubts regarding
the convergence of their calculation). 

\begin{table}[htb] 
 \centering
 \caption{Binding energies and point-proton radii of $A
   \leq 4$ nuclei using $NN$ and $NN + NNN$ pion-less EFT interactions
   at LO and defined in model space $N = 8$.}
\begin{tabular}{|c|c|D{.}{.}4|c|D{.}{.}4|c|D{.}{.}4|c|} 
\cline{1-8}
  \multicolumn{8}{|c|} {LO  $NN$} \\ 
   \cline{1-8}
$\hbar\omega$ &  $\Lambda$ & \multicolumn{1} {c|} {$E({^3}$H)}  & $ r({^3} \rm{H})$  & \multicolumn{1} {c|}{$E({^3}$He)}  & $r({^3} \rm{He})$ & \multicolumn{1}{c|}{$E({^4}$He)} & $r({^4} \rm{He})$ \\ \hline 
5& 232.35 & 8.65  & 1.78 &8.04 & 1.89&  26.81& 1.79\\ 
10& 328.59 &13.34& 1.31& 12.49&1.37 &45.45 & 1.28 \\ 
22& 487.38  &23.69 & 0.91 & 22.46 &0.95 &88.06& 0.87 \\ 
40& 657.19  &38.31 &0.69 &36.65 & 0.71&149.88& 0.65 \\ 
 \cline{1-8}
  \multicolumn{8}{|c|} {LO $NN+NNN$}\\ 
   \cline{1-8}
$ \hbar\omega$ &  $\Lambda$  &  \multicolumn{1} {c|}{$E({^3} \rm{H})$}  & $r({^3} \rm{H})$  & \multicolumn{1} {c|}{$E({^3} \rm{He})$} & $r({^3} \rm{He})$ &  \multicolumn{1} {c|}{$E({^4} \rm{He})$}  & $r({^4} \rm{He})$  \\ \hline 
5& 232.35 &8.482 & 1.79 &7.87 & 1.90& 26.05 & 1.79\\ 
10& 328.59 &8.482& 1.46 &7.71 & 1.60 &22.40& 1.44\\ 
22& 487.38  &8.482 &1.29 & 7.54 &1.46 & 17.66& 1.46 \\ 
40& 657.19  &8.482 & 1.23 & 7.41 & 1.42  &17.55& 1.42\\ \hline
\end{tabular}
\label{tab:A34}
\end{table}%

Table~\ref{tab:A34-nlo} shows our results for light nuclei at NLO.  We
note that the NLO results for $NN$ interactions alone are close to the
data, i.e., $E(^3\mbox{H})= 8.48$~MeV, $E(^3\mbox{He})= 7.5$~MeV, and
$E(^4\mbox{He})= 28.5$~MeV, and depend very weakly on the cutoff over
the considered range of cutoffs. Similar comments apply to the
radii. Including the $NNN$ contact further reduces the cutoff dependence, and
the $^4$He nucleus is close to its physical point. These results are
consistent with those by \citeauthor{platter2005}.

\begin{table}[bt] 
 \begin{center}
 \caption{Binding energies and point-proton radii of $A
   \leq 4$ nuclei using $NN$ and $NN + NNN$ pion-less EFT interactions
   at NLO and defined in model space $N = 8$.}
\begin{tabular}{|c|c|D{.}{.}4|c|D{.}{.}4|c|D{.}{.}4|c|} 
  \cline{1-8}
  \multicolumn{8}{|c|} {NLO $NN$}\\ 
   \cline{1-8} 
  $\hbar\omega$ &  $\Lambda$  & \multicolumn{1} {c|}{$E({^3} \rm{H})$}  & $r({^3} \rm{H})$ & \multicolumn{1} {c|}{$E({^3} \rm{He})$}  & $r({^3} \rm{He})$  &  \multicolumn{1} {c|}{$E({^4} \rm{He})$}  & $r({^4} \rm{He})$  \\ \hline 
5 & 232.35 & 7.94 & 1.82& 7.35 &1.97 &25.03 & 1.80\\ 
10& 328.59 &10.11& 1.49& 9.34& 1.61& 36.24& 1.34\\ 
22& 487.38 &8.62&1.62 & 7.90 & 1.82 &30.39 & 1.41\\ 
40& 657.19 &8.97&1.62 & 8.30 &1.77 &29.95 &1.53\\
  \cline{1-8}
  \multicolumn{8}{|c|} {NLO  $NN+NNN$}\\ 
   \cline{1-8} 
 $\hbar\omega$ &  $\Lambda$ &  \multicolumn{1} {c|}{$E({^3} \rm{H})$}  & $r({^3} \rm{H})$ & \multicolumn{1} {c|}{$E({^3} \rm{He})$}  & $r({^3} \rm{He})$  &  \multicolumn{1} {c|}{$E({^4} \rm{He})$} & $r({^4} \rm{He})$  \\ \hline 
5 & 232.35 & 8.482 & 1.80 & 7.88  & 1.93 & 27.52 &  1.79 \\ 
10& 328.59 & 8.482 &1.59 & 7.75 & 1.75 &  27.30& 1.43\\ 
22& 487.38& 8.482 &1.63 &7.77 & 1.83 & 29.30  & 1.44 \\ 
40& 657.19  & 8.482 & 1.65 & 7.82 & 1.82 & 27.35& 1.58 \\ \hline 
\end{tabular}
\label{tab:A34-nlo}
\end{center}
\end{table}%

Let us discuss theoretical uncertainties. The three contributions to
the error budget are (i) neglected higher-order terms of the
interaction, (ii) uncertainties in the LECs due to uncertainties of
the input, and (iii) the convergence of the calculations with respect
to the model space.  For the nuclei discussed here, only the first
contribution is relevant. The third contribution to the uncertainties
yields very small corrections as shown in App.~\ref{app-extra}.

Based on the power counting in pionless EFT, the uncertainty
for observable $X$ is expected to be of the form~\cite{furnstahl2015}
\be
\Delta X = X_0\left(c_1Q + c_2 Q^2 + \ldots\right)
 \label{err}
\ee
where $Q = p_{\rm F}/\Lambda_b$ is the typical momentum ratio,
expressed in terms of the Fermi momentum $p_{\rm F}$ and the breakdown
scale $\Lambda_b$. The coefficients $c_k$ are parameters, expected to
be of natural size. The free Fermi gas estimate
\be
\label{fermi}
{E\over A} = {3\over 10} {p_F^2\over m}
\ee
relates the average binding energy to the Fermi momentum, yielding
$p_{\rm F} \approx 150$~MeV for $^4$He. Around the UV cutoff $\Lambda
\approx 650$~MeV, we are unable to reproduce the effective range of
the $NN$ interaction and therefore we consider it to be the breakdown
scale (see App.~\ref{app-wigner} for details) giving a very
conservative $Q \approx 1/3$.  Consequently, the uncertainty in the
binding energy of ${^4}$He at LO is estimated to be about $30 \%$
i.e., $\Delta E_{\rm LO} ({^4} {\rm He}) \approx 8$ MeV. Similarly, at
NLO it is estimated to be around $10 \%$ or $\Delta E_{\rm NLO} ({^4}
{\rm He}) \approx 3$ MeV. These simple estimates are also consistent
with the change of the $\alpha$-particle binding energy resulting from
the variation of the UV cutoff $\Lambda$ at each order. Moreover, the
LO and NLO binding energies overlap after including the discussed
uncertainties. We also note that at NLO, the experimental binding
energy of $^4$He (28.3 MeV) agrees with our theoretical result within
the uncertainties.


\subsection{Lattice nuclei}
For lattice nuclei we optimize the LECs using the binding energies of
the deuteron and the di-neutron, the effective range expansion, and
the triton binding energy from lattice QCD data in
Refs.~\cite{beane2013,beane2013a}.  The relevant lattice data is
compiled in Table~\ref{data} and compared to the physical point.

\begin{table}[htb]
\begin{center}
  \caption{Relevant values of physical and lattice QCD data (all in MeV),
    namely the pion mass $m_\pi$, the nucleon mass $m$,
    the di-neutron binding energy $B_{nn}$, the deuteron binding
    energy $B_{d}$, the triton binding energy $B_{t}$, the singlet and
    triplet scattering lengths $^{np}{a_s}$ and $a_t$, respectively,
    the singlet and triplet effective ranges $^{np}{r_s}$ and $r_t$,
    respectively. \label{data}}
  \begin{tabular}{|c|dc|dc|} \hline
  & \mbox{Nature} & & \mbox{Lattice} & \\ \hline 
$m_{\pi}$ & 139.5 \pm 0.1 & \cite{olive2014} & 806. \pm 1  & \cite{beane2013} \\ 
$m$       & 939. \pm 1    & \cite{mohr2012}  & 1634. \pm 18& \cite{beane2013} \\ \hline   
$B_{nn}$ & - & & 15.9 \pm 4& \cite{beane2013}\\
$B_{d}$ & 2.2245 & &  19.5 \pm 5& \cite{beane2013} \\
$B_{t}$ & 8.482& \cite{Wapstra1985} &53.9 \pm 10.7& \cite{beane2013} \\ \hline 
$^{np}{a_s}^{-1}$ & -8.31& \cite{dumbrajs1983} & 84.7 \pm 18& \cite{beane2013a}\\
$^{np}{r_s}^{-1}$ & 71.75& \cite{dumbrajs1983} & 174.6 \pm 25& \cite{beane2013a}\\
${a_t}^{-1}$ & 36.4& \cite{dumbrajs1983} & 108. \pm 13& \cite{beane2013a} \\ 
${r_t}^{-1}$ & 112.18& \cite{dumbrajs1983} & 217.8\pm 46& \cite{beane2013a} \\ \hline
\end{tabular}
\end{center}

\end{table}

At LO, the LECs for the $NN$ contacts are adjusted to the central
values of the binding energies of the deuteron and the di-neutron. The
$NNN$ contact is adjusted to the central value of the triton binding
energy. The results are shown in Table~\ref{tab:dataLO-Lat}.

\begin{table}[htb]
\begin{center}
\caption{The leading order LECs $\Tilde{C}_{^3s_{1}}$ and
  $\Tilde{C}_{^1s_{0}}$ (both in $10^{-5}$MeV$^{-2}$), and ${c_{E}}$
  (dimensionless) for lattice nuclei (with nucleon mass $m=1634$~MeV)
  for different momentum cutoffs $\Lambda$ (in MeV) obtained from
  varying the oscillator frequency $\hbar\omega$ (in MeV), for
  interactions in a model space with $N = 8$. }
\begin{tabular}{|D{.}{.}{0}|D{.}{.}{2}|D{.}{.}{6}|D{.}{.}{6}|D{.}{.}{6}|}
 \hline 
\hbar\omega& \Lambda &  \multicolumn{1} {c|}{$\Tilde{C}_{^3s_1}$} & \multicolumn{1} {c|} {$\Tilde{C}_{^1s_0}$}  & \multicolumn{1} {c|} {$c_E$} \\ \hline 
 5& 306.52  & -1.013613 & -0.904019 & -1.731712\\
10& 433.48  & -0.502300 & -0.460312 & -0.433930 \\ 
22& 642.96  & -0.251429 & -0.236527 & -0.086293 \\ 
40& 866.97 &-0.158373   &  -0.151299&-0.025688\\  \hline
\end{tabular}
\label{tab:dataLO-Lat}
\end{center}
\end{table}%

For the NLO potential, we use the data on the effective range
expansion parameters calculated by \citeauthor{beane2013a}. In that
work, the location of the bound state was used to constrain the
effective range expansion $k\cot\delta$, followed by a two-parameter
fit to determine the scattering length and the effective range.  We
optimize the NLO interaction by performing a simultaneous fit to the
binding energy and the effective range expansion in the singlet and
triplet $S$ channels.  We determine the $NNN$ contact interaction
strength by fitting it to the triton binding
energy. Table~\ref{tab:dataNLO-Lat} contains the LECs at NLO for
different cutoffs.

\begingroup
\squeezetable
\begin{table}[tb]
\begin{center}
\caption{The next-to-leading order LECs $\Tilde{C}_{^3s_{1}}$ and
  $\Tilde{C}_{^1s_{0}}$ (both in $10^{-5}$MeV$^{-2}$), and
  ${C}_{^3s_{1}}$ and ${C}_{^1s_{0}}$ (both in $10^{-10}$MeV$^{-4}$),
  and ${c_{E}}$ (dimensionless) for lattice nuclei (with nucleon mass
  $m=1634$~MeV) for different momentum cutoffs $\Lambda$ (in MeV)
  obtained from varying the oscillator frequency $\hbar\omega$ (in
  MeV), for interactions in a model space with $N = 8$. }
\begin{tabular}{|r|D{.}{.}{2}|D{.}{.}{6}|D{.}{.}{6}|D{.}{.}{6}|D{.}{.}{6}|D{.}{.}{6}|} 
 \hline 
$\hbar\omega$& \Lambda & \multicolumn{1} {c|}{$\Tilde{C}_{^3s_1}$}&\multicolumn{1} {c|}{$C_{^3s_1}$} & \multicolumn{1} {c|} {$\Tilde{C}_{^1s_0}$} & \multicolumn{1} {c|}{$C_{^1s_0}$}& \multicolumn{1} {c|}{$c_E$} \\ \hline 
 5& 306.52 &-0.736443 &-1.034289 &-1.216789 & 1.180748 & -1.719559\\
10& 433.48 &-0.632458 & 0.246988 &-0.854814 & 0.760970 & -0.430321 \\ 
22& 642.96 &-0.449998 & 0.177545 &-0.691412 & 0.423704 & -0.076720 \\  
40& 866.97 &-0.387445 &0.118521 &-0.853482 &0.471613& -0.106436 \\ \hline
\end{tabular}
\label{tab:dataNLO-Lat}
\end{center}
\end{table}%
\endgroup

\begin{figure}[htb]
\includegraphics[width=0.48\textwidth]{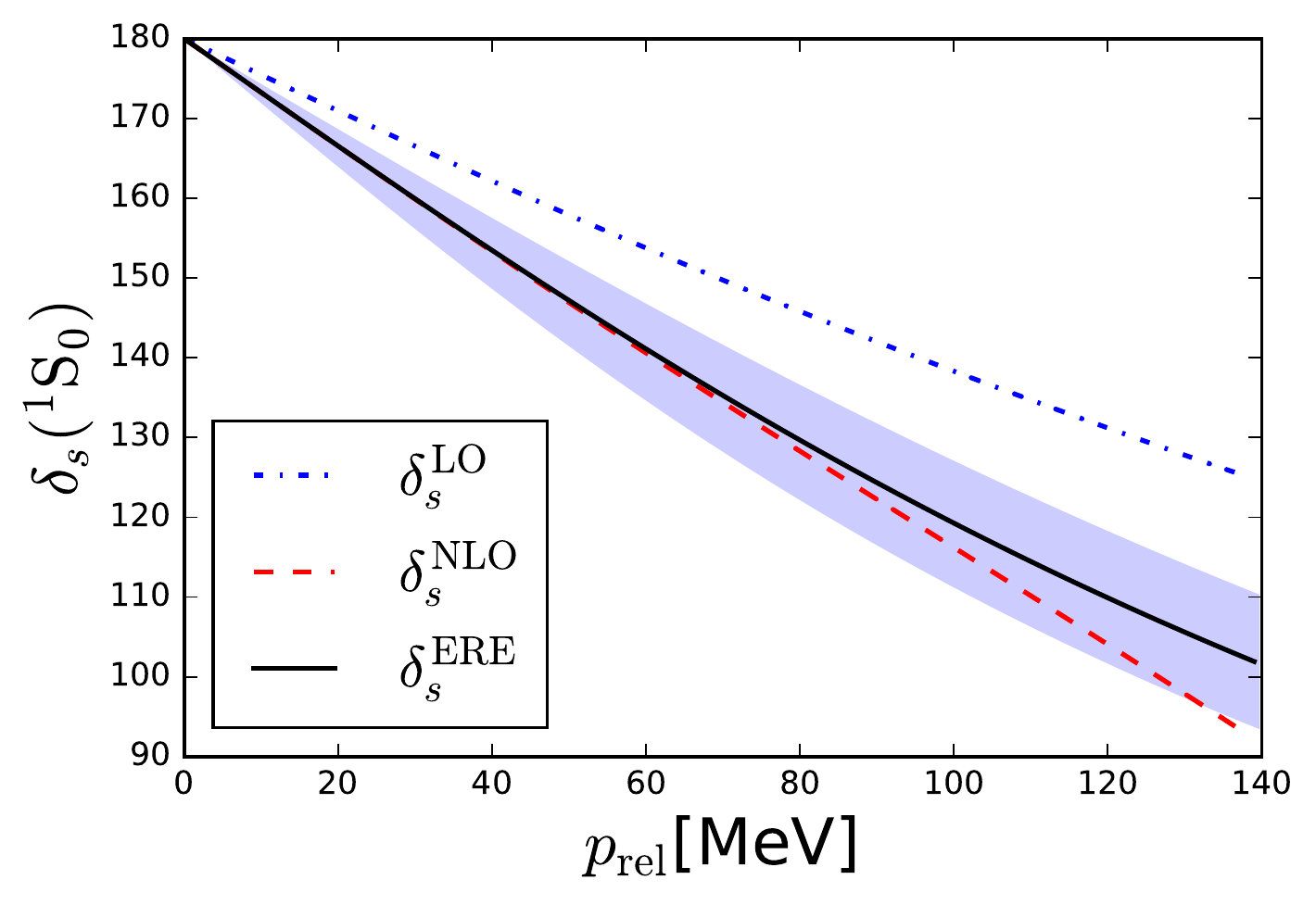}
\includegraphics[width=0.48\textwidth]{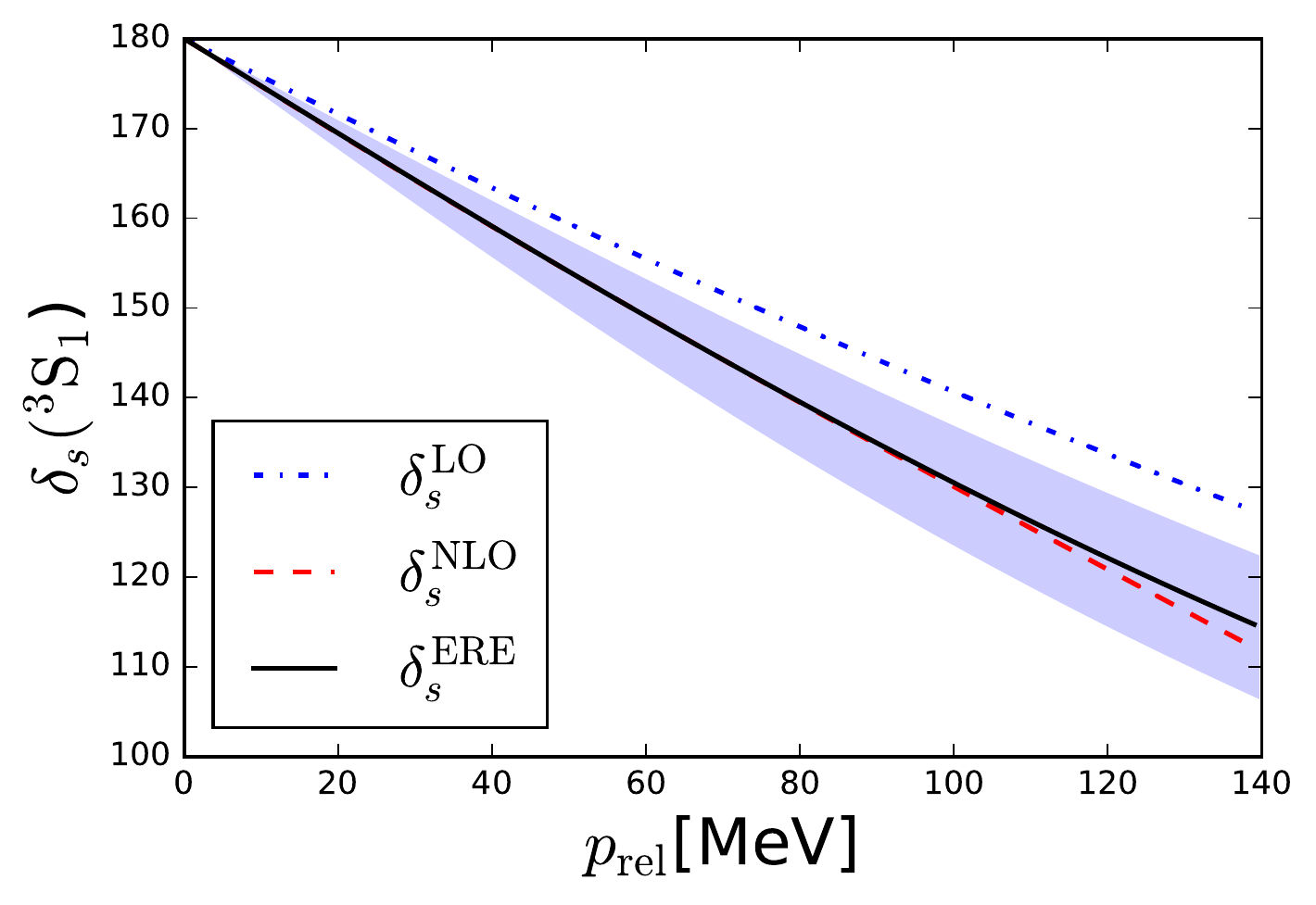}
\caption{(Color online) Phase shifts for lattice nucleons at $m_{\pi}
  = 806$~MeV in the partial waves $^1S_0$ (upper panel) and $^3S_1$
  (lower panel) from IR improved potentials at NLO (red dashed) and LO
  (blue dot-dashed), respectively, in a model space $N = 8,
  \hbar\omega = 22$~MeV. The black curves shows the effective range
  expansion from lattice QCD~\cite{beane2013a} with corresponding
  systematic plus statistical uncertainties shown as a band.}
  \label{fig:LO_NLO_lattice QCDphaseshifts}
\end{figure}

Figure~\ref{fig:LO_NLO_lattice QCDphaseshifts} shows the phase shifts
for lattice nuclei obtained at LO (dashed-dotted line) and at NLO
(dashed lines). The input from the effective range
expansion~(\ref{ere}) is shown as a solid line with uncertainty
estimates from lattice QCD. We see that the EFT agrees with the input
data at NLO over a considerable range of momenta.

We turn to the calculations of light lattice
nuclei. Table~\ref{tab:A34_lattice QCD} shows the LO results for the
binding energies and point-proton radii of lattice nuclei. At LO, the
$NN$ interaction yields binding energies for $^4$He that vary by a
factor of two over the cutoff range. This dependence is reduced once
the $NNN$ contact is added.

\begin{table}[htb] 
 \begin{center}
 \caption{Binding energies and point-proton radii of $A = 3,4$ lattice
   nuclei at $m_{\pi} = 806$~MeV using $NN$ and $NN + NNN$ pion-less
   EFT interactions at LO in model space $N = 8$.}
\begin{tabular}{|c|D{.}{.}{2}|D{.}{.}{4}|c|D{.}{.}{4}|c|D{.}{.}{4}|c|} 
\cline{1-8}
  \multicolumn{8}{|c|} {LO  $NN$} \\ 
   \cline{1-8}
$\hbar\omega$ &  \Lambda & \multicolumn{1} {c|} {$E({^3}$H)}  & $ r({^3} \rm{H})$  & \multicolumn{1} {c|}{$E({^3}$He)}  & $r({^3} \rm{He})$ & \multicolumn{1}{c|}{$E({^4}$He)} & $r({^4} \rm{He})$ \\ \hline 
5& 306.52 & 64.3 & 1.57 & 64.5  & 1.57& 142.7 & 1.62\\ 
10&  433.48 &76.6& 1.13& 75.6 &1.13 &177.5  & 1.17 \\ 
22&  642.96  & 99.1 & 0.78 & 97.6 &0.79 &249.6 & 0.80 \\ 
40& 866.97   & 127.4 & 0.60 & 125.5 &0.60 &344.7 & 0.60\\
 \cline{1-8}
  \multicolumn{8}{|c|} {LO $NN+NNN$}\\ 
   \cline{1-8}
 $\hbar\omega$ &  \Lambda  &  \multicolumn{1} {c|}{$E({^3} \rm{H})$}  & $r({^3} \rm{H})$  & \multicolumn{1} {c|}{$E({^3} \rm{He})$} & $r({^3} \rm{He})$ &  \multicolumn{1} {c|}{$E({^4} \rm{He})$}  & $r({^4} \rm{He})$  \\ \hline 
5& 306.52 &53.9 & 1.55 & 53.1& 1.55& 98.9 & 1.57\\ 
10&  433.48 &53.9& 1.13&52.9 & 1.13 &88.5 & 1.15\\ 
22&  642.96  &53.9 &0.84 & 52.5 &0.85 & 70.8 & 1.05 \\ 
40& 866.97 &53.9 &0.72 & 52.2 &0.74 & 68.3 & 1.09\\   \hline
\end{tabular}
\label{tab:A34_lattice QCD}
\end{center}
\end{table}%

We turn to NLO calculations of light lattice nuclei.  The upper and
lower parts of the Table~\ref{tab:A34_lattice QCD-nlo} show the NLO
results for binding energies and point-proton radii with $NN$
potentials only and with the $NNN$ contact included, respectively. The
cutoff dependence is strong for $NN$ forces alone and much reduced for
the complete calculation including $NNN$ forces.

\begin{table}[htb] 
 \begin{center}
 \caption{Binding energies and point-proton radii of $A = 3,4$ lattice
   nuclei at $m_{\pi} = 806$~MeV using $NN$ and $NN + NNN$ pion-less
   EFT interactions at NLO in model space $N = 8$.}
\begin{tabular}{|c|c|D{.}{.}4|c|D{.}{.}4|c|D{.}{.}4|c|} 
  \cline{1-8}
  \multicolumn{8}{|c|} {NLO $NN$}\\ 
   \cline{1-8} 
 $\hbar\omega$ &  $\Lambda$  & \multicolumn{1} {c|}{$E({^3} \rm{H})$}  & $r({^3} \rm{H})$ & \multicolumn{1} {c|}{$E({^3} \rm{He})$}  & $r({^3} \rm{He})$  &  \multicolumn{1} {c|}{$E({^4} \rm{He})$}  & $r({^4} \rm{He})$  \\ \hline 
5 & 306.52 & 65.2 & 1.57& 64.4 &1.57 &142.5 & 1.61\\ 
10&  433.48 &75.8& 1.12& 74.8& 1.13& 176.3 & 1.16\\ 
22&  642.96 &85.4&0.84& 84.0& 0.88 &217.2 & 0.82\\ 
40&  866.97 &64.6&1.07& 63.7& 1.21&139.9 & 1.23\\  
  \cline{1-8}
  \multicolumn{8}{|c|} {NLO  $NN+NNN$}\\ 
   \cline{1-8} 
 $\hbar\omega$ &  $\Lambda$ &  \multicolumn{1} {c|}{$E({^3} \rm{H})$}  & $r({^3} \rm{H})$ & \multicolumn{1} {c|}{$E({^3} \rm{He})$}  & $r({^3} \rm{He})$  &  \multicolumn{1} {c|}{$E({^4} \rm{He})$} & $r({^4} \rm{He})$  \\ \hline 
5 &306.52 & 53.9& 1.55 & 53.1  & 1.55 & 99.0&  1.55 \\ 
10&  433.48 & 53.9 &1.14 & 52.9 & 1.16 &  89.9 & 1.17\\ 
22&  642.96& 53.9 &1.04&52.7 &1.13 & 89.7  & 1.34 \\ 
40& 866.97 & 53.9 &1.17&53.1 &1.29 & 109.7  & 1.33 \\  \hline
\end{tabular}
\label{tab:A34_lattice QCD-nlo}
\end{center}
\end{table}%

Let us also discuss uncertainties for lattice nuclei. As we were not
able to fit binding energies and the effective range expansions
simultaneously at $\hbar\omega = 70$~MeV, we infer a physical
breakdown scale $\Lambda_b \approx 1150$~MeV. From the free Fermi gas
estimate~(\ref{fermi}) we find $p_{\rm F} \approx 370$~MeV based on
$^4$He. We assume a conservative $Q=p_F/\Lambda_b\approx 0.4$, and
using Eq.~(\ref{err}) yields the uncertainty $\Delta E_{\rm NLO} ({^4}
{\rm He}) \approx 15$~MeV at NLO for the binding energy of the
$\alpha$ particle. The major uncertainty, however, comes from the
large uncertainties in the input lattice QCD data, which enters the
LECs of our EFT.  For the heavier lattice nuclei discussed below, we
restrict our discussion of uncertainties to the case where LECs are
fit to central values of the lattice QCD data in Table~\ref{data}.

\section{Results for $^{16}$O and $^{40}$Ca}
\label{sec:heavierA}

We compute the nuclei $^{16}$O and $^{40}$Ca with the coupled-cluster
method~\cite{kuemmel1978,bartlett2007,hagen2014}, performed in the
coupled-cluster singles and doubles (CCSD) approximation. The
coupled-cluster method creates a similarity-transformed Hamiltonian
whose vacuum and ground state is a product state. The pionless EFT at
NLO does not include spin-orbit forces, and the coupled cluster method
produces converged results for nuclei $^{4}$He, $^{16}$O, and
$^{40}$Ca because the reference product state for these nuclei exhibit
the usual shell closures of the harmonic oscillator. At LO, the atomic
nuclei $^{16}$O and $^{40}$Ca are not bound with respect to decay into
$^4$He nuclei. This is consistent with previous results:
\citeauthor{stetcu2007} found that $^6$Li is not bound with respect to
$^4$He at LO, and similar results were also found for lattice nuclei
~\cite{contessi2017}.  For these reasons, we report only results at
NLO, which do not exhibit this shortcoming. The $NNN$ potential is
employed in the normal-ordered two-body
approximation~\cite{hagen2007a}, i.e., it contributes to the vacuum
energy of the Hartree-Fock reference, and to the normal-ordered
one-body and two-body matrix elements. This approximation is accurate
for chiral potentials where $NNN$ forces do not enter at
LO~\cite{roth2012}.

The coupled-cluster method employs a translationally invariant
intrinsic Hamiltonian
\be
\label{ham-cc}
H = T-T_{\rm cm} +V_{NN}+V_{NNN} . 
\ee
Here, $T$ denotes the total kinetic energy, $T_{\rm cm}$ the
kinetic energy of the center of mass. We note that the
Hamiltonian~(\ref{ham-cc}) does not reference the center-of-mass
coordinate. This is crucial because the many-body system is solved in
the laboratory system using second quantization.  While the
single-particle states are not eigenstates of the total momentum, the
eigenstates of the Hamiltonian~(\ref{ham-cc}) factor to a very good
approximation into an intrinsic wave function and a Gaussian for the
center-of-mass coordinate~\cite{hagen2009a}.

The number of matrix elements increases significantly when
transforming from the center-of-mass coordinates to the
single-particle basis in the laboratory system, and $NNN$ 
forces can become a bottleneck in the computation of heavy
nuclei. Therefore, in practice the number of matrix elements in the
single-particle oscillator basis needs to be limited by imposing a
truncation on the maximum energy $N_1\hbar\omega$ and
$N_3\hbar\omega$ of a single particle and three particles, respectively. Below
we will study how the results stabilize as $N_1$ and $N_3$ of the
oscillator space in laboratory coordinates are increased.

\subsection{Atomic nuclei}

As a check on the quality of the CCSD approximation, we also computed
the binding energy of $^4$He and found 27.5, 27.2, 29.0, and 27.5~MeV
for the interactions with $N=8$ and $\hbar\omega=5$, 10, 22, and
40~MeV, respectively. These results are in good agreement with the
virtually exact no-core shell-model (NCSM) results presented in
Table~\ref{tab:A34-nlo}; they suggest that the normal-ordered two-body
approximation of the $NNN$ force is accurate. The small differences of
about 1\% between CCSD and NCSM results is most likely due to
neglected triples excitations. For a light nucleus such as $^4$He, the
convergence with respect to $N_3$ is rapid and easily achieved.

The NLO results for $^{16}$O and $^{40}$Ca are shown in
Table~\ref{A1640}. For the larger cutoff values, the NLO binding
energies are within 20\% of the experimental values of about 128 and
342~MeV for $^{16}$O and $^{40}$Ca, respectively. The differences
between our NLO results and experimental data seem roughly consistent
with EFT expectations. The computation also revealed that only about
10\% of the binding energy is correlation energy, i.e., the difference
between the coupled cluster and Hartree-Fock results. This small
fraction is possibly due to the absence of any mixing between $S$ and
$D$ waves.  We note that the convergence with respect to the
three-body energy $N_3$ is excellent for $\hbar\omega=22$~MeV, but
slower for the other oscillator spacings. For these latter oscillator
spacings we also observe that the $N_3$ convergence is slower for
$^{40}$Ca than for $^{16}$O. The associated uncertainty is highest at
$\hbar\omega=40$~MeV, being about 10\%. We note that $^4$He is
virtually converged at all oscillator spacings. We can only speculate
why the $N_3$ convergence is fastest for $\hbar\omega=22$~MeV:
perhaps, this frequency is close to that of the Gaussian
center-of-mass wave function, but this warrants more investigation.

\begingroup
\squeezetable
\begin{table}[htb]
\begin{center}
  \caption{Binding energy of $^{16}$O, $^{40}$Ca for model space
    truncations as indicated, as a function of the cutoff $\Lambda$
    (or the oscillator spacing $\hbar\omega$). All quantities in units
    of MeV. A star $(^*)$ indicates that the energy is approximate and
    did not yet converge after 1000 iterations of the CCSD equations.}
  \begin{tabular}{|c|c|c|c|c|c|}
    \hline\hline
    \multicolumn{2}{|c}{}  & \multicolumn{2}{|c}{$^{16}$O} & \multicolumn{2}{|c|}{$^{40}$Ca}\\
    \cline{1-6}
    $\hbar\omega$& $\Lambda$ & $N_1,N_{3}=12$& $N_1,N_{3}=14$  & $N_1,N_{3}=12$ & $N_1,N_{3}=14$\\ \hline
    5  & $232.35$ &$174.1$ & $174.8$ & $562.5$ & $569.2$\\
    10 & $328.59$ &$136.8$ & $136.2$ & $421.8$ & $415^*{}{}$\\
    22 & $487.38$ &$143.1$ & $143.1$ & $405.8$ & $405.8$\\
    40 & $657.19$ &$144.7$ & $146.2$ & $372.2$ & $400.0$ \\ \hline \hline
  \end{tabular}
  \label{A1640}
  \end{center}
  \end{table}%
\endgroup

\begin{figure}[htb]
\includegraphics[width=0.48\textwidth]{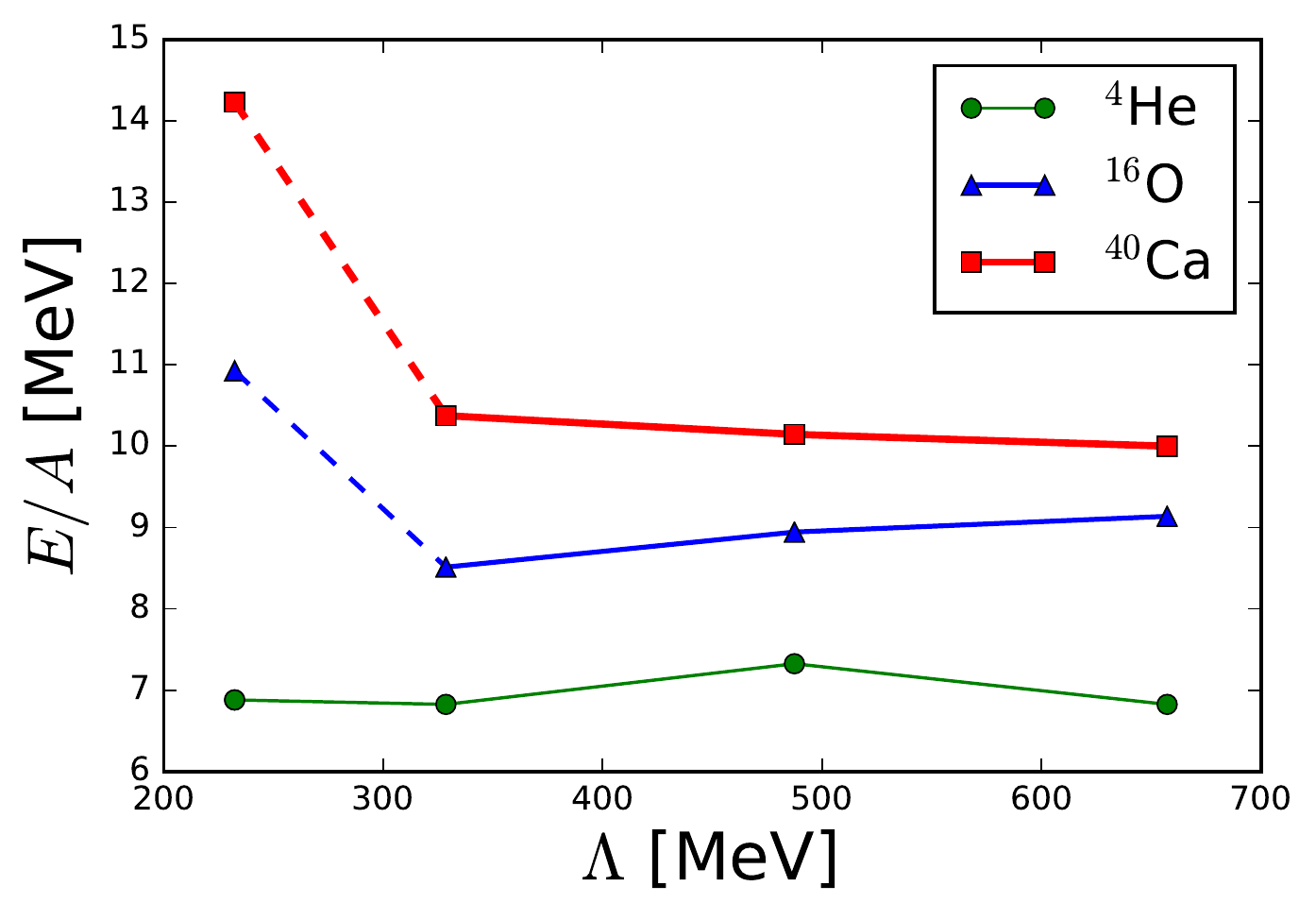}
\caption{(Color online) Binding energy per nucleon for atomic
  ${^{16}}$O (blue triangles), ${^{40}}$Ca (red squares) nuclei
  against UV cutoff of the NLO interaction in the model space $N = 8,
  l = 0$ from coupled cluster calculations.}
\label{fig:4He_O16_40Ca_real}
\end{figure}

Let us also discuss the consistency of our results. At lowest cutoffs,
binding energies are largest, and $^{40}$Ca has a binding energy per
nucleon of about $E/A\approx 14$~MeV at $\hbar\omega=5$~MeV. In a free
Fermi this leads to a Fermi momentum of $k_F\approx 210$~MeV. This is
marginally below the cutoff of $\Lambda\approx 232$~MeV at
$\hbar\omega=5$~MeV. Thus, it is probably safest to limit our
discussion of results to the calculations involving oscillator
spacings $\hbar\omega\ge 10$~MeV. We note that these results also
exhibit a smaller cutoff dependence. Our results show that pionless
EFT binds $^{16}$O and $^{40}$Ca at about 9 and 10~MeV per nucleon,
respectively. Interestingly, these binding energies are close to
results from a chiral EFT at NLO~\cite{binder2016}.
Figure~\ref{fig:4He_O16_40Ca_real} shows binding energies as a
function of the UV cutoff, and the results at the smallest cutoff are
probably inconsistent because of the proximity of the Fermi momentum.

Let us again discuss uncertainties. We adopt the estimates made in
light nuclei to the present case.  Thus, the uncertainty from the EFT
interaction is about $10\%$ at NLO, implying $\Delta E_{\rm
  NLO}({^{16}} {\rm O}) \sim 15$ MeV and $\Delta E_{\rm NLO}({^{40}}
{\rm Ca}) \sim 40$ MeV. The variation of binding energies with UV
cutoff at fixed $N$ and $N_3$ is in this range, and so are the
uncertainties from the $N_3$ convergence of coupled cluster results at
fixed $\Lambda_{\rm UV}$.

\subsection{Lattice nuclei}

We re-compute $^4$He with the coupled-cluster method and at NLO we
find binding energies $E=98.0$, 89.0, and 88.1~MeV for the
interactions with $\hbar\omega=5$, 10, and 22~MeV, respectively. This
is in agreement with the NCSM results of Table~\ref{tab:A34_lattice
  QCD-nlo} and suggests that the normal-ordered two-body approximation
of the $NNN$ potential is accurate also for lattice nuclei. Again we
find only a small amount of about 10\% for the correlation energy. The
small differences of about 1\% between CCSD and NCSM results is due to
neglected triples excitations.  In contrast, at $\hbar\omega = 40$~MeV
we find a $^4$He binding energy of 99.5~MeV, which differs from the
NCSM result by about 10\%. Closer inspection and varying the strength
of the $NNN$ interaction suggests that this discrepancy is due to the
normal-ordering two-body approximation of the $NNN$ interaction at
this frequency. As the normal-ordered two-body approximation is
expected to improve with increasing mass number~\cite{roth2012}, we
will also compute $^{16}$O and $^{40}$Ca at $\hbar\omega=40$~MeV,
keeping in mind a conservative 10\% uncertainty due to the normal
ordering approximation. 

Our results for $^{16}$O and $^{40}$Ca are shown in
Table~\ref{A1640-lat}. We observe that lattice nuclei are bound with
approximately 30~MeV per nucleon at $\hbar\omega=22$~MeV.  In a free
Fermi gas this corresponds to a Fermi momentum $p_F\approx
400$~MeV. This is well below the pion mass employed in the lattice QCD
calculations and also below the corresponding cutoff $\Lambda =
642.96$~MeV of the EFT.  For the smaller oscillator spacings
$\hbar\omega = 5$ and $10$~MeV (and correspondingly smaller cutoffs),
however, the Fermi momentum is above the cutoff. Therefore, our
calculations are probably meaningful only for $\hbar\omega=22$~MeV and
40~MeV.

  \begingroup
\squeezetable
  \begin{table}[htb]
\begin{center}
  \caption{Binding energies of the lattice nuclei $^{16}$O, $^{40}$Ca
    for model space truncations as indicated, as a
    function of the cutoff $\Lambda$ (or the oscillator spacing
    $\hbar\omega$). All quantities in units of MeV. }
 \begin{tabular}{|c|c|c|c|c|c|}
    \hline\hline
    \multicolumn{2}{|c}{}  & \multicolumn{2}{|c}{$^{16}$O} & \multicolumn{2}{|c|}{$^{40}$Ca}\\
    \cline{1-6}
    $\hbar\omega$& $\Lambda$ & $N_1,N_{3}=12$& $N_1,N_{3}=14$  & $N_1,N_{3}=12$ & $N_1,N_{3}=14$\\ \hline
    22 & $642.96$ &$429.5$ & $429.5$ & $1187.0$ & $1168.5$\\
    40 & $866.97$ &$547.8$ & $546.0$ & $1252.0$ & $1422.0$ \\ \hline \hline
  \end{tabular}
  \label{A1640-lat}
  \end{center}
  \end{table}%
\endgroup

As for the light latice nuclei, the error in the binding energy at NLO
for lattice nuclei is of the order of 15\% which leads to $\Delta
E_{\rm NLO}({^{16}} {\rm O}) \approx 75$~MeV and $\Delta E_{\rm
  NLO}({^{40}} {\rm Ca}) \approx 200$~MeV for ${^{16}} {\rm O}$ and
${^{40}} {\rm Ca}$ lattice nuclei. We remind the reader that this
estimate excludes the dominant uncertainties due to the limited
precision of the lattice QCD results that are input. At $\hbar\omega =
40$~MeV there also is an additional 10\% uncertainty estimate due to
the normal ordering approximation of $NNN$ forces.

We use the results at $\hbar\omega=22$~MeV to compute the volume and
surface terms $a_V$ and $a_S$, respectively, of the Bethe-Weizs\"acker
formula
\be
E(A) = a_V A - a_S A^{2/3} .
\label{eq:L-D}
\ee
Here, $E(A)$ is binding energy of an $A$-nucleon system.  We find
$a_V\approx 35$~to~40~MeV and $a_S \approx 14$~to~22~MeV.

\section{Summary}
\label{sum}
We implemented pionless EFT as a DVR in the harmonic oscillator
basis. The DVR formulation has several advantages over traditional
approaches that transform momentum-space interactions to the
oscillator basis: (i) The UV cutoff and regulator are tailored to the
underlying basis; (ii) the DVR facilitates the computation of matrix
elements as this becomes essentially a function call; (iii) the IR
improvement allows one to optimize interactions directly in the
harmonic oscillator basis. We showed that the DVR formulation indeed
yields an EFT with the correct low-momentum behavior.

To put the DVR in the context of momentum-space EFTs, we performed
many checks and tests, and reported them in a set of Appendices. The
Thomas effect~\cite{thomas1935} and the Tjon line~\cite{tjon1975} can
be understood analytically from scaling arguments that connect the
potential matrix elements at different UV cutoffs. Different
implementations of the EFT -- at constant UV cutoff -- yield results
that differ by small amounts, consistent with expectations regarding
regulator dependencies. 

We calibrated the pionless EFT for atomic nuclei and for lattice
nuclei (at an unphysical pion mass) in $A=2,3$ systems and make
predictions for $^4$He, $^{16}$O, and $^{40}$Ca. At LO $^{16}$O and
$^{40}$Ca are not bound with respect to decay into $\alpha$ particles;
this deficiency is remedied at next-to-leading order. Varying the UV
cutoff by about a factor of two suggests that pionless EFT at
next-to-leading order yields meaningful results for the binding
energies of medium-mass nuclei that are consistent with chiral EFT
calculations at that order.

Our results also suggest that medium-mass nuclei can be connected to
lattice QCD input. To make further progress in this direction,
however, requires a resolution of the controversy between the
different lattice QCD approaches to light nuclei, increasing the
precision of the lattice QCD results that are input to EFTs, and
finally, moving towards the physical pion mass.

\begin{acknowledgments}
  We thank Bijaya Acharya, Dick Furnstahl, Lucas Platter, and Gautam
  Rupak for useful discussions.  We also thank Bira van Kolck, Daniel
  Phillips, and Ionel Stetcu for questions and discussions during the
  program INT 17-1a ``Toward Predictive Theories of Nuclear Reactions
  Across the Isotopic Chart,'' and Daniel Phillips for several useful
  communications.  This material is based upon work supported in part
  by the U.S.  Department of Energy, Office of Science, Office of
  Nuclear Physics, under Award Numbers DE-FG02-96ER40963 (University
  of Tennessee), DE-SC0008499, DE-SC0018223 (SciDAC NUCLEI
  Collaboration), the Field Work Proposal ERKBP57 at Oak Ridge
  National Laboratory (ORNL), the Leadership Computing Facility at
  ORNL under contract number DEAC05-00OR22725 (ORNL), Swedish Research
  Council under Grant No. 2015- 00225 and Marie Sklodowska Curie
  Actions, Cofund, Project INCA 600398.  S.B. gratefully acknowledges
  the financial support from the Alexander-von-Humboldt Foundation
  (Feodor-Lynen fellowship).
\end{acknowledgments}

\appendix


\section{Overview of Appendices}
The formulation of pionless EFT as a DVR in the oscillator basis
invites questions regarding details of the implementation and its
relation to established results. In these Appendices, we address a few
relevant points. In App.~\ref{app-dvr} we show that a continuous
family of DVR formulations exists, including one that exhibits a
zero-momentum point. In App.~\ref{app-er} we show that the IR
improvement of the LO two-body contact exhibits effective-range
corrections that are parametrically small and inverse proportional to
the number of DVR states. In App.~\ref{sec:scale} we derive simple
scaling laws that govern the potential matrix elements as the
oscillator frequency or the nucleon mass is varied. This makes it
particularly simple to relate matrix elements corresponding to
different UV cutoffs and to different nucleon masses. It also allows
us to derive known relations such as the Thomas
effect~\cite{thomas1935} or the Tjon~\cite{tjon1975} correlations.  In
App.~\ref{app-wigner} we confirm that our formulation of pionless EFT
obeys the Wigner bound~\cite{wigner1955}. In App.~\ref{harmonic
  oscillatorEFT} we study the regulator dependence of our EFT by
comparing different combinations of $(N,\hbar\omega)$ that yield
similar UV cutoffs. In App.~\ref{trunc-E3max} we discuss the effects
of oscillator basis truncation on $NNN$ contact with cutoff in Jacobi
momenta. Finally, App.~\ref{app-extra} is dedicated to IR
extrapolations. There, we show that L\"uscher-like~\cite{luscher1985}
formulas account for finite-size corrections that stem from finite
harmonic oscillator spaces

\section{DVR with a zero-momentum point} 
\label{app-dvr}

A discrete variable representation (DVR) in momentum space consists of basis
functions $\tilde{\phi}_{\mu,l}(k)$ that are orthogonal to each other
and localized around certain discrete momentum points. Let us start by expressing the 
DVR basis in terms of oscillator wave functions 
\be
\label{phi_mu_kappa}
\tilde{\phi}_{\kappa,l}(k) = d_{\kappa,l} \sum_{n=0}^{N_l}
\tilde{\psi}_{n,l}(\kappa)\tilde{\psi}_{n,l}(k) .
\ee
Here $\kappa$ is a discrete momentum (to be determined) and
$d_{\kappa,l}$ is a normalization constant. The DVR wave function
$\tilde{\phi_{\kappa,l}}$ is the projection of a spherical wave with
momentum $\kappa$ onto the finite harmonic oscillator basis. To see
this, we start from the completeness relation
\be
\sum_{n=0}^\infty \tilde{\psi}_{n,l}(\kappa)\tilde{\psi}_{n,l}(k) = {\delta(k-\kappa)\over k\kappa} , 
\ee
and note that this is also the orthogonality condition for spherical
waves with momenta $k$ and $\kappa$, respectively.

We need to determine the DVR points $\kappa = \kappa_\mu$ in the wave
function~(\ref{phi_mu_kappa}) such that wave functions belonging to
different $\kappa_\mu$ are orthogonal to each other. Here $\mu$
enumerates the discrete set of momenta (the DVR points).  The overlap
between two such wave functions is
\ba
\lefteqn{\int\limits_{0}^{\infty} dk k^2 \tilde{\phi}_{\kappa_\mu,l}(k) \tilde{\phi}_{\kappa_\nu,l}(k) }\nonumber\\
&=& d_{\kappa_\mu,l} d_{\kappa_\nu,l} \sum\limits_{n=0}^{N_l} \tilde{\psi}_{n,l}(\kappa_\mu) \tilde{\psi}_{n,l}(\kappa_\nu) \\
&=& d_{\kappa_\mu,l} d_{\kappa_\nu,l} \sqrt{(N_l + 1)(N_l + l + 3/2)}  \nonumber\\
&\times& \frac{\tilde{\psi}_{N_l,l}(\kappa_\mu) \tilde{\psi}_{N_l + 1,l} (\kappa_\nu) - \tilde{\psi}_{N_l +1,l}(\kappa_\mu) \tilde{\psi}_{N_l,l}(\kappa_\nu)} {b^2\left(\kappa_\mu^2 - \kappa_\nu^2 \right)} \nonumber.
\label{ortho_eqn}
\ea
For $\kappa_{\mu} \neq \kappa_{\nu}$, orthogonality implies
\be
\frac{\tilde{\psi}_{N_l+1,l}(\kappa_{\mu})}{\tilde{\psi}_{N_l,l}(\kappa_{\mu})} = \frac{\tilde{\psi}_{N_l+1,l}(\kappa_{\nu})}{\tilde{\psi}_{N_l,l}(\kappa_{\nu})} ,
\ee
and we can solve for DVR points $\kappa_{\mu}$ by demanding that 
\be
\label{ratio_R}
\frac{\tilde{\psi}_{N_l+1,l}(\kappa_\mu)}{\tilde{\psi}_{N_l,l}(\kappa_\mu)}  = R , 
\ee
with $R$ being a constant.  

\begin{figure}[htb]
\includegraphics[width=0.48\textwidth]{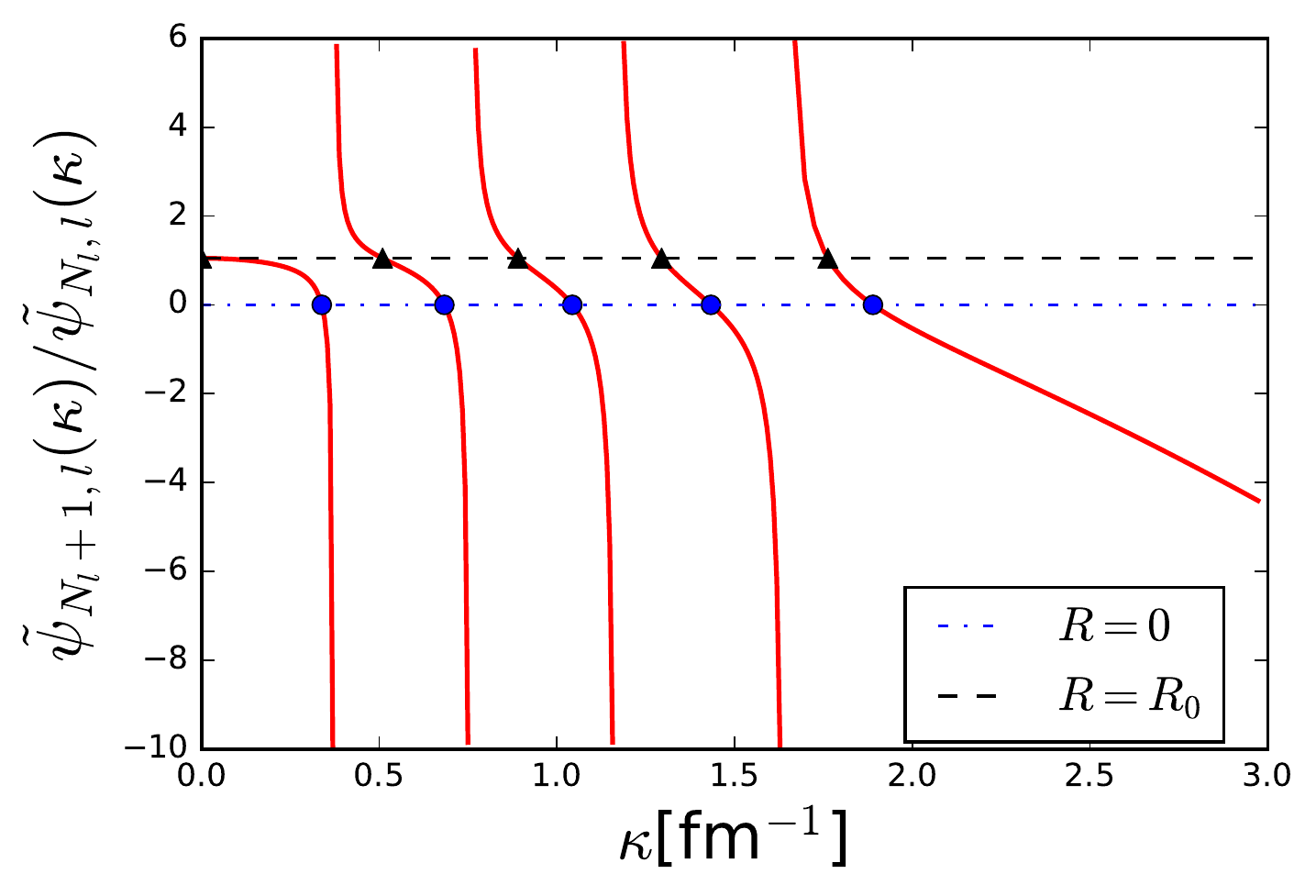}
\caption{(Color online) Solid red curve: The ratio $R$ of
  Eq.~(\ref{ratio_R}) as a function of the momentum in a model space
  $N = 8$, $\hbar\omega = 22$~MeV, and $l=0$.  The dashed-dotted
  horizontal line corresponds to $R= 0$, and its intersection with the
  red curve, denoted by solid blue dots, yields the DVR points we used
  in the main text. The dashed line corresponds to the ratio $R=R_0$
  in Eq.~(\ref{ratio}), and the intersection of this line with red
  curve, denoted by solid black triangles, marks the DVR points of the
  DVR discussed in this Appendix.}
\label{possible_DVRs}
\end{figure}

Figure~\ref{possible_DVRs} shows the ratio in Eq.~(\ref{ratio_R}) as a
a function of momentum (red curve) for a model space with $N = 8$ and
$l = 0$ . The dashed-dotted horizontal line $R=0$ yields the blue
circles as intersection points; these are the DVR points we employed
in the main text of this paper. The dashed horizontal line
\be
\label{ratio}
R = \frac{\tilde{\psi}_{N_l+1,l}(0)}{\tilde{\psi}_{N_l,l}(0)} = \sqrt{\frac{N_l+l+3/2}{N_l+ 1}}
\ee
yields the black triangles as intersection points. This is the DVR we
seek as it contains the point $k=0$. We note that there is a
continuous set of DVRs, each being identified by the value of $R$.

To find the DVR points $\kappa_\mu$, we solve
\be
0 = \sqrt{\frac{N_l+l+3/2}{N_l+ 1}}
\tilde{\psi}_{N_l,l}(\kappa) - \tilde{\psi}_{N_l + 1,l}(\kappa) , 
\ee
which is equivalent to
\begin{align}
\label{DVR2_k}
0 &= (N+l+3/2)L_{N_l}^{l+ 1/2} (\kappa^2 b^2) - (N_l + 1)L_{N_l + 1}^{l+1/2} (\kappa^2 b^2) \nonumber \\ 
&= \kappa^2 b^2 L_{N_l}^{l+3/2}(\kappa^2 b^2) .
\end{align}
In the last step we used formula 8.971(4) of
Ref.~\cite{gradshteyn2007}. Thus, the DVR points are $\kappa = 0$ and
the $N_l$ roots of the polynomial $L_{N_l}^{l+3/2}(\kappa^2 b^2)$.  It
is understood that discrete momentum points are different for each
partial wave and to keep our notation simpler we denote them by
$\kappa_\mu$ instead of $\kappa_{\mu,l}$.

We note that Eq.~(\ref{ratio_R}) only exhibits $N_l$ solutions for $R >
\tilde{\psi}_{N_l+1,l}(0)/\tilde{\psi}_{N_l,l}(0)$.  For $R \to
+\infty$, for instance, the solutions are $N_l$ zeros of the
generalized Laguerre polynomial $L_{N_l}^{l+1/2}(\kappa^2 b^2)$. This
yields only $N_l$ DVR functions. The remaining basis function is
$\tilde{\psi}_{N_l + 1,l}(k)$, but the resulting set of $N_l + 1$
basis functions is no longer a DVR.

We return to Eq.~(\ref{ortho_eqn}) and compute the normalization for
DVR wave functions whose momenta fulfill Eq.~(\ref{DVR2_k}). This
yields
\ba
\label{dmu}
d_{\kappa_\mu,l}^{-2} &=& -\sqrt{N_l(N_l+1)(N_l+l+3/2)}
\nonumber\\ &\times&{\tilde{\psi}_{N_l-1,l+1}(\kappa_\mu)
  \tilde{\psi}_{N_l+1,l}(\kappa_\mu)\over \kappa_\mu b}
\nonumber\\ &=&
(N_l+l+3/2)\left[\tilde{\psi}_{N_l,l}(\kappa_\mu)\right]^2\nonumber\\ &=&
(N_l+1)\left[\tilde{\psi}_{N_l+1,l}(\kappa_\mu)\right]^2 .  \ea To
derive this result, we employ the rule of l'Hospital, Eq. 8.971(2)
from Ref.~\cite{gradshteyn2007} and the recurrence relations between
Laguerre polynomials. Returning to Eq.~(\ref{phi_mu_kappa}) we compute
\be
\label{wave3}
\tilde{\phi}_{\kappa_\mu, l}(k) = {k\over b(k^2-\kappa_\mu^2)}\tilde{\psi}_{N, l+1}(k) \ .
\ee

We note that the norm $d_{0,l}$ diverges as $(kb)^l$ for $\kappa =0$
and $l > 0$.  The corresponding localized eigenfunction in
Eq.~(\ref{phi_mu_kappa}) remains finite because $\tilde{\psi}_{n,l}(0)
\propto (kb)^l$ and we have
\begin{align}
\label{wave0}
\tilde{\phi}_{0, l}(k) &= \sqrt{\frac{N_l!\Gamma(l+5/2)}{\Gamma(N_l + l+ 5/2)\Gamma(l+3/2)}} \nonumber \\
&\times \sum \limits_{n = 0}^{N_l}\sqrt{\frac{\Gamma(n+l+3/2)}{n!}}\tilde{\psi}_{n,l}(k^2 b^2) .
\end{align}

We want to compare the DVR of this Appendix to the one we used
in the main text of the paper. In the large $N_0$ limit, the $(l=0)$
wave functions of the latter DVR are essentially $j_0(k_{\mu,0}r)$ with
$k_{\mu,0}\approx \mu\pi/L$. In contrast, the DVR points $\kappa_\mu$ of the
DVR developed in this Appendix which explicitly include $ k =0$ momentum
satisfy $\kappa_\mu \approx (2\mu+1)\pi/(2L)$, i.e., the DVR wave functions 
approach a Neumann boundary condition close to $r=L$. 
For other values of the ratio $R$ of Eq.~(\ref{ratio}), one obtains mixed 
boundary conditions close to $r=L$.

\begin{figure}[htb]
   \includegraphics[width=0.48\textwidth]{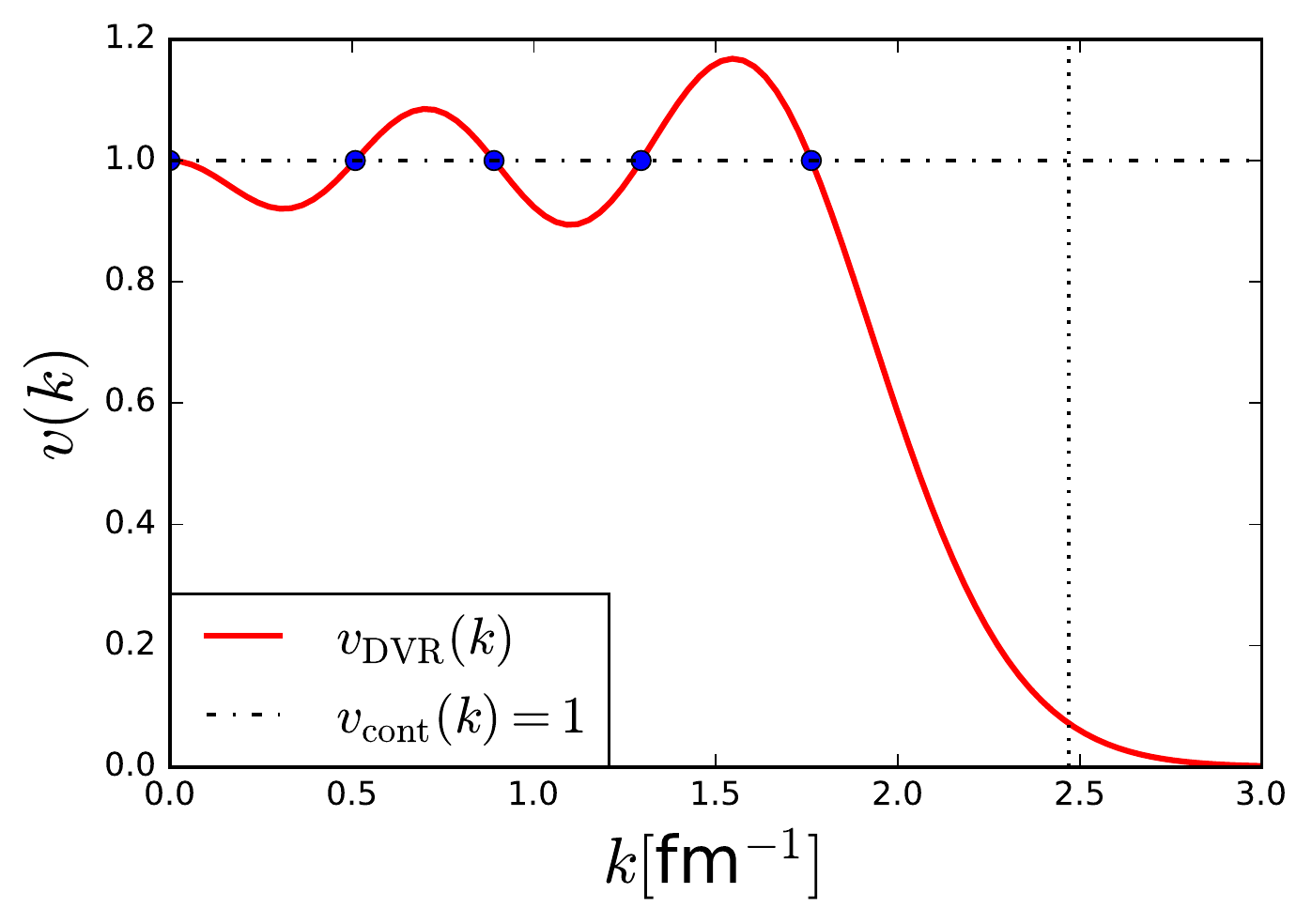}
   \includegraphics[width=0.48\textwidth]{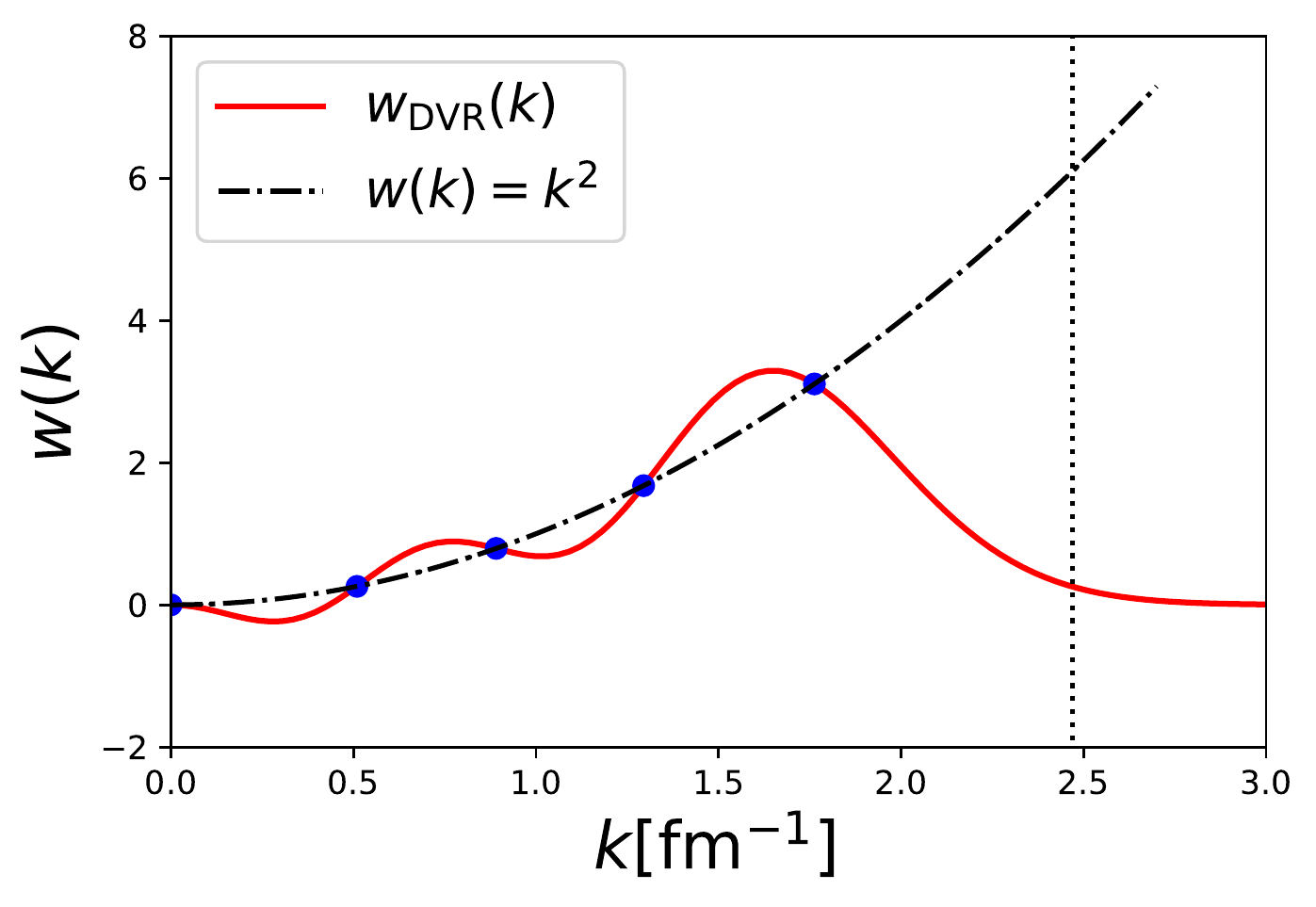}
\caption{(Color online) The red curve shows the contact realized in a
  DVR with a zero-momentum point, in comparison with the original
  momentum-space contact shown as a dash-dotted black line.  The solid
  dots represent the DVR momenta.  Note that $k=0$ is a DVR point.
  The upper (lower) panel is for $NN$ LO (NLO) interaction in
  pionless EFT.  The vertical black dotted line depicts the UV
  cutoff introduced by finite harmonic oscillator basis space with
  $l=0$, $N = 8$, and $\hbar\omega = 22$~MeV.}
  \label{fig:DVR2_2N_V}
  \end{figure}

Let us compute the $S$-wave $NN$ contacts $v(k)=1$ at LO and $w(k)=k^2$ at NLO
in this DVR. The results are shown as solid red lines in the upper
and lower panels of Fig.~\ref{fig:DVR2_2N_V}, respectively.  The
original momentum-space interaction is plotted as a black dashed line.
Dots representing the new discrete momenta, now including $k=0$.
The model-space parameters are $N = 8$ and $\hbar\omega = 22$~MeV.

While these DVR potentials are slightly more oscillatory than the
IR-improved DVR potential in Figs.~\ref{fig:4} and \ref{fig:5}, they
reproduce the original momentum-space interaction much better than the
other DVR without IR improvement (red dashed curves in
Figs.~\ref{fig:4} and \ref{fig:5}).

\begin{figure}[htb]
   \includegraphics[width=0.48\textwidth]{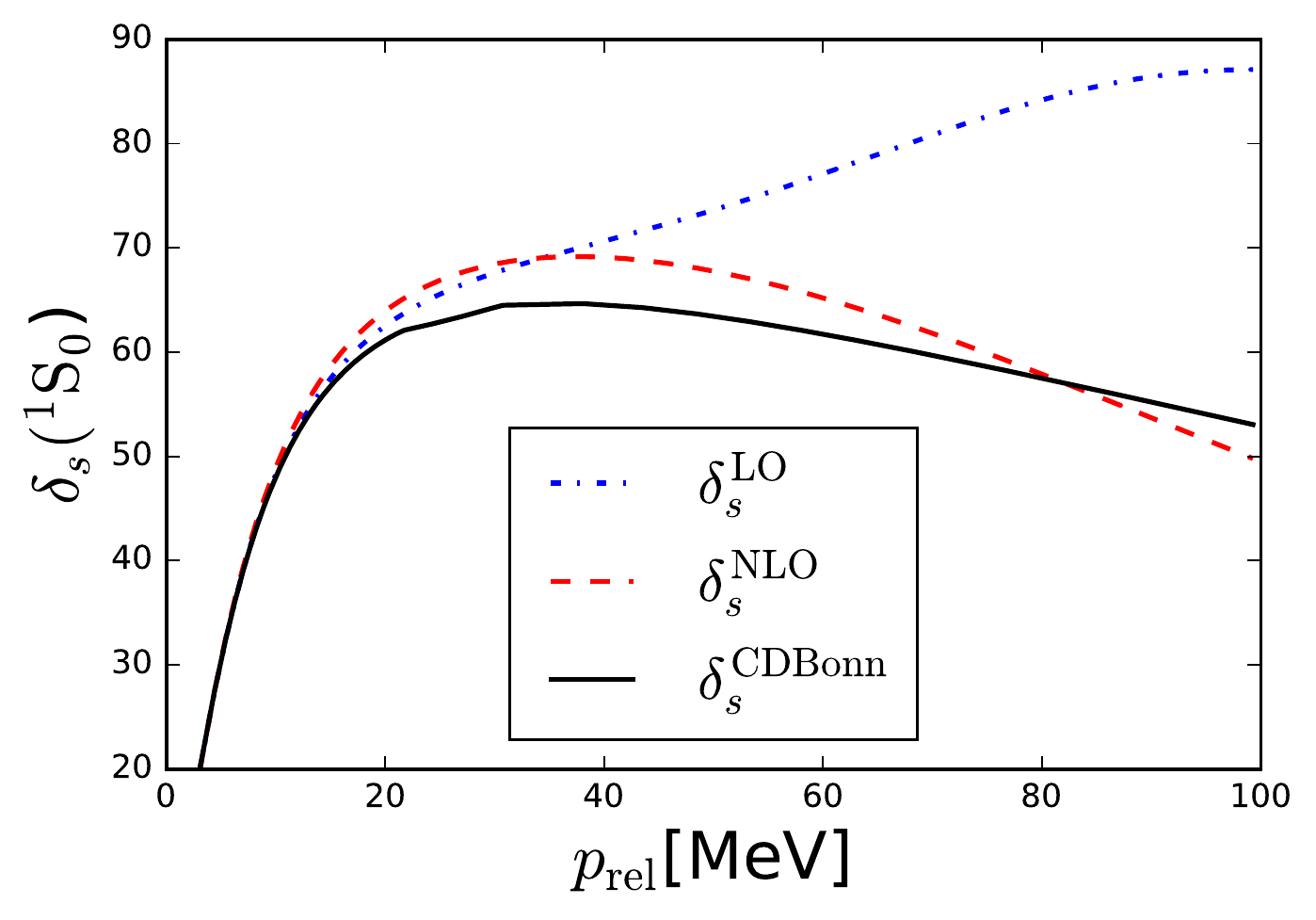}
   \includegraphics[width=0.48\textwidth]{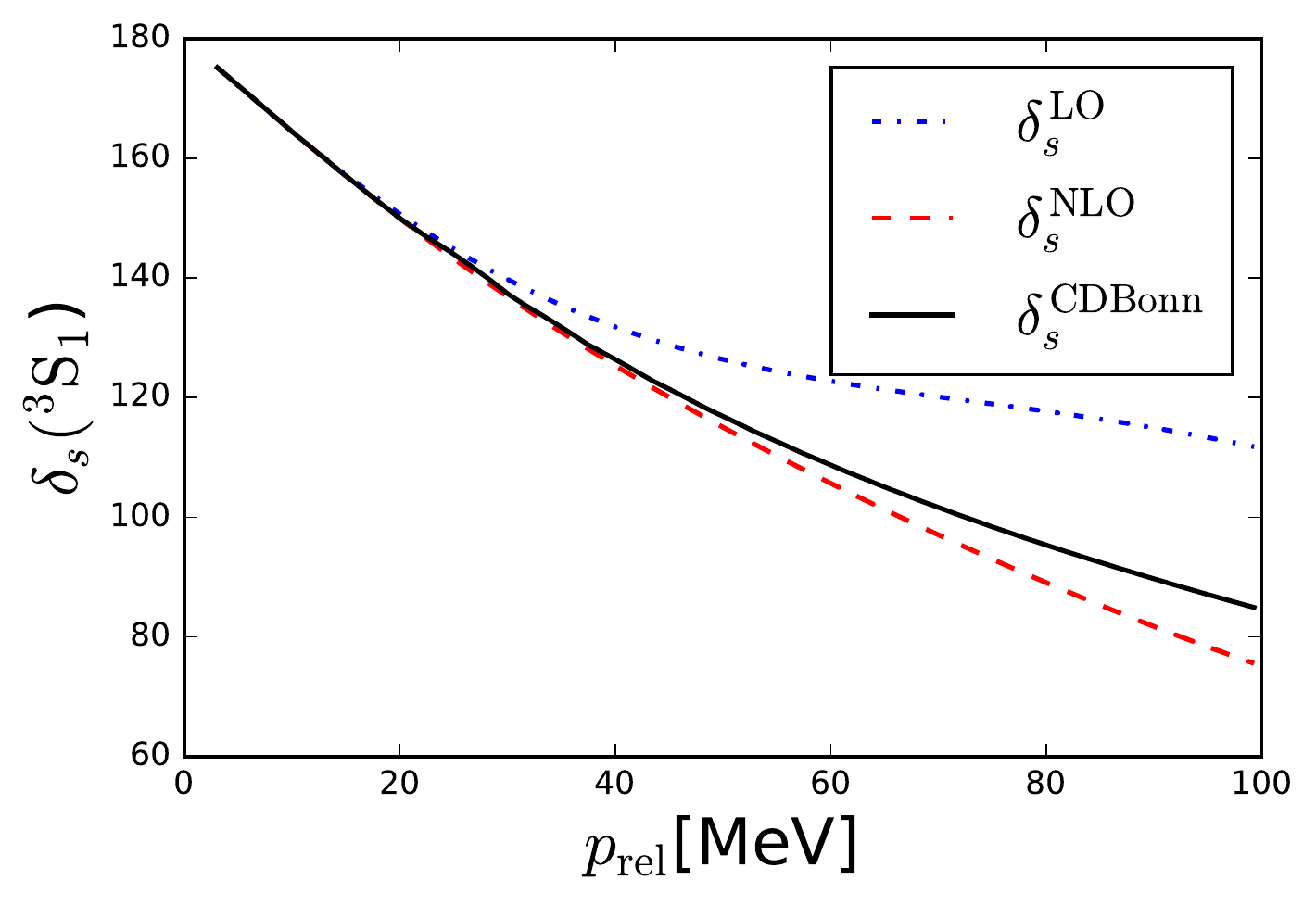}
\caption{(Color online) The $^1S_0$ (upper panel) and $^3S_1$ phase
  shifts (lower panel) from a DVR potential at NLO (LO) in pionless
  EFT in a model space $N = 8, \hbar\omega = 22$~MeV at NLO (red
  dashed line) and LO (blue dot-dashed line).  The black curve shows
  the neutron-proton phase shifts of the CD-Bonn potential.}
  \label{DVR2_LO_NLO}
  \end{figure}

This makes it interesting to compute phase shifts with the DVR of this
Appendix.  Figure~\ref{DVR2_LO_NLO} shows the LO and NLO $np$ phase
shifts from the DVR interaction in $^1S_0$ (top) and $^3S_1$ (bottom)
partial wave channels. Since at NLO $NN$ interaction in the DVR representation 
has incorrect curvature at $k=0$, the NLO phase shifts are slightly
oscillatory in both channels in comparison to phase shifts form IR
improved interaction in the other DVR. Even so, we find it to be a
simple alternative to the IR improvement.

\section{IR improvements and effective range}
\label{app-er}

We want to understand the quality of the IR improvement of the $NN$
contact.  As the number of DVR states $N_0$ is finite, we have to
understand finite-size effects.  Here, we focus on the curvature of
the function~(\ref{v-dvr-lo}) at $k=0$, as this introduces a finite
range correction. To understand the finite size effects, we recall
that -- at low momenta and long wave lengths -- the spherical harmonic
oscillator basis is indistinguishable from a spherical cavity with
radius $L=\pi/k_{0,0}$. This allows us to understand finite-size
effects in the oscillator DVR by studying corresponding effects in a
spherical cavity.

We therefore consider a spherical cavity of radius $L$.  Eigenfunctions for
$S$ waves with momentum $k_\mu$ are spherical Bessel functions
$j_0(\kappa r)$. In momentum space, the corresponding wave function
results from a Fourier-Bessel transform
\ba
\label{finmom}
\tilde{\psi}_\kappa(k) &\equiv& {2\over \pi}\int\limits_0^L dr r^2 j_0(\kappa r)j_0(kr) \nonumber\\
&=& {1\over \pi \kappa k}  \left({\sin{(k-\kappa)L}\over k- \kappa} -
      {\sin{(k+ \kappa)L}\over k+ \kappa} \right).
\ea
The momentum-space function $\tilde{\psi}_\kappa(k)$ is a smeared
Dirac-$\delta$ function with a peak at $k= \kappa$ and also exhibits
oscillations.  As a check, we see that $\tilde{\psi}_\kappa(k)\to
\delta(k- \kappa)/(\kappa k)$ for $L\to\infty$. The
expression~(\ref{finmom}) can be simplified when it is evaluated at
the quantized momenta of 
      \be
      k_\mu \equiv {\mu\pi\over L} \ .
      \ee
Then we have
      \be
      \tilde{\psi}_{k_\mu}(k)= {2L(-1)^\mu\over \pi}{j_0(kL)\over {k^2-k_\mu^2}} \ .
      \ee
In particular, the DVR property is 
\be
\tilde{\psi}_{k_\mu}(k_\nu)=\delta_\mu^\nu c_\mu^{-1}
\ee
with
\be
c_\mu^{-1} \equiv {L^3\over  \pi^3\mu^2} \ .
\ee

To see the analogy with the oscillator DVR, we note that
$\tilde{\psi}_{k_\mu}(k)\leftrightarrow \tilde{\psi}_{\mu,0}(k)$,
and that $c_\mu \leftrightarrow c_{\mu,0}$.  In an EFT based on $N$
spherical Bessel functions, we would approximate the contact function
$v_{\rm DVR}(k)$ of Eq.~(\ref{v-dvr}) as
\be
\tilde{v}(k) \approx \sum_{\mu=1}^{N}
      c_\mu \tilde{\psi}_{k_\mu}(k) \ .
\ee
Here, the tilde indicates that this function exhibits oscillations.
By construction, $\tilde{v}(k_\mu)=1$, but this function is
certainly not a constant. It has an oscillatory component, and at zero
momentum we have
      \ba
      \tilde{v}(0)=\left\{\begin{array}{ll}
      2 & \mbox{for $N$ odd} \ , \\
      0 & \mbox{for $N$ even} \ . \end{array}\right.
      \ea

This suggests to make an IR improvement by adding one more basis
function with momentum $k_{N+1}$, and with half the usual
amplitude. (Alternatively, we could reduce the amplitude at $k_N$
by a factor two as is approximately done for the oscillator DVR, see
Fig.~\ref{fig:4}.)  This yields
      \ba
      \label{gimp}
      \overline{v}(k) &=& \sum_{\mu=1}^{N} c_\mu \tilde{\psi}_{k_\mu}(k)
      +{c_{N+1}\over 2} \tilde{\psi}_{k_{N+1}}(k)
      \nonumber\\
      &=& 2j_0(kL) \sum_{\mu=1}^{N} {(-1)^\mu \mu^2\over \left({kL\over \pi}\right)^2-\mu^2} \nonumber\\
      &&+ j_0(kL) {(-1)^{N+1} (N+1)^2\over \left({kL\over \pi}\right)^2-(N+1)^2} \ .
      \ea
By construction, $\overline{v}(k_\mu)=1$ for $\mu=1, \ldots, N$ and
$\overline{v}(0)=1$. The function $\overline{v}$ exhibits oscillations
with a much reduced amplitude in comparison to $\tilde{v}$, and it is an
even function in $k$. To gauge its quality in the IR, we compute its
curvature at $k=0$. For $k\to 0$ we find
      \ba
      \label{gimpexpand}
      \overline{v}(k) \approx 1- {k^2L^2\over \pi^2}
      \left[{\pi^2\over 6} + 2\sum_{\mu=1}^N {(-1)^\mu\over \mu^2} - {(-1)^{N}\over (N+1)^2}\right] \ .\nonumber\\
      \ea
We use
      \be
      \sum_{n=1}^K\frac{(-1)^n}{n^2} = {1\over 2}\sum_{n=1}^{\left[{K\over 2}\right]} {1\over n^2} - \sum_{n=1}^K {1\over n^2} \ .
      \ee
Here, $[x]$ denotes the integer part of $x$. We rewrite
      \ba
      \sum_{n=1}^K {1\over n^2} &=& \sum_{n=1}^\infty {1\over n^2} -\sum_{n=K+1}^\infty {1\over n^2}\nonumber\\
      &=& {\pi^2\over 6} -\sum_{n=K+1}^\infty {1\over n^2} \ ,
      \ea
and employ the Euler-Maclaurin summation formula
      \be
      \sum_{n=K+1}^\infty {1\over n^2}= {1\over K}-{1\over 2K^2} + {\cal O}(K^{-3}) \ .
      \ee
Thus, the expansion~(\ref{gimpexpand}) becomes
      \be
      \overline{v}(k) \approx 1+ {\cal O} \left({(kL)^2\over N^3}\right) \ .
      \ee
This result is also confirmed numerically.  Using
 $L\Lambda\propto N$, we see that the quadratic correction scales
      as
      \be
      \label{scale}
            {1\over N} \left({k\over\Lambda}\right)^2 \ .
\ee

Thus, the effective range correction of the IR-improved contact is
parametrically small as the number $N$ of DVR points increases. This
is an interesting and encouraging result. The IR improved contact in
an EFT based on the lowest $N$ discrete momentum states of a spherical
cavity exhibits small effective-range corrections proportional to
$1/N$. This correction vanishes as $N\to\infty$ and is clearly a
finite-size effect.

We also note here the IR improvement of the contact essentially
reduces the weight of the eigenfunction corresponding to the largest
momentum by a factor of about 0.5.  This suggests a simple way to
perform IR improvements.  In the partial wave with angular momentum
$l$ we introduce non-local regulators for the potential via
\ba
V(p',l';p,l) \to e^{-\left({p'\over k_{N_{l'},l'}}\right)^{2n}}  V(p',l';p, l) e^{-\left({p\over k_{N_l, l}}\right)^{2n}} . \nonumber
\ea 
This widely used regulator approximately introduces the factor
one-half reduction at about the right momentum. In practice we find
that this simple procedure works quite well, in particular for chiral
interactions where analytical IR improvements might be more tedious.

\section{Thomas effect and Tjon line}
\label{sec:scale}

In this Appendix, we derive simple scaling relations that hold at fixed
$N$. We will use them to explain the key results the Thomas
effect~\cite{thomas1935} (i.e., the increase of binding in the
three-nucleon system with increasing cutoff of the $NN$ interaction)
and the Tjon line~\cite{tjon1975} (i.e., the correlation between
binding energies of the $A=3$ and $A=4$ bound states.  These results
suggest that the EFT as a DVR in the oscillator basis is also useful
to obtain analytical insights.

In what follows we vary the UV cutoff~(\ref{lambdaUV}) at fixed number
of oscillator shells $N$ by changing the oscillator length
$b$, i.e., the oscillator frequency $\hbar\omega$. 
We also allow the nucleon mass to vary, as this will be useful with
view on lattice nuclei. As we will see, varying $\hbar\omega$ or nucleon mass $m$
simply rescales the matrix elements of the contact interactions and
kinetic energy in the oscillator EFT. 

From Eqs.~(\ref{cmu}) and (\ref{harmonic oscillatorwavefunction}) we find
$\tilde{\psi}_{n,l}(k) \propto b^{3/2}$, $c_{\mu,l} \propto b^{-3/2}$,
and $k_{\mu,l} \propto b^{-1}$. Thus, the roots of the generalized
Laguerre polynomial $L_{N+1}^{l+1/2}(k^{2} b^{2})$ do not change, and
a rescaling of $b$ and $m$ simply changes the matrix elements of the
LO contact, the NLO contact, and the three-nucleon force as
\begin{eqnarray}
  \label{scaling}
  V_{\rm LO}  &\propto& C_{\rm LO}b^{-3} \propto C_{\rm LO}\left(m \hbar\omega\right)^{3/2} , \nonumber\\
  V_{\rm NLO}&\propto& C_{\rm NLO}b^{-5} \propto C_{\rm NLO}\left(m \hbar\omega\right)^{5/2} , \nonumber\\
  V_{NNN}&\propto& C_{NNN}b^{-6} \propto C_{NNN}\left(m \hbar\omega\right)^3 , 
\end{eqnarray}
respectively. The Schr\"odinger equation for two nucleons at leading
order in either the $^1S_0$ or the $^3S_1$ partial wave is
\be
\label{schrodinger2}
{\hbar^2\over m b^2}\left(\hat{t}_2 + {{m\hbar}} {C_{\rm LO} \over b}\hat{v}_2\right)|\psi\rangle = E_2|\psi\rangle
\ee
Here, $\hat{t}_2$ and $\hat{v}_2$ are dimensionless matrices of the
kinetic and potential energies, respectively. Thus,
\be
\left(\hat{t}_2 + {{m\hbar}}{C_{\rm LO}\over b} \hat{v}_2\right)|\psi\rangle = {E_2\over \hbar\omega}|\psi\rangle \approx 0. 
\ee
The last approximation is exact in the case of an infinite scattering
length or a zero-energy bound state. It is a good approximation in
general as most model spaces of {\it ab initio} calculations have
$E_2/(\hbar\omega) \ll 1$. Thus,
\be
\label{scaleCLO}
{C_{\rm LO}\over b} {m\hbar} = {\rm const} . 
\ee
This relation implies $C_{\rm LO}\propto (\hbar\omega)^{-1/2}$ and is
the oscillator-EFT equivalent of the well-known relation $C_{\rm
  LO}\propto (m\Lambda)^{-1}$ in the momentum-space formulation of
pionless EFT at infinite scattering length (or zero-energy bound
states).

\begin{figure}[htb]
\includegraphics[width=0.48\textwidth]{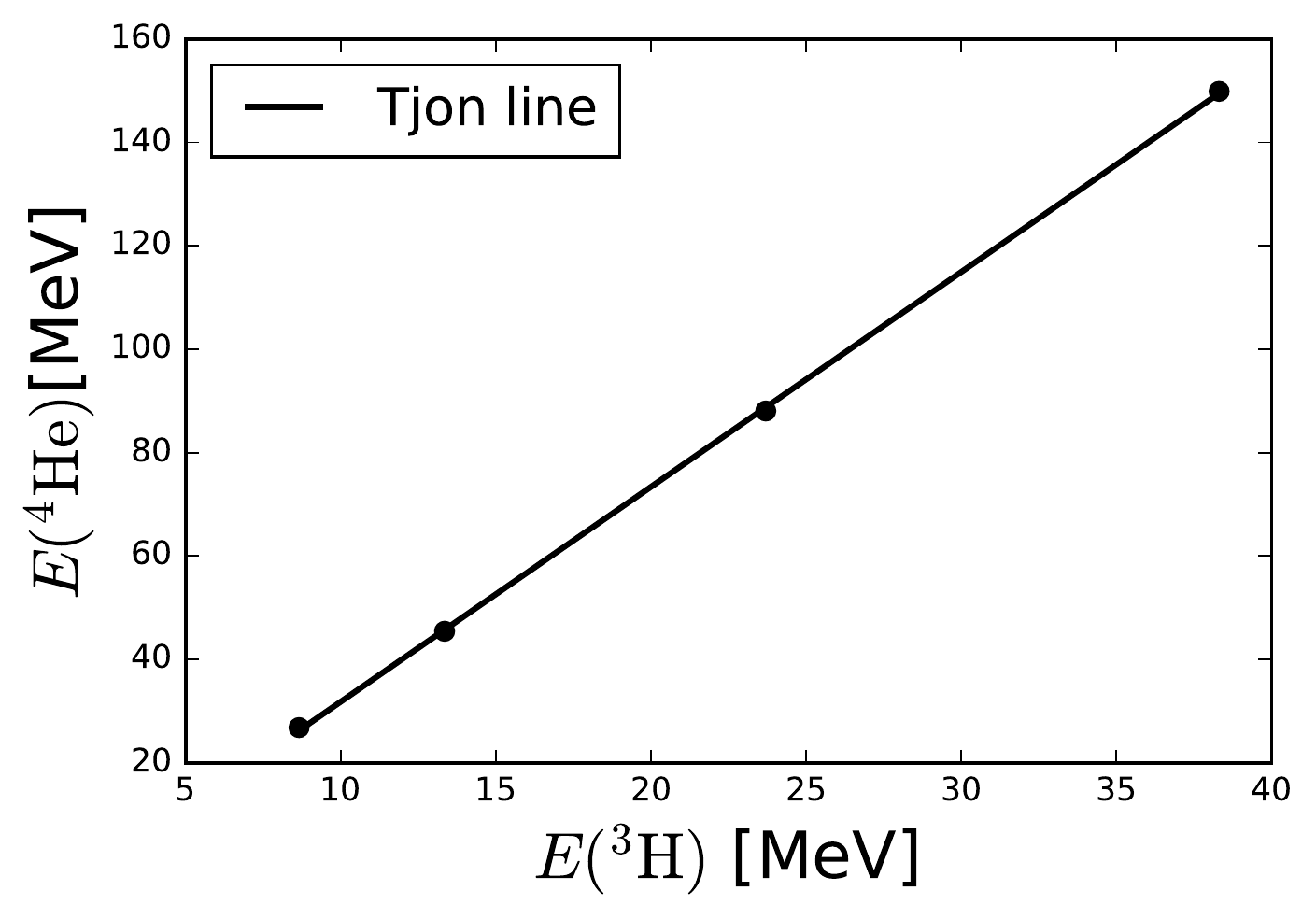}
\caption{Correlation between the triton and ${^4}$He
  binding energies, computed in LO with $NN$ interactions from
  pionless EFT. Different points correspond to different UV cutoffs.}
\label{tjon_line}
\end{figure}

Let us now consider the Schr\"odinger equation for the $A$-body
system, based on $NN$ interactions at LO. We find similar to
Eq.~(\ref{schrodinger2}) that
\ba
\label{schrodingerA}
E_A|\psi\rangle &=&
{\hbar^2\over m b^2}\left(\hat{t}_A + {{m\hbar}} {C_{\rm LO}\over b} \hat{v}_A\right)|\psi\rangle \nonumber\\
        &=& {\hbar^2\over m b^2}\hat{h}_A|\psi\rangle .
\ea
Here, $\hat{t}_A$, and $\hat{v}_A$ are the dimensionless matrices for
the kinetic and potential energy in the $A$-body system,
respectively. These quantities do not depend on the oscillator
length. We note that $\hat{h}_A$ is a dimensionless matrix that is
independent of $b$ because of the scaling
relation~(\ref{scaleCLO}). Thus,
\be
\label{scaleA}
E_A\propto {\hbar^2\over m b^2} = \hbar\omega .
\ee
This scaling relation explains the Thomas effect~\cite{thomas1935}: the binding energy
of the $A=3$ system increases with decreasing range of the potential,
i.e., with increasing cutoff or increasing $\hbar\omega$. It also
explains the Tjon line~\cite{tjon1975}, i.e., the correlation between the binding
energies of the $A=3$ and $A=4$ nuclei. Of course, both effects led to
beautiful insights regarding the renormalization of the $A=3$ body
system via a three-body force~\cite{bedaque1999} and the Tjon line as
a generic property of systems with large scattering
lengths~\cite{platter2005}.  To illustrate our analytical insights we
use the results obtained for $NN$ potentials alone (see, e.g.,
Table~\ref{tab:A34}) and show the Tjon correlations, i.e., the
proportionality of the binding energies for $A=3$ nuclei and $^4$He in
Fig.~\ref{tjon_line}.

\section{Large UV cutoffs and the Wigner bound}
\label{app-wigner}

Based on \citeauthor{wigner1955}'s bound on the derivative of phase
shifts, \citeauthor{phillips1996} showed that the 
effective range $r_e$ of the potential obeys the inequality
\begin{equation} \label{Wigbound}
r_e \leq 2\left(R - \frac{R^{2}}{a} + \frac{R^{3}}{3a^{2}}\right) .
\end{equation}
Here, $R$ is the physical range of the potential, i.e., the radius
beyond which the potential is zero and $a$ is the scattering length. 
As the physical range scales as
$R\propto \Lambda^{-1}$ for interactions with a UV cutoff
$\Lambda$, it is clear that the effective range
expansion~(\ref{ere}) cannot be reproduced at sufficiently large UV
cutoffs. How does the EFT employed in this work reflect this behavior?

\begin{figure}[htb]
\includegraphics[width=0.48\textwidth]{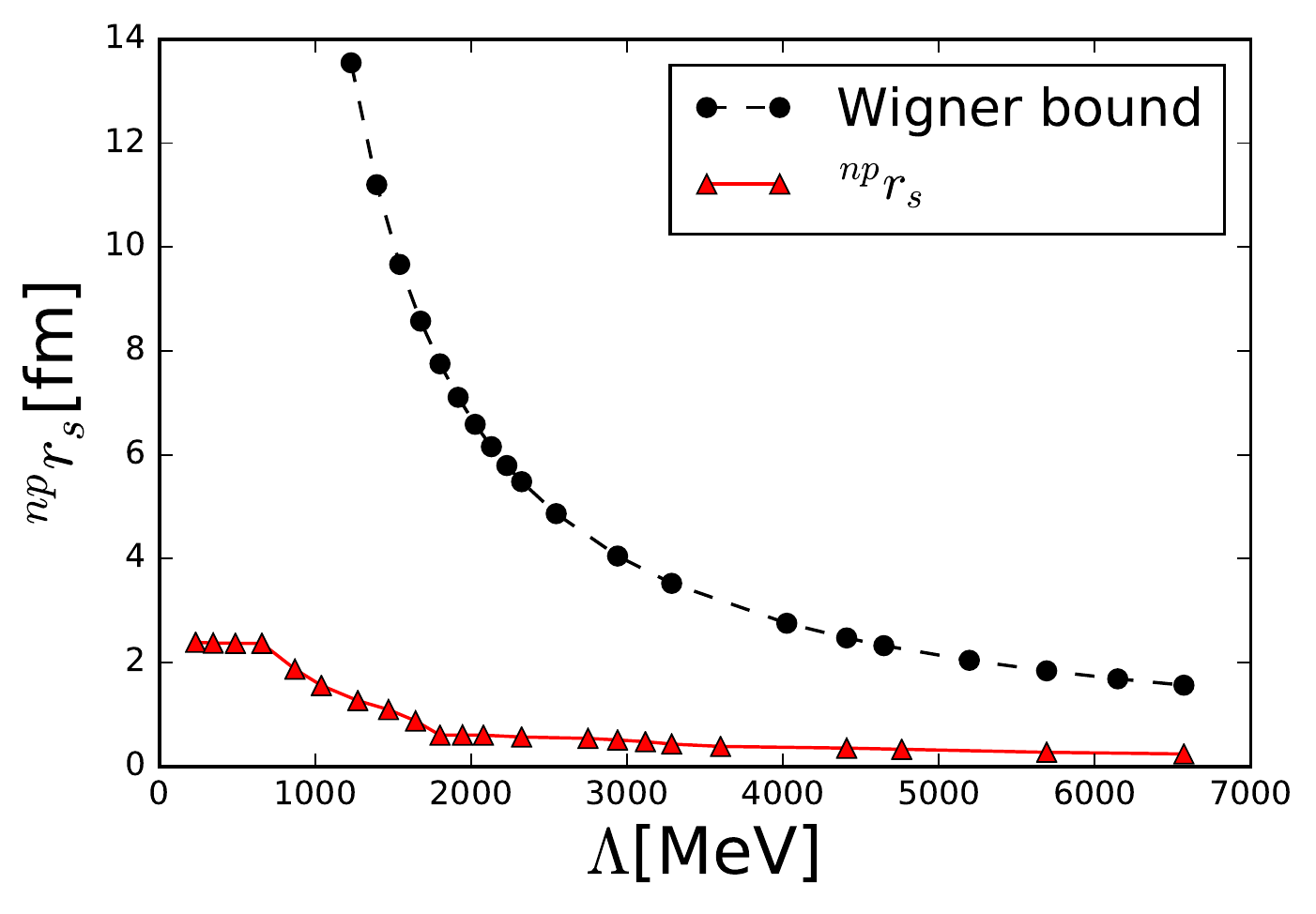}
\caption{(Color online) The effective range in the singlet $S$ wave (red line) as a
  function of the UV cutoff. The blue line shows Wigner's bound.}
\label{fig:Wignersbound}
\end{figure}

Figure~\ref{fig:Wignersbound} shows the effective range in the singlet
$S$ wave (red curve) obtained from a fit to the effective range
expansion~(\ref{ere}) for NLO interactions regularized in a finite harmonic oscillator
basis with $N=8$. The UV cutoff is increased by increasing the
oscillator spacing $\hbar\omega$. Beyond $\Lambda \approx
650$~MeV, we are unable to reproduce the effective range of the $NN$
interaction. The dashed black curve shows the Wigner bound, i.e., the equality
sign holds in Eq.~(\ref{Wigbound}). We see that our EFT obeys the
Wigner bound. We also note that the effective range seems to approach
zero for very large cutoffs. Negative effective ranges (as discussed
in Ref.~\cite{phillips1996}) are not realized in our EFT.

\section{Regulator effects}
\label{harmonic oscillatorEFT}

In the DVR implementation of pionless EFT, the UV
cutoff~(\ref{lambdaUV}) can be varied at fixed $N$ by changing the
oscillator frequency $\hbar\omega$. Strictly speaking the variation of
$\hbar\omega$ also changes the IR cutoff, but the IR improvement
essentially eliminates the effect of this variation on the potential.

In this Appendix we will consider different combinations of
$(N,\hbar\omega)$ that keep the UV cutoff constant and thus correspond
to different regulators. In an EFT, regulator dependencies are
expected to be higher-order effects.  Thus, we expect that IR improved
interactions with an identical UV cutoff but different
$(N,\hbar\omega)$ combinations should yield similar results for finite
nuclei.  How small can $N$ be chosen? Semiclassical arguments indicate
that the number $N$ should scale as $N\propto A^{1/3}$ so that all
nucleons are indeed interacting. But besides this, there seems to be
little to be gained by considering (unnecessary) large interaction
spaces.

\begin{figure}[htb]
\includegraphics[width=0.48\textwidth]{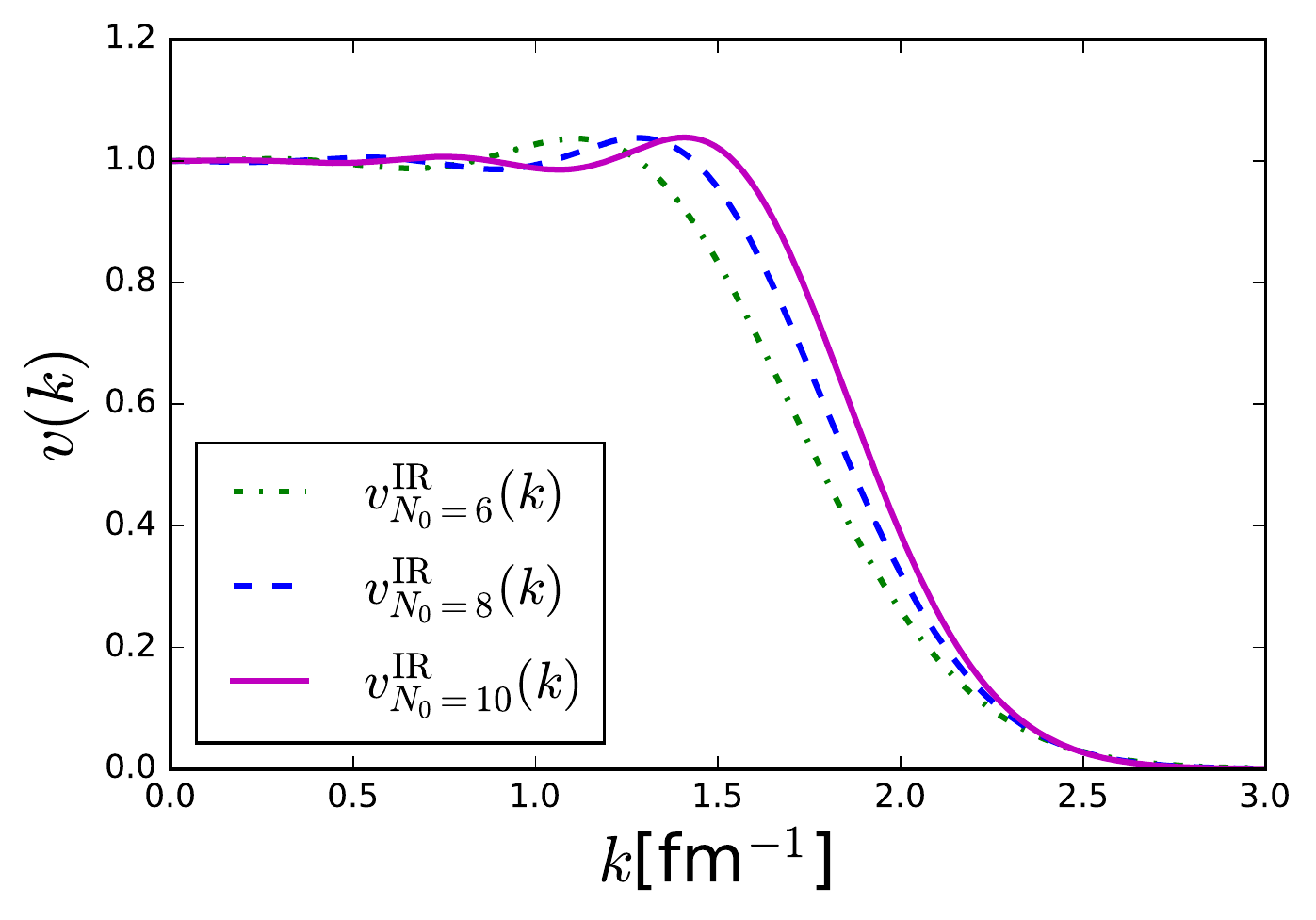}
\includegraphics[width=0.48\textwidth]{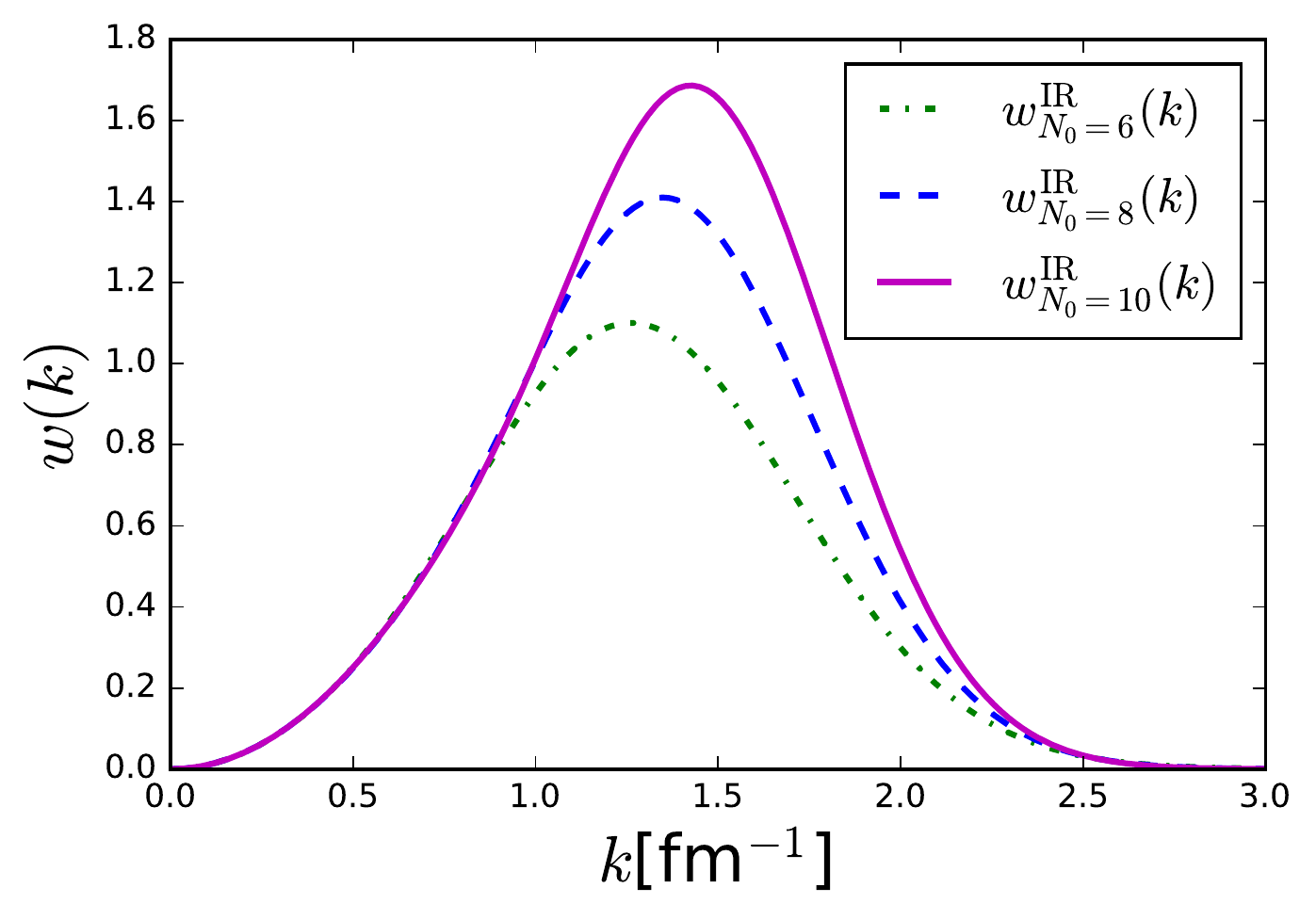}
\caption{(Color online) The dash-dotted green (solid magenta) curve
  shows the $NN$ interaction in model space $N = 6, \hbar\omega =
  26.63$~MeV, ($N = 10, \hbar\omega = 18.74$~MeV).  The dashed blue
  curve shows the same interactions in model space $N = 8, \hbar\omega
  = 22$~MeV. All three cases have a momentum cutoff $\Lambda \approx
  487$~MeV. }
\label{fig:differentNmax_interaction}
\end{figure}

To probe regulator dependencies, we consider model spaces with
combinations $N = 6$, $\hbar\omega = 26.63$~MeV, $N = 8$, $\hbar\omega
= 22$~MeV, and $N = 10$, $\hbar\omega = 18.74$~MeV; these have a
similar UV cutoff $\Lambda \approx 487$~MeV.
Figure~\ref{fig:differentNmax_interaction} shows that the IR-improved
potentials $v(k) = 1$ at LO and $w(k) = k^2$ at NLO are similar for
the different model spaces. Due to the IR improvement, the effective
UV cutoff decreases somewhat with decreasing $N$, but the differences
are small, particularly at low momenta. This suggests that the
different model spaces translate into small differences in the
effective regulator functions.

We fit the $NN$ potential at LO to the scattering lengths and the
deuteron binding energy. The resulting LECs are shown in
Table~\ref{tab:differentNmax_LOresults}. We note that the LECs exhibit
only a small dependency on the model space, in keeping with EFT
expectations that regulator dependencies at similar cutoffs are
higher-order effects.

\begin{table}[htb]
\begin{center}
\caption{The LECs of the $NN$ potential at LO for physical nuclei at
  constant $\Lambda \approx 487$~MeV and varying model space
  size.}
\begin{tabular}{|c|d|c|c|}
 \hline 
$N$ & \hbar\omega & $\Tilde{C}_{^3s_1}$ & $\Tilde{C}_{^1s_0}$  \\ \hline 
6   & 26.63 &$-0.407880$   & $-0.313361$ \\ 
8   & 22      &$-0.379465$   & $-0.296100$ \\ 
10 & 18.74 &$-0.360988$   & $-0.284491$\\ \hline
\end{tabular}
\label{tab:differentNmax_LOresults}
\end{center}
\end{table}%

We turn to the $NN$ interaction at NLO and employ the effective ranges
as additional constraints to determine the LECs.
Table~\ref{tab:differentNmax_NLO} shows the results.  Again we observe
a mild dependence of the model space, and this is again consistent
with EFT expectations that regulator dependencies are higher-order
effects.

\begin{table}[htb] 
\begin{center}
\caption{ $NN$ LECs at NLO  for physical nuclei at
  constant $\Lambda \approx 487$~MeV and varying model space
  size.}
\begin{tabular}{|c|d|c|c|c|c|} 
 \hline 
$N$ & \hbar\omega &  $\Tilde{C}_{^3s_1}$ &$C_{^3s_1}$ & $\Tilde{C}_{^1s_0}$ &$C_{^1s_0}$\\ \hline 
 6   & 26.63 & $-0.792415$ & $0.834806$ &$-0.571535$ &  $0.469715$  \\ 
 8   & 22      & $-0.809378$ & $0.772254$ & $-0.612966$ & $0.691221$ \\ 
 10 & 17.84 & $-0.798677$ & $0.693435$ & $-0.587451$ & $0.614043$ \\ \hline 
\end{tabular}
\label{tab:differentNmax_NLO}
\end{center}
\end{table}%

\begin{figure}[tbh]
\includegraphics[width=0.5\textwidth]{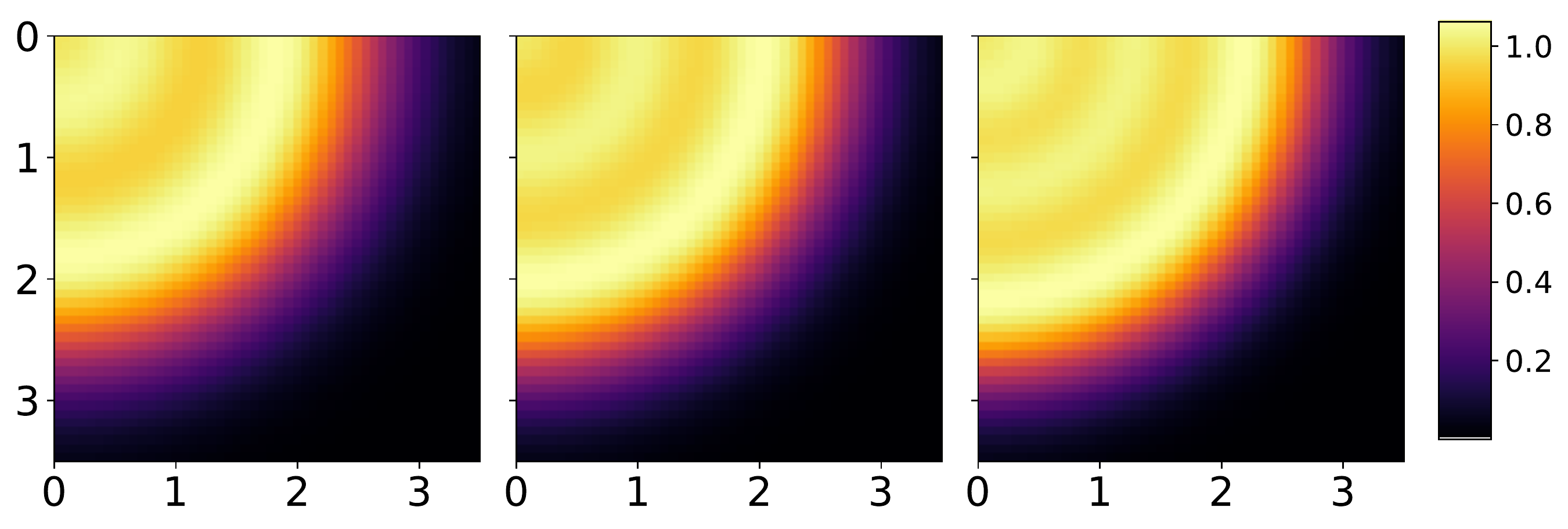}
\caption{(Color online) Momentum space matrix elements $\bar{u}(k,p) =
  c_{\mu,0}c_{\nu,0}u_{\rm DVR}^{\rm IR}\big{(}\sqrt{k^2 +
    p^2}\big{)}$ in harmonic oscillator basis with $N = 6$ (left), 8
  (center) and10 (right) with identical UV cutoff. The $x$ and $y$
  axes represent Jacobi momenta in fm$^{-1}$.}
\label{nnn_diffNmax}
\end{figure}

We turn to the $NNN$ contact. Figure~\ref{nnn_diffNmax} compares $NNN$
function $\bar{u}(k,p)$ regulated in hyperradial momentum for three
model spaces of interest.  All three interactions are quite similar,
particularly at low momenta.  We note that this observation also
extends to $NNN$ contact when regulated in each Jacobi momentum.

We include the $NNN$ contact and determine its LEC by adjusting to
the triton binding energy. We perform two independent computations of
the ground state energies and matter radii of $A = 3,4$ nuclei (at a
physical pion mass) from these $NN$ interactions at LO and NLO, and
present the results in Table~\ref{tab:differentNmax_NLOresults}. The
Coulomb interaction was included. 

\begin{table}[htb] 
\begin{center}
\caption{Binding energies and radii for $A = 3,4$ nuclei at constant
  $\Lambda \approx 487$~MeV and different model spaces employing a
  hyperspherical regulator for the $NNN$ contact. The $NNN$ coupling
  $c_E$ is adjusted to reproduce the triton binding energy $B_t =
  8.482$ MeV.}
\begin{tabular}{|c|c|c|c|c|c|c|} 
 \cline{1-7} \hline \hline
 \multicolumn{7}{|c|}{$E_{3{\rm max}} = N$  (triangular)} \\
 \cline{1-7}
\cline{1-7}
 \multicolumn{7}{|c|}{LO} \\
  \cline{1-7}
 $N$& $c_E$& $r(^3$H)  &  $E(^3$He) & $r(^3$He) & $E(^4$He) & $r(^4$He) \\ \hline 
 6  & $-0.269308$ & 1.30 & 7.55 & 1.47 & 18.28 & 1.45 \\ 
 8  & $-0.238514$ & 1.29 & 7.52 & 1.46 & 17.66 & 1.46  \\ 
10 & $-0.218702$ & 1.28 & 7.50 & 1.45 & 17.27 & 1.46  \\ 
\cline{1-7}
\multicolumn{7}{|c|}{NLO} \\
\cline{1-7}
 $N$& $c_E$ & $r(^3$H)  &  $E(^3$He) & $r(^3$He) & $E(^4$He) & $r(^4$He) \\ \hline 
 6  & $-0.073289$ & 1.58 & 7.71 & 1.77 & 28.39 & 1.36 \\ 
 8  & $-0.008170$ & 1.63 & 7.77 & 1.83 & 29.30 & 1.44  \\ 
10 & $-0.024851$ & 1.63 & 7.77 & 1.82 & 27.90 & 1.51  \\    \hline\hline
\end{tabular}
\label{tab:differentNmax_NLOresults}
\end{center}
\end{table}%

We also performed calculations where the $NNN$ interaction is
regulated in each of the Jacobi
momenta. Table~\ref{tab:differentNmax_NLOresults-sq} shows the
results. The comparison with Table~\ref{tab:differentNmax_NLOresults}
shows that regulator differences in the $NNN$ contact are small, as
expected in an EFT.

\begin{table}[htb] 
\begin{center}
  \caption{Same as Table~\ref{tab:differentNmax_NLOresults} but for
    regulators in each Jacobi coordinate of the $NNN$ force.}
    \begin{tabular}{|c|c|c|c|c|c|c|} 
\cline{1-7} \hline \hline 
\multicolumn{7}{|c|}{$E_{3{\rm max}}= 2N$ (square)} \\
\cline{1-7}
\cline{1-7}
\multicolumn{7}{|c|}{LO } \\
\cline{1-7}
 $N$& $c_E$ & $r(^3$H)  &  $E(^3$He) & $r(^3$He) & $E(^4$He) & $r(^4$He) \\ \hline 
 6  & $-0.224040$& 1.33 & 7.58 & 1.49 & $21.28$ & $1.51$ \\ 
 8  & $-0.191847$ & 1.32 & 7.56 & 1.48 & $20.82$ & 1.40  \\ 
10 & $-0.171713$ & 1.31 & 7.55 & 1.47 & $23.07$ &$1.38$  \\ 
\cline{1-7} 
\multicolumn{7}{|c|}{NLO} \\
\cline{1-7}
 $N$& $c_E$& $r(^3$H)  &  $E(^3$He) & $r(^3$He) & $E(^4$He) & $r(^4$He) \\ \hline 
 6  & $-0.059819$ & 1.58 & 7.71  & 1.78 & 28.34 & 1.40 \\ 
 8  & $-0.006553$ & 1.68 & 7.77  & 1.83 & 29.27 & 1.45  \\ 
10 & $-0.020162$ & 1.63 & 7.77 & 1.66 & 28.13 & 1.51  \\  \hline \hline
\end{tabular}
\label{tab:differentNmax_NLOresults-sq}
\end{center}
\end{table}%

\section{Effects of oscillator basis truncation on $NNN$ contact}
\label{trunc-E3max}
In this Appendix we discuss the effects of an oscillator basis
truncation where the $NNN$ interaction matrix elements of the
oscillator states with $n_1 + n_2 > N_3$ are set to zero. Here and
throughout this Appendix, $n_1$ and $n_2$ are principal harmonic
oscillator quantum numbers for a three-nucleon system in intrinsic
Jacobi coordinates.

\begin{figure}[htb]
\includegraphics[width=0.45\textwidth]{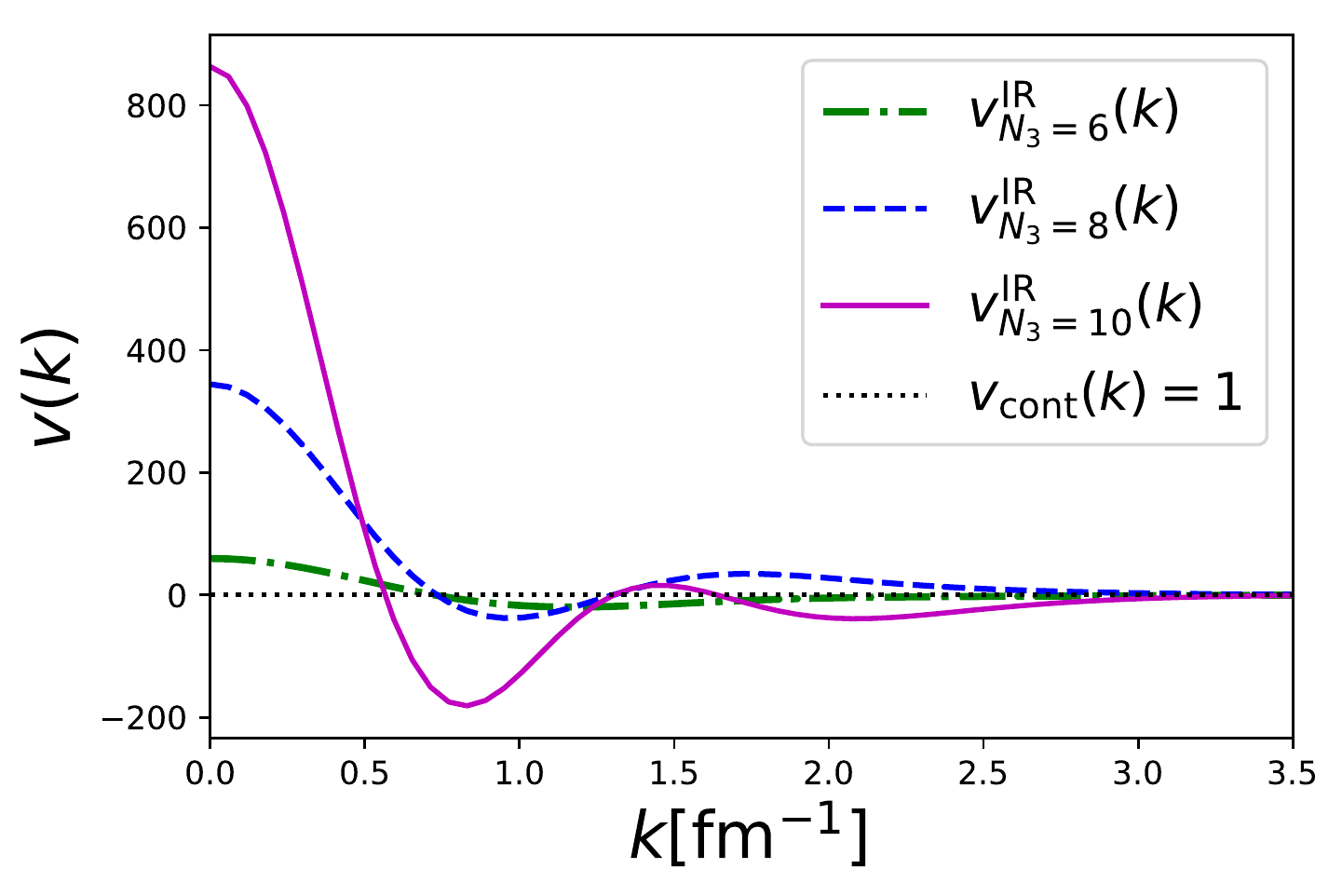}
\caption{(Color online) The $v_{\rm DVR}$ of $NNN$ interaction
  ~(\ref{3nfdvr}) reconstructed only from interaction matrix elements
  in remaining harmonic oscillator basis states after $n_1 + n_2 \leq
  N_3$ truncation.  Dashed-dotted green line: $NNN$ interaction in
  model space $N_3 = 6$ and $\hbar\omega = 26.63$~MeV; dashed blue
  line: $N_3 = 8$ and $\hbar\omega = 22$~MeV; solid magenta line:
  $N_3= 10$ and $\hbar\omega = 18.74$~MeV.  All three cases have the
  same momentum cutoff $\Lambda \approx 487$ MeV. The dotted black
  line shows the original momentum space interaction $v=1$. }
\label{3nf_truncated}
\end{figure}

The $NNN$ contact with the hyperspherical cutoff in
Eq.~(\ref{Vtr_3nf}) remains unaffected by this truncation for $N_3\ge
N$, because it fulfills $n_1 + n_2 \leq N$ by construction. This is
the key reason why we chose to work with the hyperspherical regulator
in this paper. On the other hand, the IR improved $NNN$
interaction~(\ref{3nfdvr}) with cutoff in Jacobi momenta is affected
by this truncation once $N_3 < 2N$.  As shown in
Fig.~\ref{3nf_truncated}, lowering $N_3$ below $2N=16$ significantly
modifies the $NNN$ contact with cutoff in Jacobi momenta after
truncation.  Here, $N_3 = 6, 8$, and 10 are shown by the dash-dotted
green, dashed blue, and solid magenta lines respectively.

Not surprisingly, in the truncated bases with $N_3< 2N=16$ the
ground-state energy of $^4$He exhibits a strong dependence on the
$N_3$ truncation. In an effort to reduce the number of matrix elements
of the $NNN$ force, we also employed the $NNN$ contact~(\ref{3nfdvr})
such that the interaction vanishes for $n_1,n_2 > N/2$. (This would
still keep $NNN$ excitations up to $N\hbar\omega$ in the potential.)
Choosing combinations of $N$ and $\hbar\omega$ that exhibit similar UV
cutoffs, we found that the $^4$He binding energy increases with
increasing $N$ for this truncation.

\section{IR extrapolations}
\label{app-extra}
The EFT formulation in the harmonic oscillator basis provides us with
a UV cutoff that is tailored to the model space, and this makes UV
extrapolations~\cite{konig2014} unnecessary. To overcome finite-volume
effects, one can employ IR extrapolations. The corresponding
extrapolation~\cite{furnstahl2012} formulas generalize L\"uscher's
approach~\cite{luscher1985} to the harmonic oscillator.

The EFT potential is defined in a model space of size $N$. For the
Hamiltonian matrix we choose $N_{\rm max}\ge N$ such that the
potential is active only between states with energy $E\le
N\hbar\omega$, while the kinetic energy is active in the full space,
i.e., in all states with energy $E\le N_{\rm max}\hbar\omega$. (Here,
we neglected the zero-point energy.) As $N_{\rm max}$ increases the
radius $L$ associated with the harmonic oscillator basis also
increases, and the tail of the bound-state wave function becomes
increasingly accurate. For energies, we have~\cite{furnstahl2012}
\be
\label{IR-E}
E(N_{\rm max})= E_\infty +a e^{-2k_\infty L}
\ee
as the leading correction for $k_\infty L\gg 1$. For the deuteron,
$k_\infty$ is the bound-state momentum~\cite{more2013} and 
$L$ is calculated using the Eq.~(\ref{L}). In general,
$k_\infty$ is the separation momentum of the lowest breakup
channel~\cite{konig2017,forssen2017}, i.e.
\be
S={\hbar^2k_\infty^2\over 2m}
\ee
is the separation energy of the lowest-lying breakup channel. This
suggests that the relevant small momentum scale $k_{\rm sep}$ might be
much larger than the low-momentum scales encountered in the deuteron
and in the effective range expansion of the nuclear force. A
separation energy of 8~MeV, for instance, corresponds to a separation
momentum of about 120~MeV.

\begin{figure}[htb]
\includegraphics[width=0.48\textwidth]{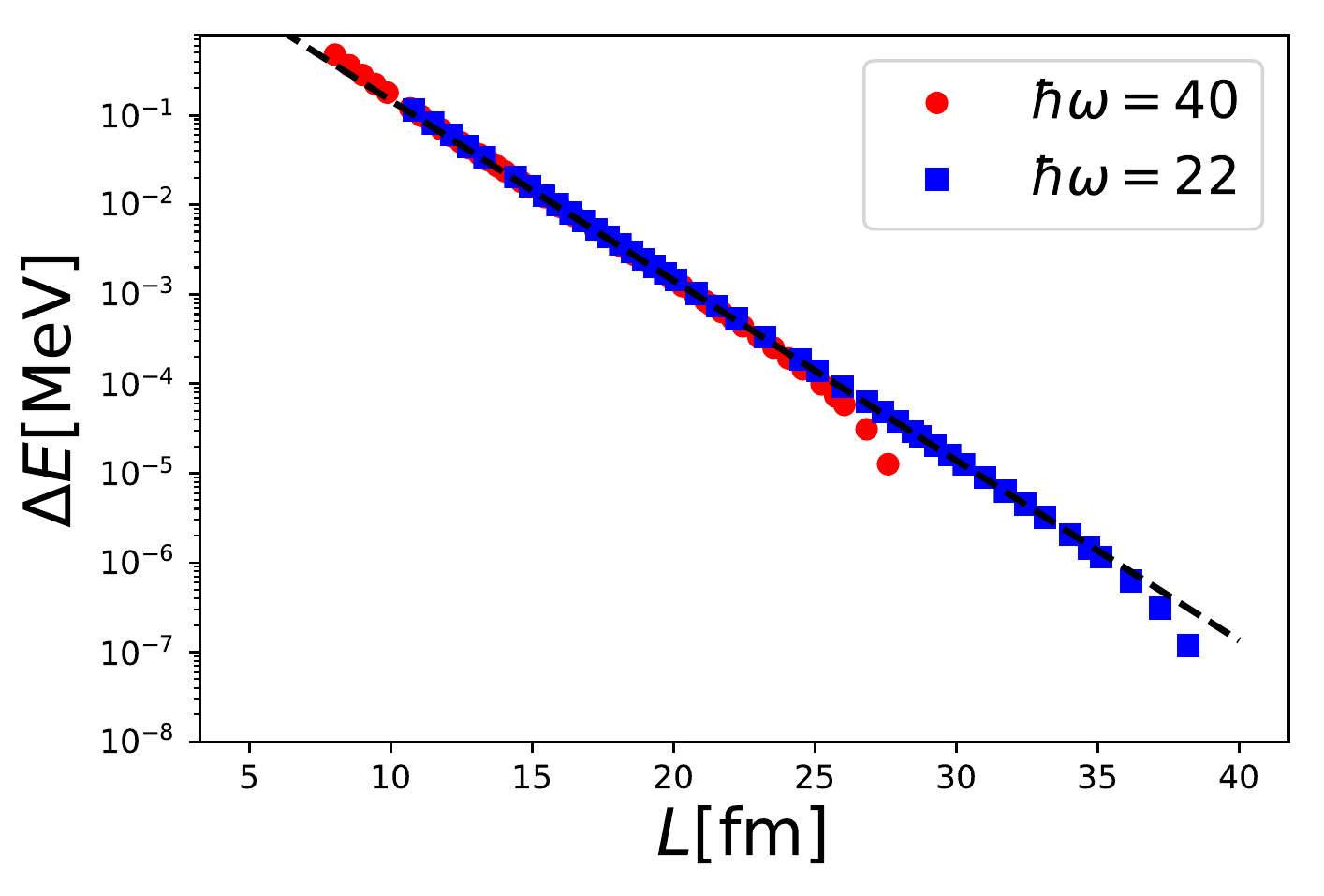}
\caption{Difference of the deuteron binding energy in a finite space
  of size $L$ and the infinite-space result for $\hbar\omega=40$~MeV
  (solid red dots) and 22~MeV (solid blue squares) for our NLO
  oscillator EFT potential in model space $N = 8$. The dashed black
  line shows $a\exp{(-2 k_\infty L)}$ where $k_\infty \equiv 0.2316$
  fm$^{-1}$ is the separation momentum.}
 \label{H2-kinf}
\end{figure}

Let us illustrate the extrapolation using the example of the deuteron
at NLO and in a model space $N=8$ for the
potential. Figuer~\ref{H2-kinf} shows that the energy difference
$\Delta E \equiv E(N_{\rm max}) - (E_\infty)_{\rm actual}$ converges
exponentially fast as a function of $L$. Solid red dots (solid blue
squares) $\hbar\omega = 40$~MeV (22~MeV), and the dashed black line is
the function $a\exp{(-2 k_\infty L)}$ with $a \approx 15$~MeV and the
separation momentum $0.2316$~fm$^{-1}$.  We note that the exponential
decay is indeed governed by the separation momentum and that the
equality of this momentum and $k_\infty$ is much more accurate here
than reported in Ref.~\cite{more2013}. The reason is presumably the
fully achieved UV convergence in the present approach.

\begin{figure}[htb]
  \includegraphics[width=0.45\textwidth]{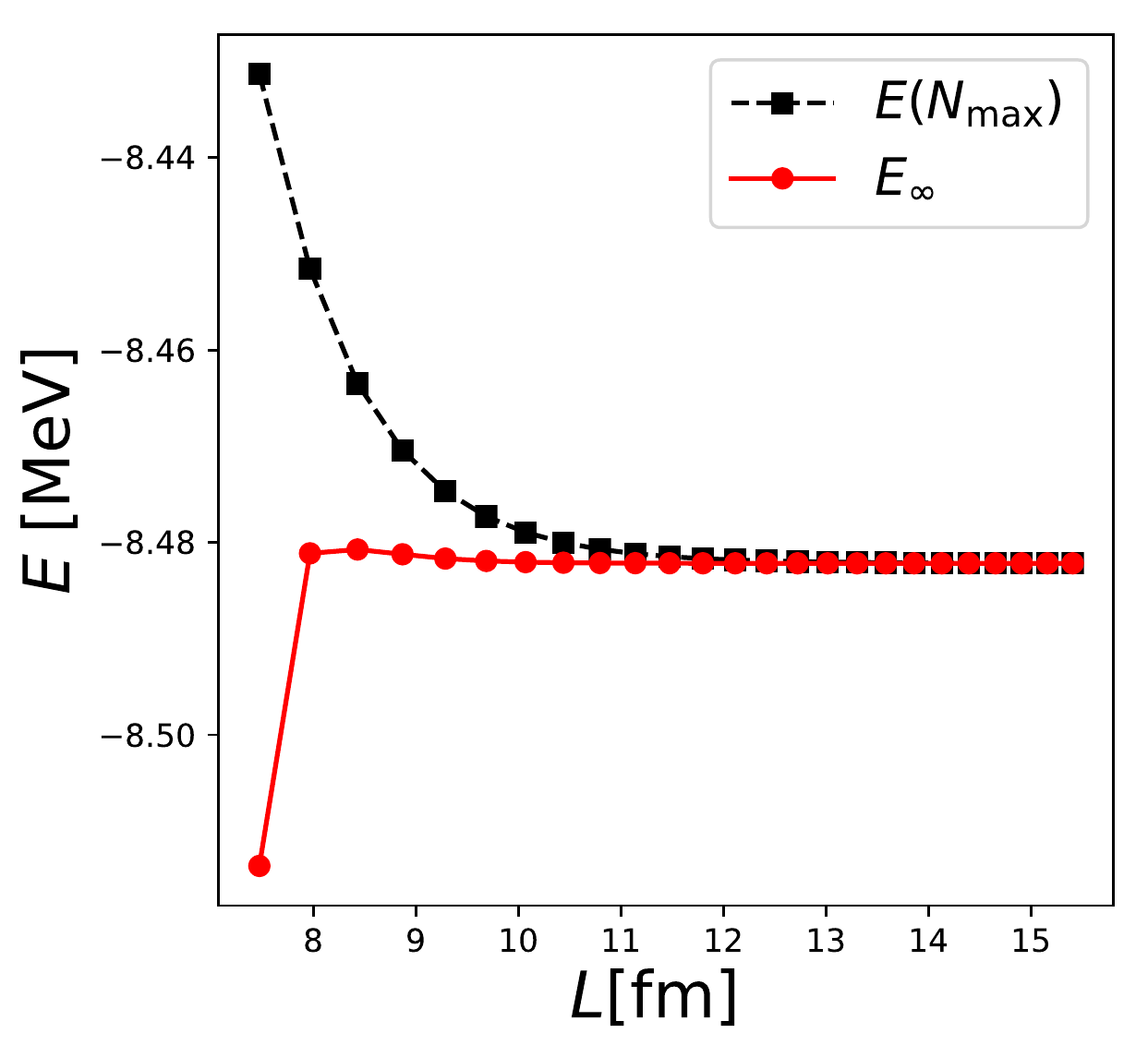}
  \caption{(Color online) Ground-state energy (black squares) of $^3$H
    computed in a model space of $N_{\rm max}+1$ shells and compared to the IR
    extrapolated result $E_\infty$ (red circles). }
  \label{3H-Einf}
\end{figure}

Though the no-core shell-model calculations for $A=3,4$ nuclei are
virtually converged with respect to the model space, it is still
useful to consider IR extrapolation. At low energies, the harmonic
oscillator is indistinguishable from a spherical cavity of radius
$L$. For the no-core shell model, the radius $L$ is a known function
of the number of shells $N$ and the frequency $\hbar\omega$ of the
employed basis~\cite{wendt2015}. The NLO calculation of $^3$H with an EFT
potential of $N =8$ and $\hbar\omega=22$~MeV. As the
formula~(\ref{IR-E}) depends on the three parameters
$(E_\infty,k_\infty,a)$, extrapolations start from three data points of
the ground-state energy $E(L) = E(N_{\rm max})$ computed in
$N_{\rm max}=8,10,12$. Figure~\ref{3H-Einf} compares $E(N_{\rm max})$ with the
extrapolation result $E_\infty$. From $N_{\rm max}=14$ and higher, the
extrapolated result is much more accurate than the finite-volume
result. 

\begin{figure}[htb]
  \includegraphics[width=0.45\textwidth]{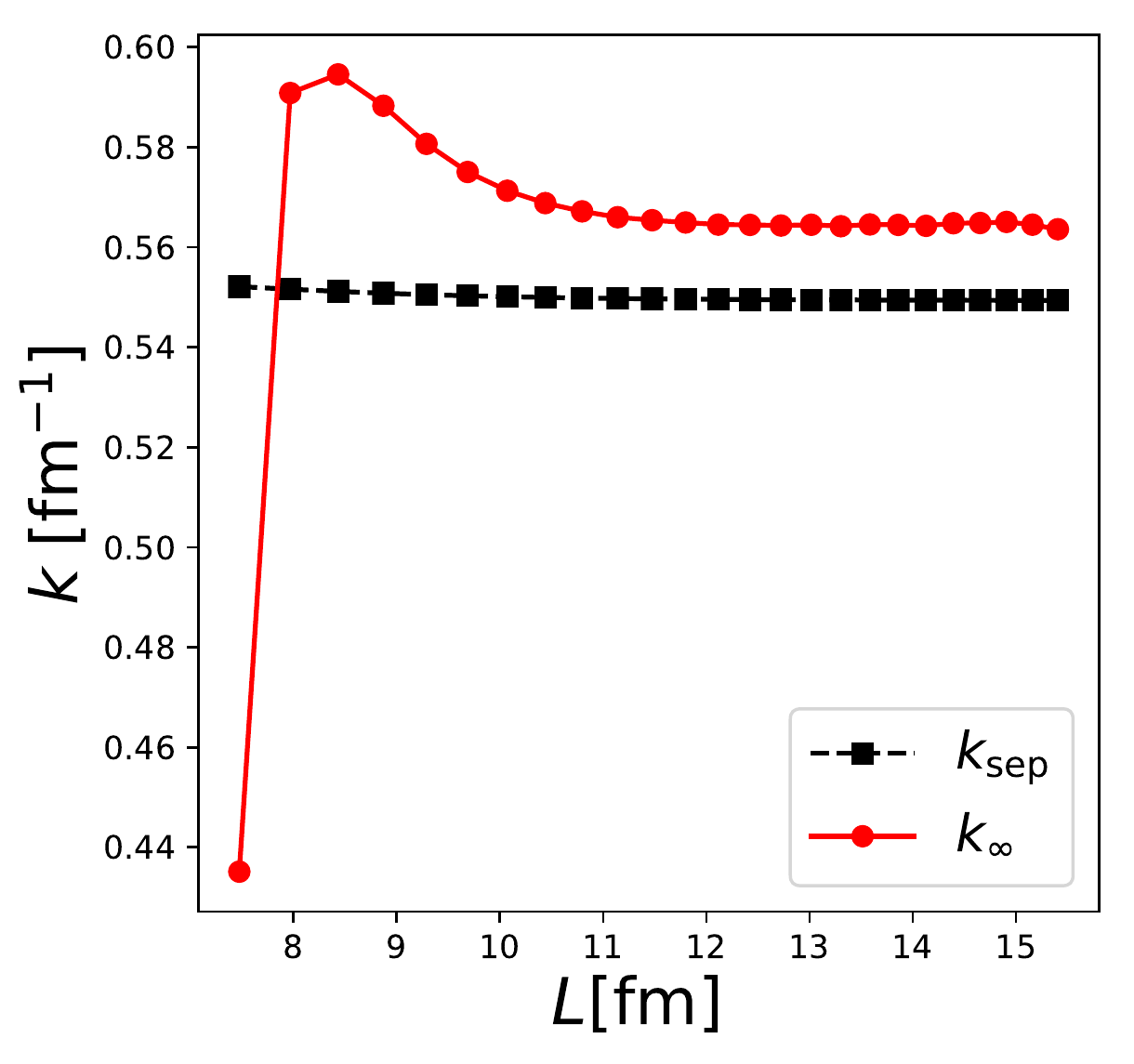}
  \caption{(Color online) Separation momentum (black squares) of $^3$H
    computed in a model space of $N_{\rm max}+1$ shells and compared
    to the IR extrapolated result $k_\infty$ (red circles). }
  \label{3H-kinf}
\end{figure}

For the triton, the lowest open decay channel is $t\to d+n$, with a separation momentum
fulfilling
\be
{\hbar^2k_{\rm sep}^2\over 2m} = B_t -B_d ,
\ee
where $B_t$ and $B_d$ are the binding energies of the triton and
deuteron, respectively. Figure~\ref{3H-kinf} compares the theoretical
value of $k_{\rm sep}$, computed from the theoretical energy
differences, with the results $k_\infty$ from the extrapolation. Both
quantities become close, but not identical, as the model space is
increased. We do not completely understand the reason for the
difference between $k_{\rm sep}$ and $k_\infty$. However, at LO and
using $NN$ forces only, the triton is strongly bound, and the
agreement between $k_\infty$ and $k_{\rm sep}$ is much better.
 
\begin{figure}[htb]
  \includegraphics[width=0.45\textwidth]{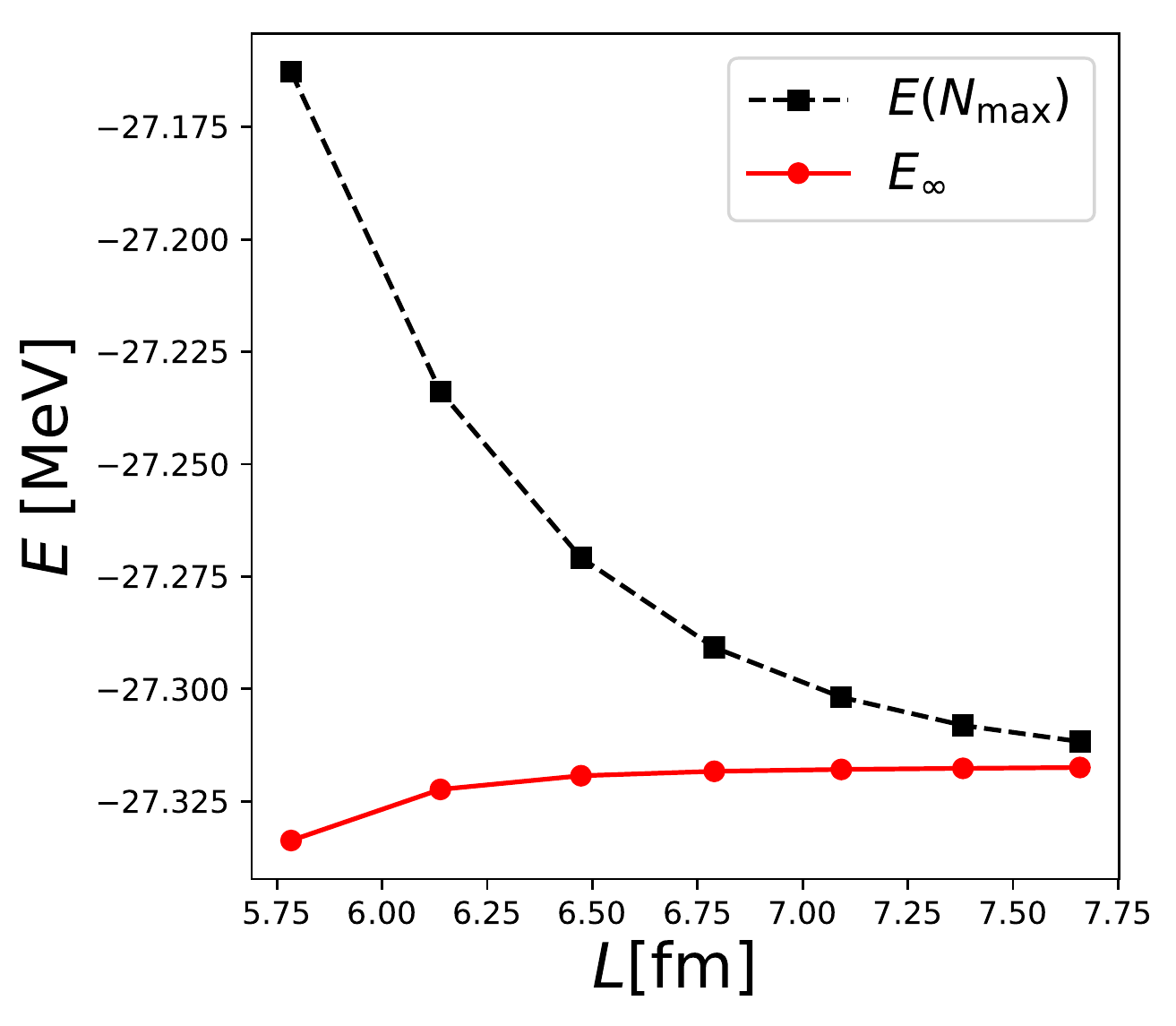}
  \caption{(Color online) Ground-state energy (black squares) of
    $^4$He computed in a model space of $N_{\rm max}+1$ shells and compared to
    the IR extrapolated result $E_\infty$ (red circles). }
  \label{4He-Einf}
\end{figure}

We turn to $^4$He, where the lowest-energetic breakup channel is
$\alpha\to t+p$. We consider the case of the NLO calculation with a
potential defined in $N=8$ and
$\hbar\omega=40$~MeV. Figure~\ref{4He-Einf} shows the convergence of
the energy as the model space is increased and compares it to the
extrapolated result.

\begin{figure}[htb]
  \includegraphics[width=0.45\textwidth]{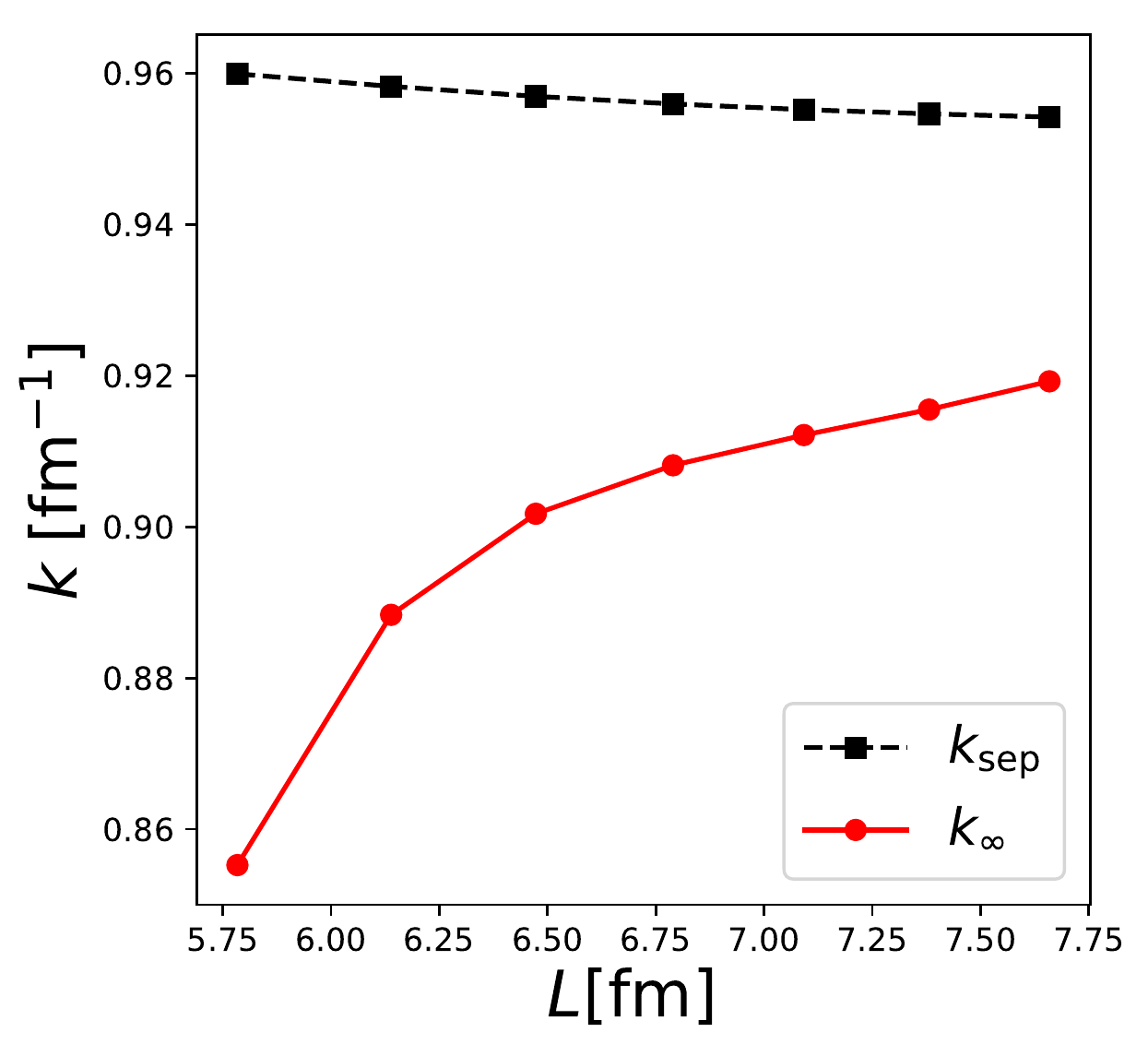}
  \caption{(Color online) Separation momentum (black squares) of
    $^4$He computed in a model space of $N_{\rm max}+1$ shells and compared to
    the IR extrapolated result $k_\infty$ (red circles). }
  \label{4He-kinf}
  \end{figure}

For this case, we can also compare the value of the extrapolated
momentum $k_\infty$ with that of the corresponding separation
momentum. The results are shown in Fig.~\ref{4He-kinf}.  Here, the
extrapolated $k_\infty$ is somewhat smaller than the separation
momentum $k_{\rm sep}$, but the results are not yet converged as the
model space is increased.

Overall, the results of this Appendix show that the IR extrapolations
of the EFT realized as a DVR in the harmonic oscillator basis work
quite well and agree with expectations.


\clearpage 

\end{document}